\documentclass[a4paper,fleqn,usenatbib]{mnras}

\usepackage{newtxtext,newtxmath}
\usepackage{subfloat} 
\usepackage{subcaption}		
\usepackage[T1]{fontenc}
\usepackage{ae,aecompl}

\usepackage{graphicx}	
\usepackage{amsmath}	
\usepackage{amssymb}	
\usepackage{listings}
\usepackage[dvipsnames]{xcolor}

\title[Hypercompact stellar clusters]{Hypercompact stellar clusters: morphological renditions and spectro-photometric models}

\author[Davide Lena]{
D. Lena,$^{1,2}$\thanks{E-mail: d.lena@sron.nl}
P. G. Jonker$^{1,2}$,
J. P. Rauer,$^{1,2}$
S. Hernandez$^{3}$  
\newauthor
and Z. Kostrzewa-Rutkowska$^{4,1,2}$
\\
$^{1}$SRON, Netherlands Institute for Space Research, Sorbonnelaan 2, NL-3584 CA Utrecht, the Netherlands\\
$^{2}$Department of Astrophysics/IMAPP, Radboud University, PO Box 9010, NL-6500 GL Nijmegen, the Netherlands\\
$^{3}$Space Telescope Science Institute, 3700 San Martin Drive, Baltimore, MD 21218, USA\\
$^{4}$Leiden Observatory, Leiden University, PO Box 9513, NL-2300 RA Leiden, the Netherlands
}

\date{Accepted for publication on MNRAS}

\pubyear{2020}

\begin{document}
\label{firstpage}
\pagerange{\pageref{firstpage}--\pageref{lastpage}}
\maketitle

\begin{abstract}
Numerical relativity predicts that the coalescence of a black hole-binary causes the newly formed black hole to recoil, and evidence for such recoils has been found in the gravitational waves observed during the merger of stellar-mass black holes. Recoiling (super)massive black holes are expected to reside in hypercompact stellar clusters (HCSCs). Simulations of galaxy assembly predict that hundreds of HCSCs should be present in the halo of a Milky Way-type galaxy, and a fraction of those around the Milky Way should have magnitudes within the sensitivity limit of existing surveys. However, recoiling black holes and their HCSCs are still waiting to be securely identified. With the goal of enabling searches through recent and forthcoming databases, we improve over existing literature to produce realistic renditions of HCSCs bound to black holes with a mass of 10$^{5}$ M$_{\odot}$. Including the effects of a population of blue stragglers, we simulate their appearance in Pan-STARRS and in forthcoming \textit{Euclid} images. We also derive broad-band spectra and the corresponding multi-wavelength colours, finding that the great majority of the simulated HCSCs fall on the colour-colour loci defined by stars and galaxies, with their spectra resembling those of giant K-type stars. We discuss the clusters properties, search strategies, and possible interlopers. 
\end{abstract}




\section{Introduction}
General relativity predicts that a burst of gravitational waves (GW) is emitted during the coalescence of two compact objects \citep[e.g.][]{Fitchett83,RedmountR89,Wiseman92}. The first direct observation of such an event took place on 2015 September 14, marking the advent of a new era. \citet{AbbottAA16} inferred that the emission originated from the merger of two black holes with masses $29_{-4}^{+4}$ and $36_{-4}^{+5}$ M$_{\odot}$. 

Asymmetries in the merging objects (i.e. different masses and spins) are expected to produce asymmetries also in the GW emission, leading to a net flux of linear momentum. To conserve linear momentum, the merging binary and the resulting object recoil accordingly. They get a ``kick''. The amplitude of such kick is largest for black hole (BH) binaries, and it must be computed numerically. When this became possible, the expected amplitudes of order 10$^2$ km s$^{-1}$ were confirmed \citep{Pre05,BakerCC06,camp06,Pretorius06}. Moreover, much larger values, as high as 5000 km s$^{-1}$ for special configurations, were also obtained \citep[e.g.][]{CampA,TM07,BrugmannGH08,RezzollaDR08,LZ11,LoustoZ13}. 
Via the modelling of GW waveforms, recoil velocities have been estimated for GW170104 \citep[][]{HealyLO18}, and for GW150914 \citep[][]{HealyLL19,LoustoH19}, where the inferred kick attains $-1500$ km s$^{-1}$.

As the escape velocities from the most massive galaxies are estimated to be below 3000 km s$^{-1}$ \citep[e.g.][]{Mer04}, the predictions outlined above attracted a great deal of interest: merging galaxies can bring two supermassive black holes (SMBHs) together \citep[e.g.][]{BBR80}, and the resulting SMBH could experience a recoil large enough to be appreciably displaced from the center of the host galaxy, or even ejected in intergalactic space. The first scenario, where the black hole (BH) receives a moderate kick and is not ejected from its host galaxy, is the one expected to be the most common: the recoiling BH would oscillate about the galaxy center loosing energy via dynamical friction at each passage through the core of an early-type galaxy \citep[e.g.][]{gualam}, or through the disk and bulge of a late-type galaxy \citep[e.g.][]{Korn08,BlechaL08}; bursts of accretion would follow the passage through a gas-rich disk \citep[e.g.][]{BlechaL08}. The second scenario, where recoiling velocities are high enough to remove a SMBH from its host galaxy, is much more unlikely than the first: attaining recoiling velocities in excess of a few hundreds km s$^{-1}$ requires special configurations for the BH-binary \citep[e.g.][]{LZ11,LoustoHangUp}. For instance, gas accretion onto the merging SMBHs could align their individual spins with the binary orbital angular momentum, heavily hampering the recoil velocity, and producing kicks which are, for the most part, below 100 km s$^{-1}$ \citep[e.g.][]{BogdanovicRM07,Dotti2010}. 

Depending on the environment where it is born, the recoiling BH carries along a mixture of gas and stars, with the amount of matter bound to the recoiling BH being inversely proportional to the kick velocity, V$_{k}$. When the recoiling BH originates in a gas-rich environment (the ``wet'' merger scenario), for example from a binary embedded in a gaseous disk, then, assuming a disk with a mass much smaller than the binary, the recoiling BH is expected to carry with it a punctured disk with outer radius $r\propto V_{k}^{-2}$, and mass $M_{\mathrm{disk}} \propto V_{k}^{-2.8}$. This gas will be accreted within a few million years \citep[e.g.][]{Loeb07, Bon07}.

When gas is not present (the ``dry'' merger scenario), then the recoiling SMBH will still carry with it a retinue of stars: those located within a distance $r_{k} \equiv GM_{\bullet}/V_{k}^{2}$ from the SMBH will remain bound after the kick, and the predicted stellar mass of the cluster is M$_{\star} \propto V_{k}^{-2(3-\gamma)}$, with $\gamma$ the slope of the stellar density distribution before the kick \citep{KomossaM08, MSK2009,OLearyL09}. The effective radius for these clusters is predicted to depend on a number of parameters: the central velocity dispersion of the host galaxy, the stellar density distribution prior to the kick, and the dynamical status of the nucleus (collisional or collisionless, \citealt{MSK2009}); while the largest clusters could extend as much as a few tens of parsecs (similarly to globular clusters and ultra-compact dwarf galaxies), the great majority is predicted to have sizes below 1 pc (hence the appellative "hypercompact") and velocity dispersion in excess of a few tens of km s$^{-1}$ \citep{MSK2009, OLearyL09}, much larger than the typical velocity dispersion of globular clusters \citep[approximately 10 km s$^{-1}$, e.g.][]{PryorM93}.

The observed velocity dispersion of an hypercompact stellar cluster (HCSC) is of primary importance: as simulations predict a simple proportionality with the kick velocity ($\sigma_{obs} \approx V_{k}/3.3$, \citealt{MSK2009}), it is clear that a population of HCSCs observed in the halo of a galaxy would open the door to a direct determination of the kick velocity distribution, therefore constraining the merger history of the host galaxy, the models of galaxy-assembly, the simulations of merging BHs and the assumptions upon which they rest. From a determination of M$_{\star}$ and V$_{k}$ one could also derive $\gamma$, gaining insights on the distribution of stars in the nucleus at the time of the merger.

The predicted number of HCSCs bound to the Milky Way (MW) ranges from a few tens to a few thousands:
\citet{OLearyL09} estimated that a MW-like galaxy which undergoes a hierarchical assembly, with no major mergers since redshift $z=1$, should retain in its halo hundreds of HCSCs bound to BHs with masses in the range $10^{3} \leq \mathrm{M}_{\bullet} \leq 10^{5}$ M$_{\odot}$, and tens of clusters bound to BHs with M$_{\bullet} \gtrsim 10^{5}$ M$_{\odot}$. These HCSCs were ejected from the shallow potential-well of the building blocks of the main galaxy, and they were trapped in the region which collapsed to make the MW-like galaxy. 
Later, \citet{RashkovM14} used the cosmological simulation \textit{Via Lactea II} \citep{DiemandKM08} to predict the properties of a population of relic intermediate-mass BHs (IMBHs) in the halo of a MW-type galaxy. These too are leftovers of the galaxy hierarchical assembly. They identified a population of ``naked'' IMBHs (their sub-haloes were destroyed during infall) and ``clothed" IMBHs (residing in the nuclei of stripped galaxies). The naked BHs make up 40 to 50 per cent of the total population and are mostly located within 50 kpc of the halo center, where tidal stripping is more effective. These BHs are also associated with compact stellar clusters, not necessarily bound to recoiling BHs, but simple residuals of a stripped nuclear star-cluster. The total number of the relic population ranges between 70 and 2000, depending on the BH seeding scenario and the steepness of the M$_{\bullet}$-$\sigma$ relation \citep{FerrareseM2000,GBBetAl00} adopted to populate the nuclei. They would be spatially-resolved and with apparent magnitude as bright as $m_{V} \approx 16$, in the scenario producing the most massive IMBHs. 

Considering the challenges that an all-sky survey would imply to search for HCSCs in our own galaxy, \citet{MSK2009} produced a prediction for nearby galaxy clusters, where the limited angular extension would ease the task: they estimated that Virgo should contain one HCSC with apparent magnitude $K \leq 20$ and up to 150 with $K \leq 26$.

Observational evidence for the existence of HCSCs and recoiled SMBHs is scant. \citet{OLearyL12} mined SDSS DR7 finding 100 HCSC candidates with photometric properties consistent with theoretical models, however, to date none of these has been confirmed as a genuine HCSC. \citet{PostmanLD12} argued that the SMBH of the Abell 2261 brightest cluster galaxy was ejected via gravitational recoil, and they identified four HCSC candidates located in the vicinity of the galactic nucleus - possible carriers of the putative ejected SMBH; later, \citet{BurkeSpolaorGP17} showed that two of them are consistent with either dwarf galaxies or stripped nuclei, while the nature of the other two remains poorly constrained. 
The puzzling object HVGC-1 was tentatively interpreted as a hypervelocity globular cluster ejected from the Virgo cluster in the aftermath of a three-body interaction; the HCSC scenario was dismissed because of the low metallicity and the low velocity dispersion ($\sigma_{\star} \leq 80$ km s$^{-1}$, too low when compared with the putative kick velocity, V$_k \geq 1025$ km s$^{-1}$, \citealt{CaldwellSR14}). \citet{BoubertSA19} considered the possibility that the fastest star in \textit{Gaia} DR2 is bound to a recoiling IMBH, but they favoured an interpretation where the measurement is spurious.
Several other sources have been proposed as candidate recoiling SMBHs (presumably clothed with an HCSC), but alternative interpretations remain often equally viable and difficult to rule out conclusively \citep[e.g.][]{Bon07,KZL08,SRS09,batch,Andy10,JonkerTF10,Civano10,Tsal11,Erac11,KossBM14,lenaRMA14,MenezesSR14,MarkakisDE15,ChiabergeEM17,MakarovFB17,KalfountzouST17,LopezNP18}.

With the aim of facilitating the task of their identification, we present spectroscopic and photometric renditions of HCSCs bound to recoiling BHs with a mass of $10^{5}$ M$_{\odot}$. Photometric renditions are built upon the dynamical simulations of \citet{MSK2009} and \citet{OLearyL12}. The effects of a population of blue stragglers are accounted for, in both photometry and spectroscopy. Methods and results are presented in Sec.\ref{sec: methods_results}, where we provide details on spectroscopic simulations, on the derivation of colours for a number of publicly available datasets, and on the rendition of the cluster morphology, which includes the effects of kick velocity and dynamical ageing. In Sec.\ref{sec_discussion} we discuss the derived properties of HCSCs, along with search strategies and challenges in their identification. We sum up and conclude in Sec.\ref{sec_conclusions}.

Through the paper we assume the cosmological parameters H$_{0} = 69.6$, $\Omega_\mathrm{M} = 0.286$, and $\Omega_{\mathrm{vac}} = 0.714$.

\section{Methods and results}
\label{sec: methods_results}
In this section we provide details on the methods and tools used to simulate spectra, colours, and morphology of HCSCs. Results are also shown. 

\subsection{Synthetic spectra}  \label{sec_spectral}
We simulated a set of hypothetical HCSC spectra for clusters with about 10000 stars - the number of stars expected for a cluster bound to a $10^{5}$ M$_{\odot}$ black hole ejected at low velocity (v$_{k}$ = 150 km s$^{-1}$), and relatively young (time since the kick $\tau_{k} < 10^{7}$ yr). 
For each cluster we assumed a single stellar population and a single metallicity. However, the grid of models presented here allowed to explore the effects of different metallicities and ages of the stellar population on the cluster properties.

To simulate the integrated spectra of HCSCs we generated atmospheric models for the individual stars which make up the cluster, we derived the corresponding spectra via a spectral synthesis software, and we co-added the individual spectra to produce an integrated spectrum for the HCSC as a whole. Additional details on the steps of this process are given below.

As a starting point, we created a Hertzsprung-Russell diagram to represent every evolutionary stage present in the chosen stellar population. These diagrams were produced using the web interface \textsc{cmd}\footnote{http://stev.oapd.inaf.it/cmd} v3.3 in conjunction with the ``PAdova and TRieste Stellar Evolutionary Code'' \citep[\textsc{parsec} v1.2S,][]{BressanMG12,TangBR14,ChenBG15}; the evolutionary tracks provided us with the physical parameters for each of the stars present in the star cluster. In the \textsc{cmd} interface we chose the YBC version of the bolometric correction \citep{ChenGF19}, and we included the effects of circumstellar dust adopting the following composition: 60 per cent silicate plus 40 per cent Aluminum Oxide for M stars, and 85 per cent Amorphous Carbon plus 15 per cent Silicon Carbide for C stars \citep{Groenewegen06}.

To generate the isochrones we adopted a Kroupa Initial Mass Function following a two-part power law \citep{kroupa2001,kroupa02} and we generated a grid of models with metallicities Z = 0.0002, 0.002, 0.02 (solar), 0.03, 0.07, and ages of the stellar population $\tau_{\star} = $ 1, 7, and 13 Gyr. 
To extract stellar parameters we inverted the cumulative mass function generated via \textsc{cmd}, extracting randomly the parameters until we reached the expected total mass bound to the recoiling BH. 
The stellar parameters were then used to create a series of atmospheric models using \textsc{atlas9} \citep{Kurucz70}. \textsc{atlas9} is a local thermodynamic equilibrium one-dimensional plane-parallel atmospheric modeling software which uses opacity distribution functions to reduce the computational time. These \textsc{atlas9} atmospheres are used to generate synthetic spectra for each stellar evolutionary stage using \textsc{synthe}, a suite of programs requiring input model atmospheres, chemical abundances, and a list of atomic and molecular species \citep{KuruczF79,KuruczA81}; we adopted the atomic and molecular lines lists provided in the Castelli website\footnote{http://wwwuser.oats.inaf.it/castelli/}. Finally, individual stellar spectra were co-added to create a synthetic integrated-light spectrum for each of the star clusters of a given age and metallicity. The synthetic spectra were created with a wavelength coverage of 3000--24000 \AA\; at high resolutions (R$\sim 500,000$), and then degraded to the desired velocity dispersions.

Blue stragglers were also included in the stellar population of the cluster with the following approach: 
for each isochrone of a given age, we compiled a separate set of theoretical models with ages in the range of 10-90 per cent the age of the isochrone in question. We then randomly extracted stars with masses 1 - 2 times the mass of the main sequence turn-off, with the number of blue stragglers satisfying the relation by \citet{XinDG11} for Galactic open clusters:

\begin{align}\label{eq: nbs}
N_{BS} = (0.114 \pm 0.006) N_{2} - (1.549 \pm 0.731),
\end{align}

\noindent where $N_{2}$ is the number of stars within two magnitudes below the main sequence turn-off. With this approach the fraction of blue stragglers is approximately 0.3, 1, and 2 per cent in clusters with a stellar population of age $\tau_{\star}$ = 1, 7, and 13 Gyr, respectively. Here we note that this approach naturally yields a population of yellow and red stragglers, i.e. evolved blue stragglers which left the main sequence \citep[e.g.][]{Kaluzny03,LeinerMS16}. 

The resulting spectra are shown in Fig.\ref{fig: hcss_stellar_library} for a range of metallicities and ages, along with the best-matching spectra from the Pickles Atlas \citep{Pickles98}. 

\subsection{Colours} \label{sec_colours}
Using instrument-specific transmission curves and the synthetic spectra described in Sec.\ref{sec_spectral}, we derived the expected colours, in the different filter bands, for a number of publicly available datasets, namely SDSS \citep{YorkAA00,BlantonBA17}, KIDS \citep{deJongKK13,deJongKA13}, VIKING \citep{EdgeSK13}, CFHTLS \citep[MEGACAM,][]{HudelotCW12}, Pan-STARRS \citep{ChambersMMF16}, \textit{Gaia} \citep{PerrymanBG01,BrownVP16}, NGVS \citep{FerrareseCC12}, and 2MASS \citep{SkrutskieCS06}. 
Results are presented in Fig.\ref{fig: colors_sim}, where colours are shown for different stellar ages and metallicities, and compared with the observed colours of stars, galaxies, and globular clusters. 

Transmission curves were taken from the \textit{Spanish Virtual Observatory Filter Profile Service}\footnote{\url{http://svo2.cab.inta-csic.es/svo/theory/fps3/}. For \textit{Gaia} we adopted the filter dubbed Gaia2m for faint sources (G $>10.87$), which is based on the revision of \citet{MaizApellnizW18}.}. Best-fitting curves to the predicted colour-colour loci are given in Appendix \ref{sec: fit_color}.

\begin{figure*}
\begin{center}$
\begin{array}{ccc}
\includegraphics[trim= 0cm 0cm 3.5cm 0cm, clip=true, scale=0.37]{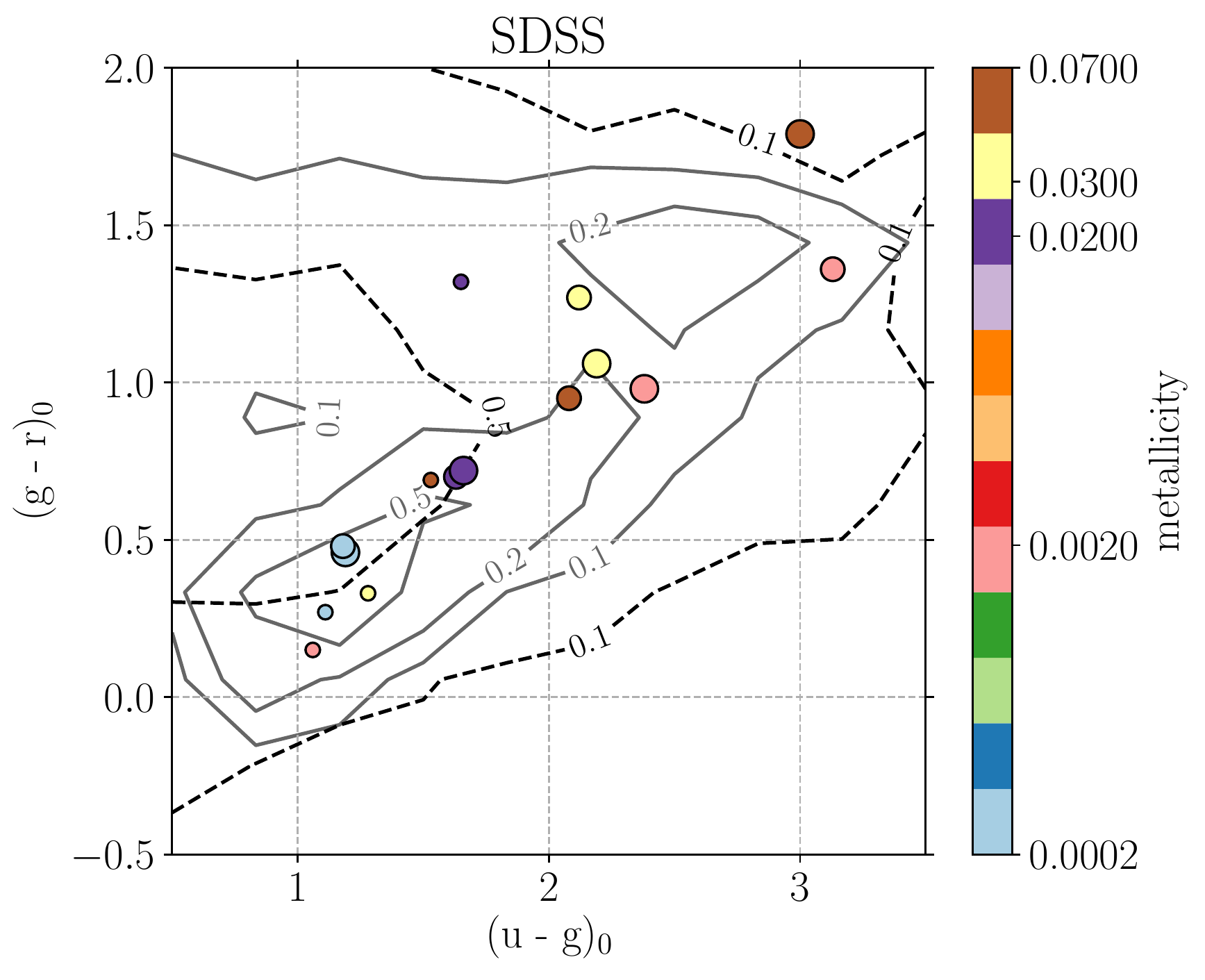}  & 
\includegraphics[trim= 0cm 0cm 3.5cm 0cm, clip=true, scale=0.37]{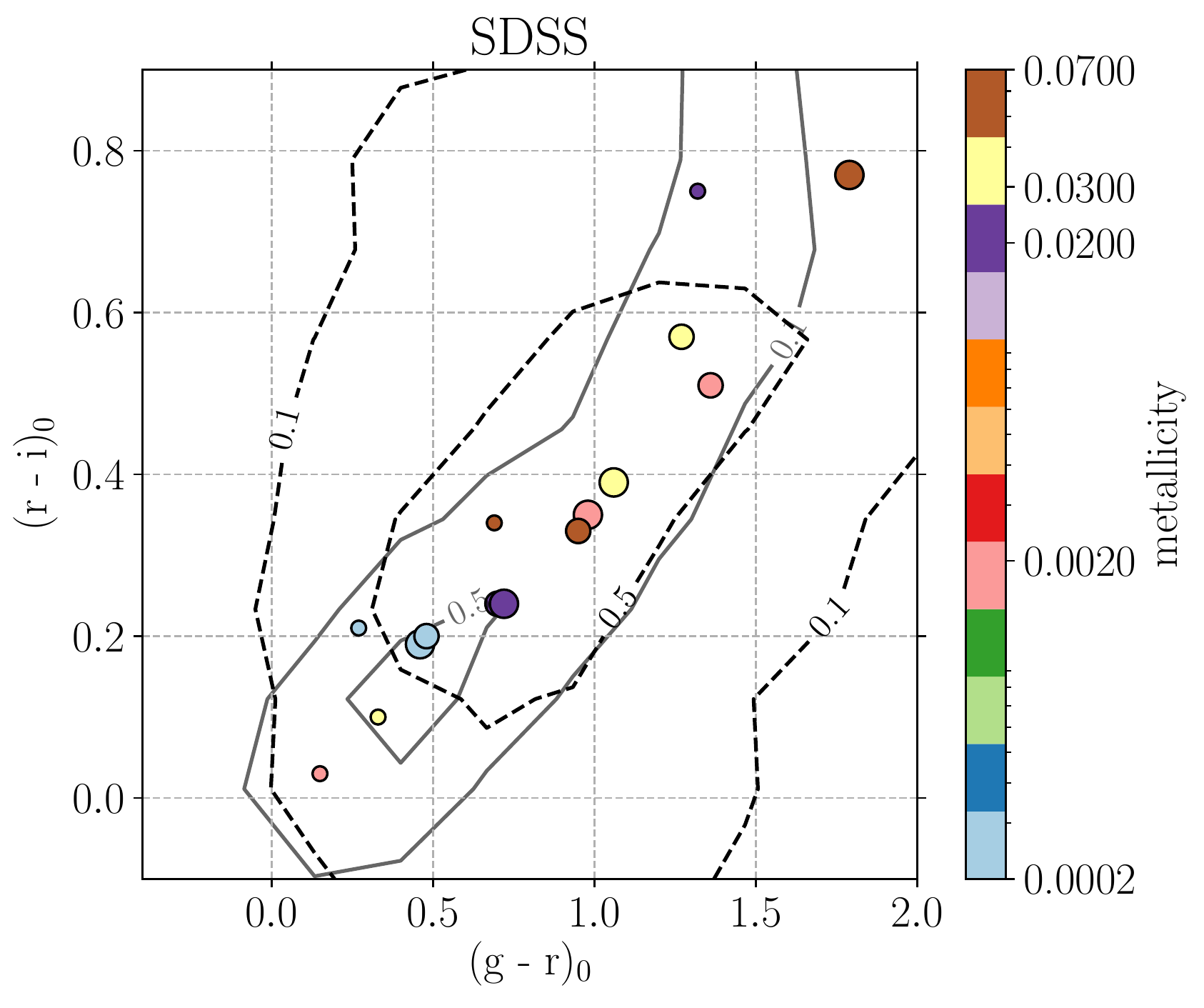} &
\includegraphics[trim= 0cm 0cm 0cm 0cm, clip=true, scale=0.37]{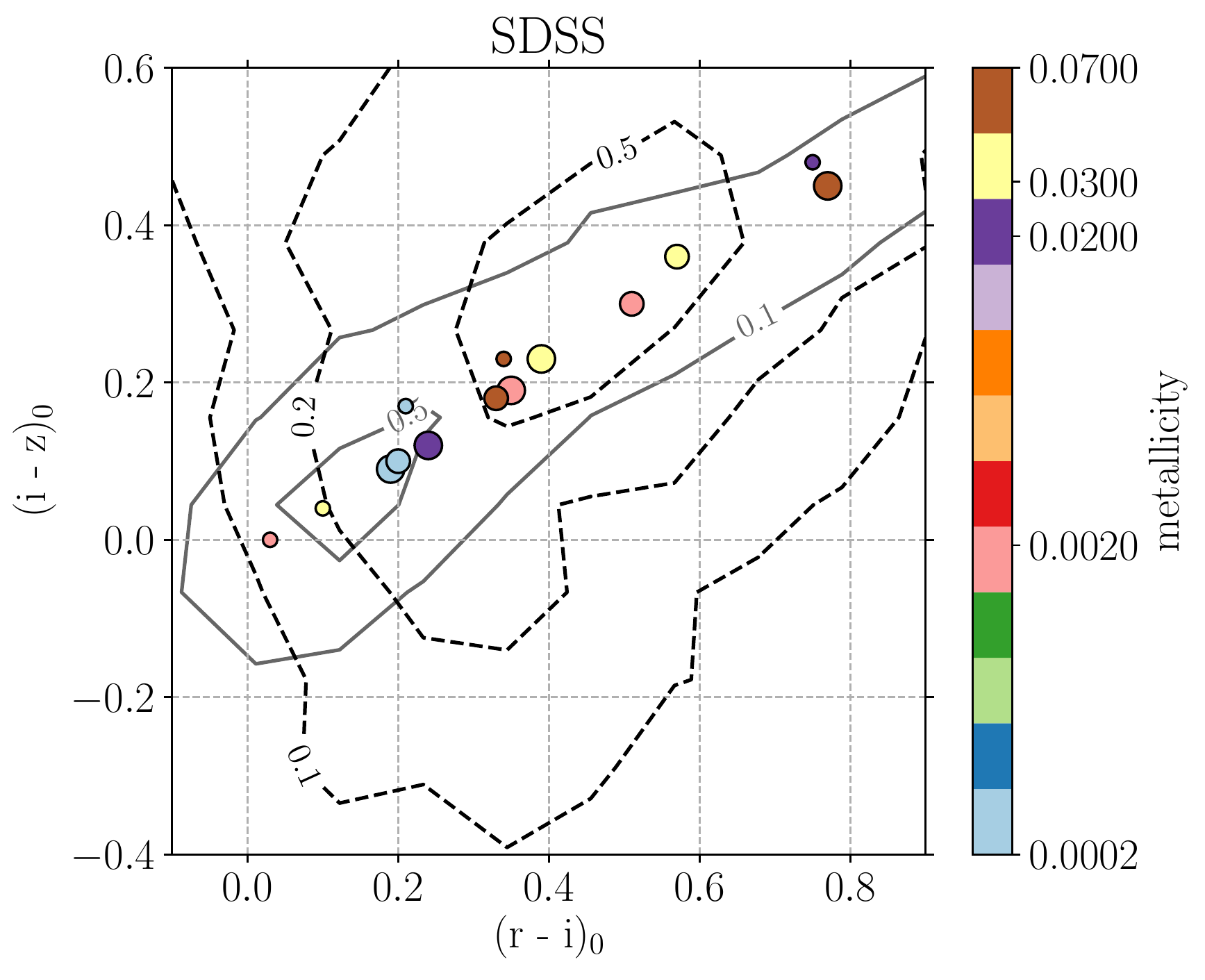} \\

\includegraphics[trim= 0cm 0cm 3.5cm 0cm, clip=true, scale=0.38]{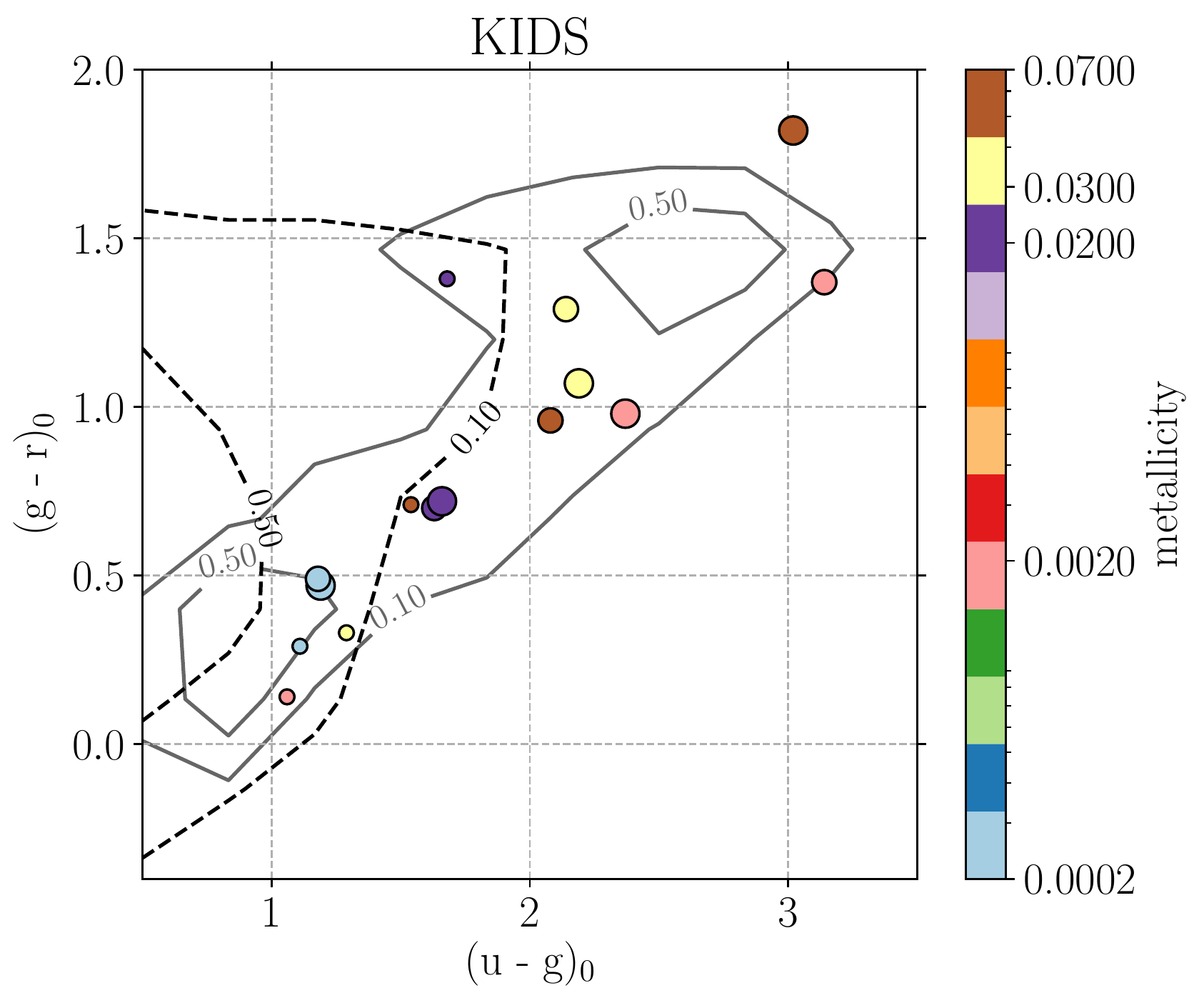} & 
\includegraphics[trim= 0cm 0cm 3.5cm 0cm, clip=true, scale=0.38]{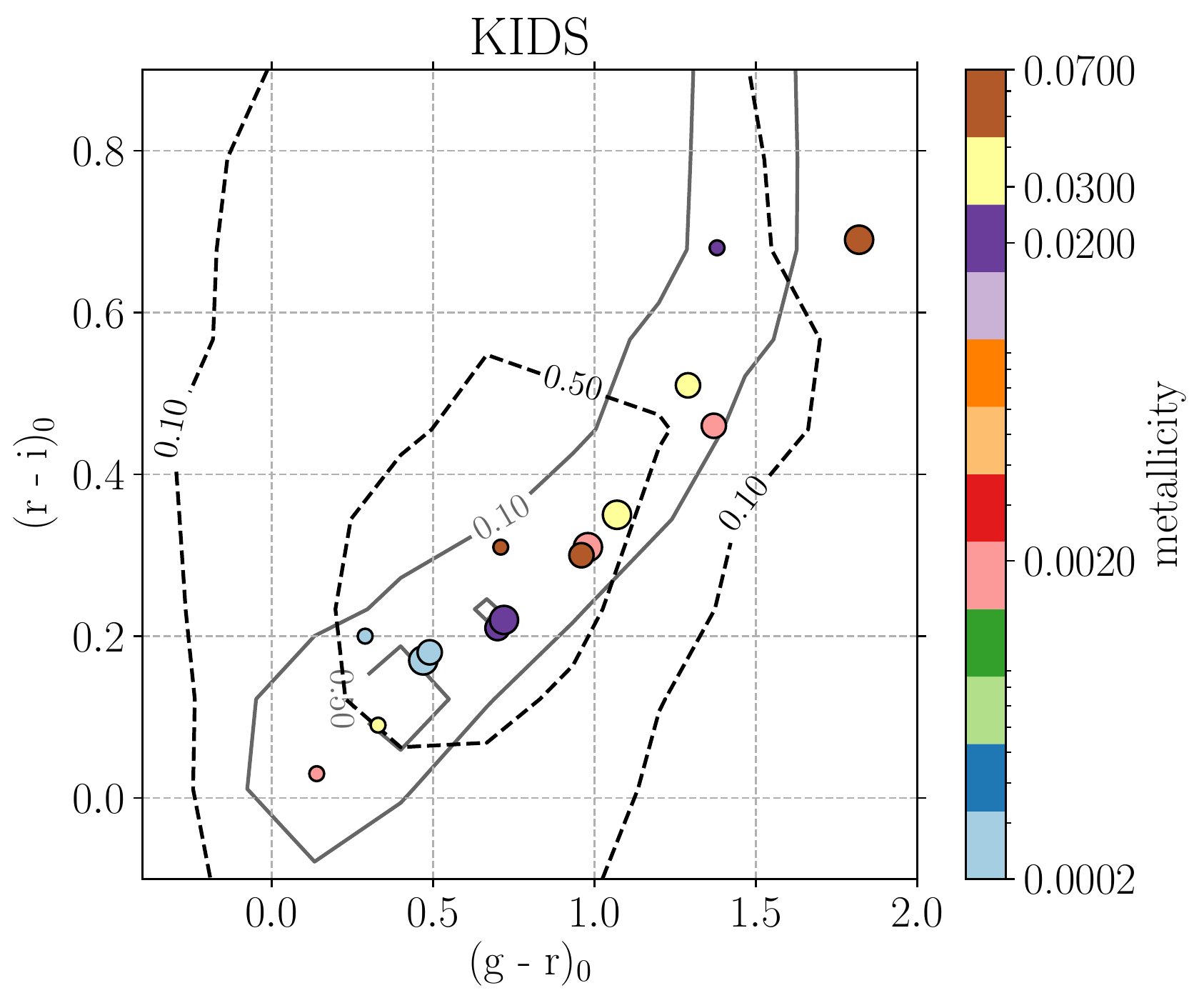}  &
 \includegraphics[trim= 0cm 0cm 3.5cm 0cm, clip=true, scale=0.38]{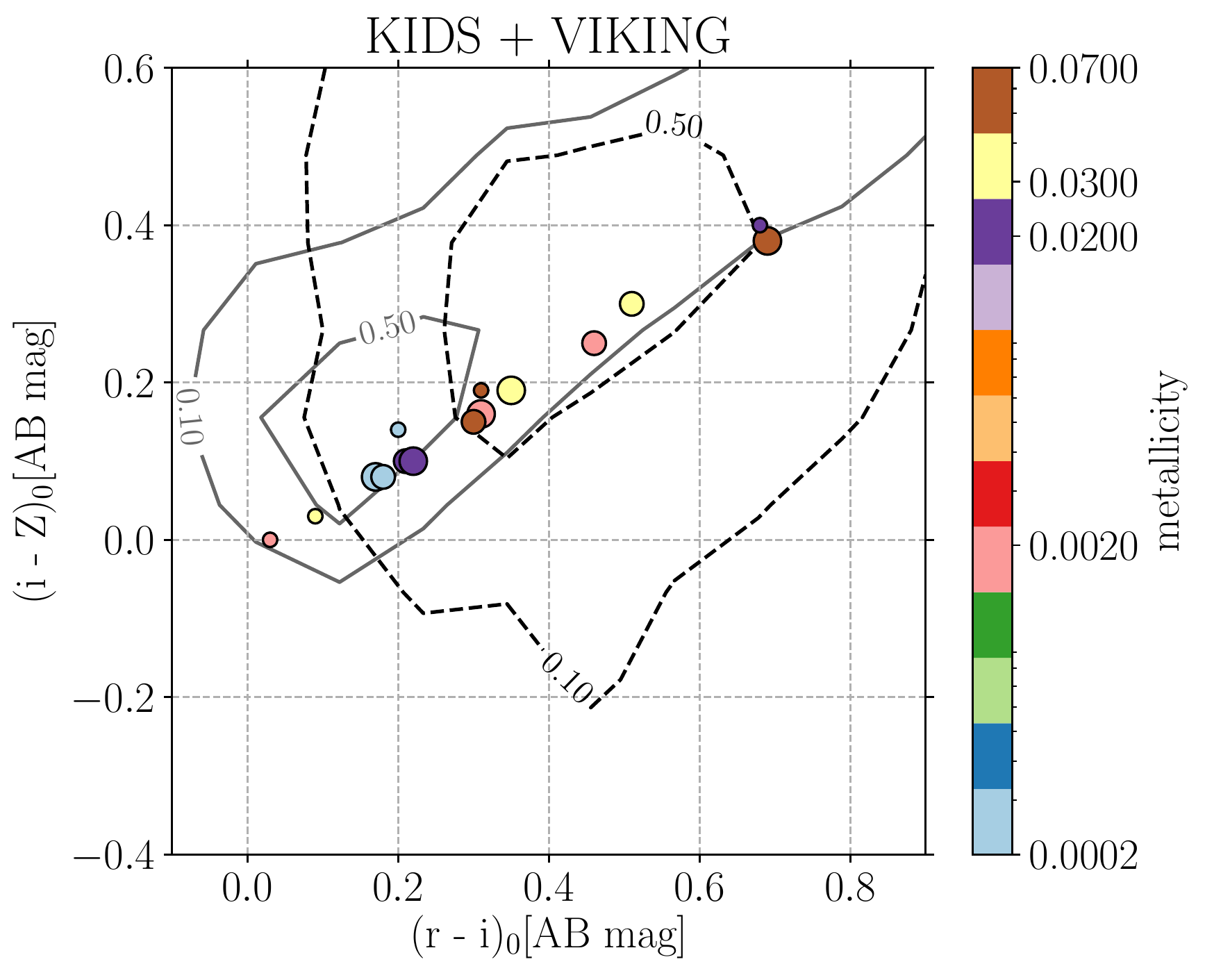} \\
 
\includegraphics[trim= 0cm 0cm 3.5cm 0cm, clip=true, scale=0.38]{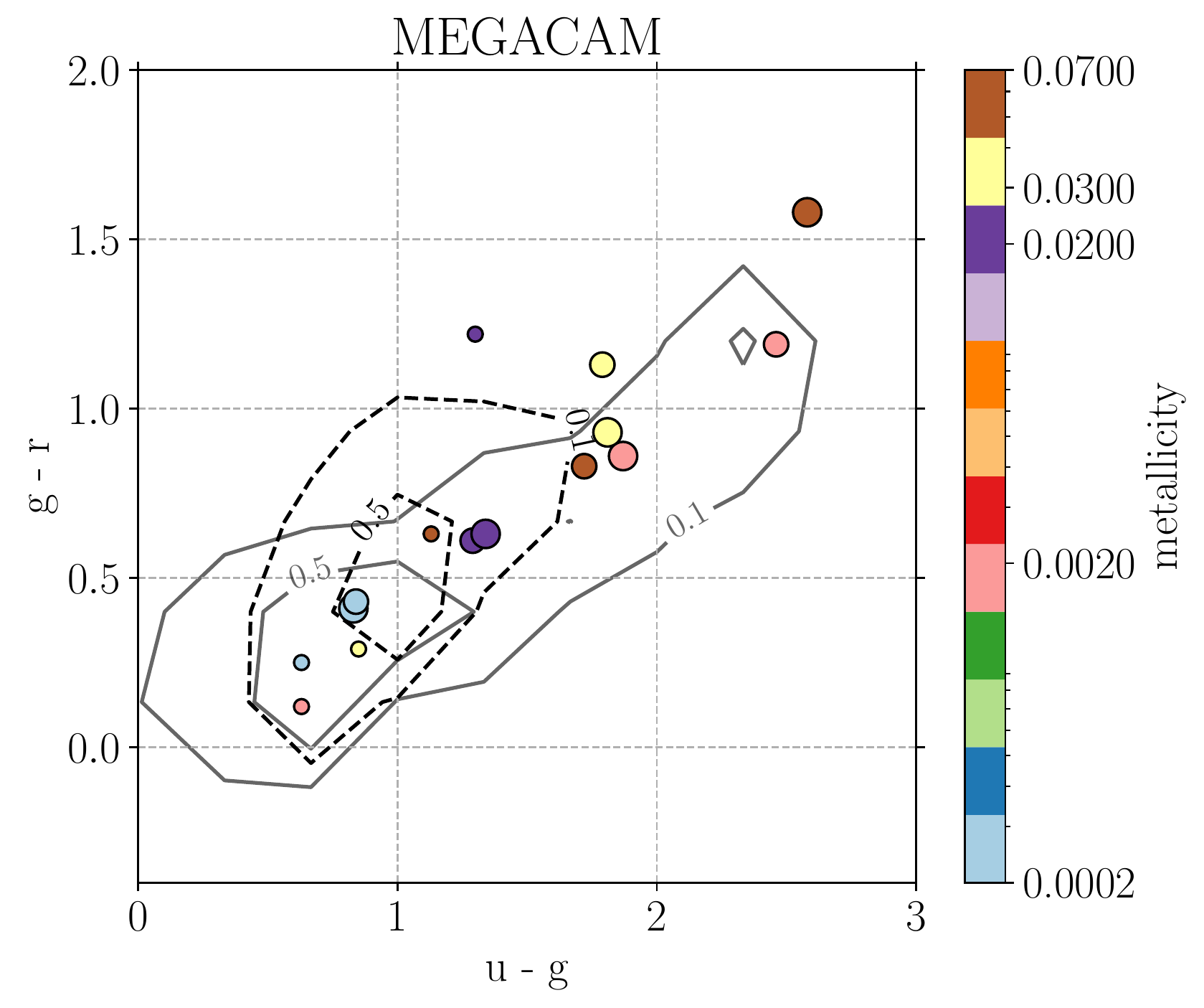} & 
\includegraphics[trim= 0cm 0cm 3.5cm 0cm, clip=true, scale=0.38]{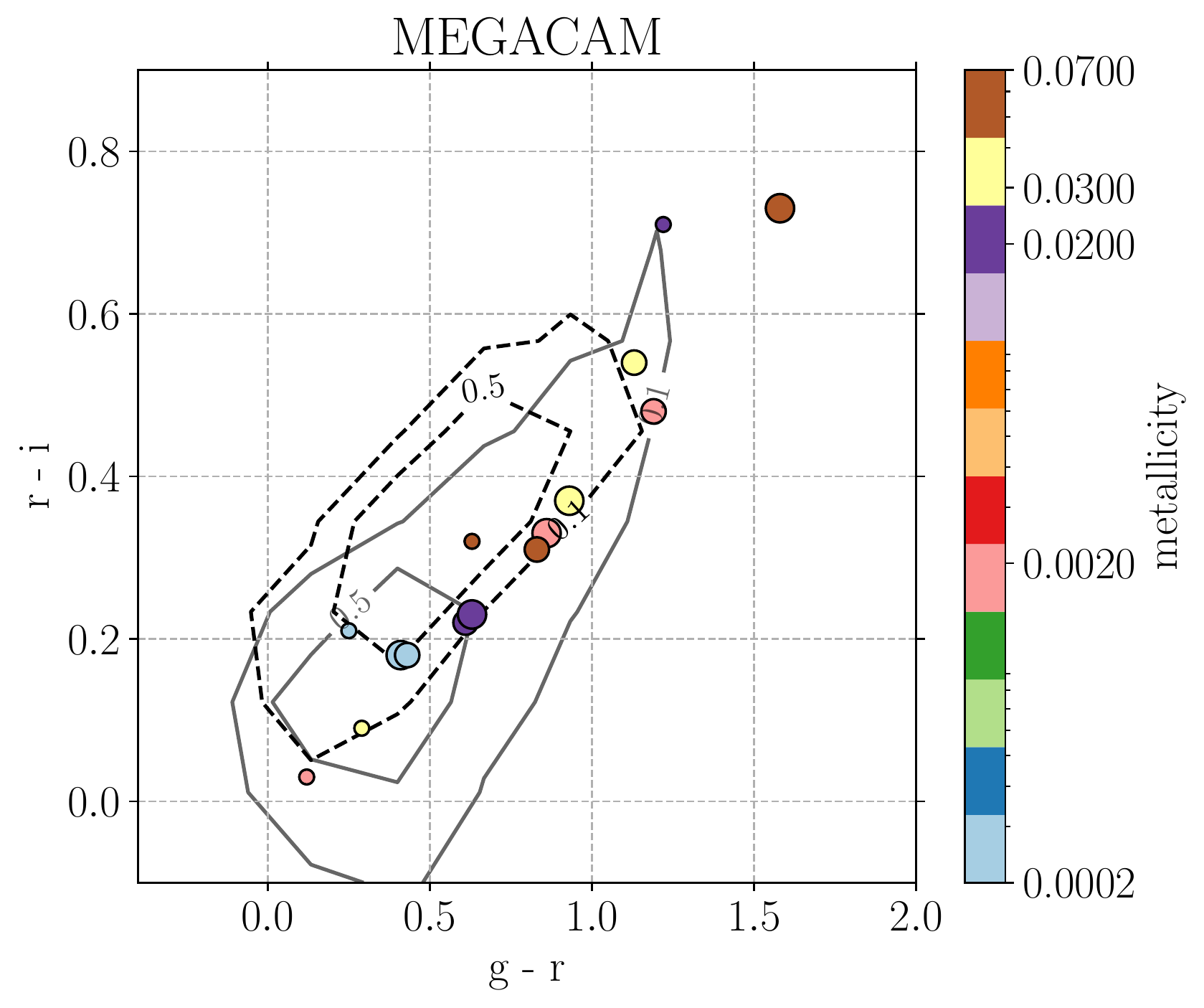}  & 
\includegraphics[trim= 0cm 0cm 3.5cm 0cm, clip=true, scale=0.38]{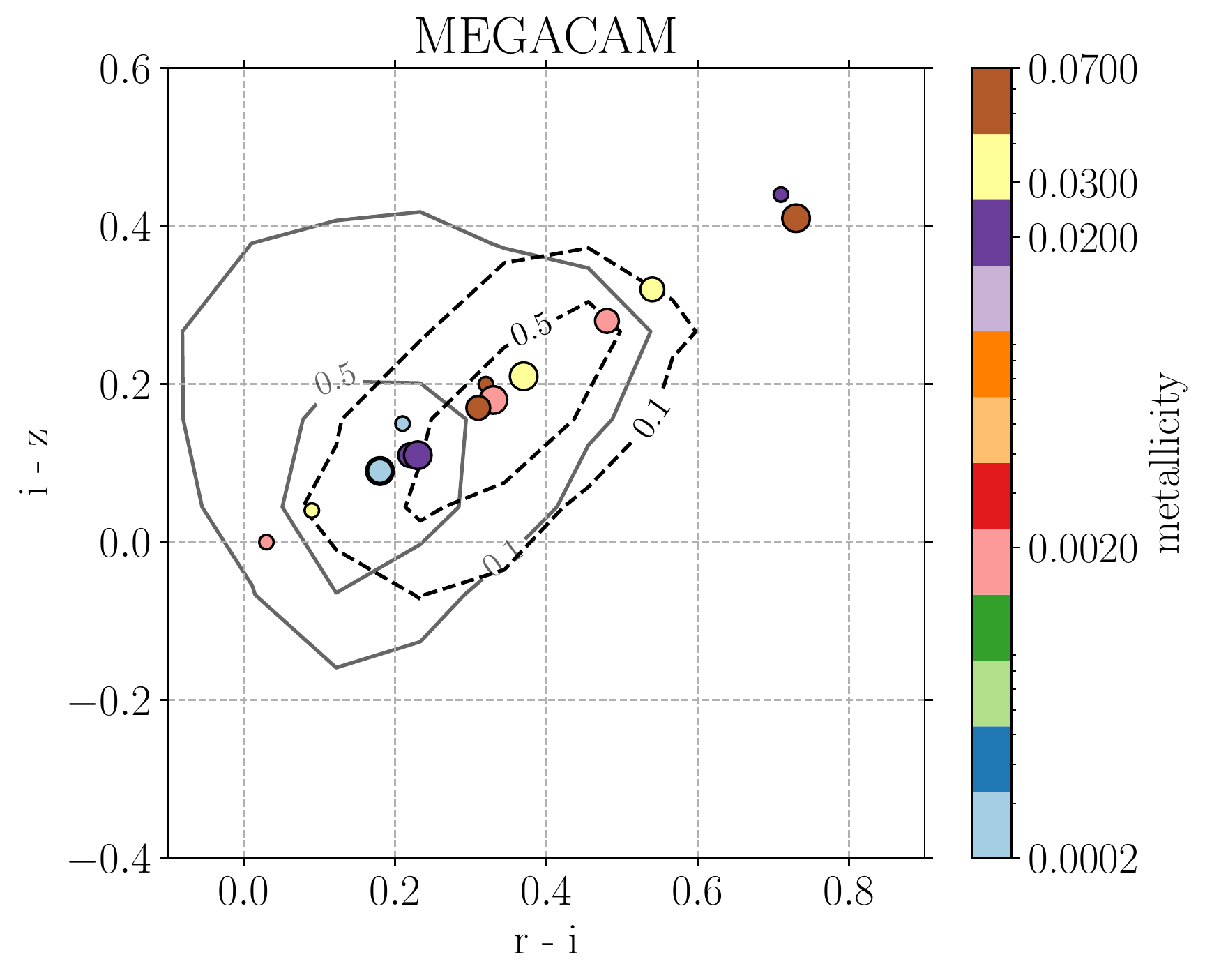} \\

\includegraphics[trim= 0cm 0cm 3.5cm 0cm, clip=true, scale=0.38]{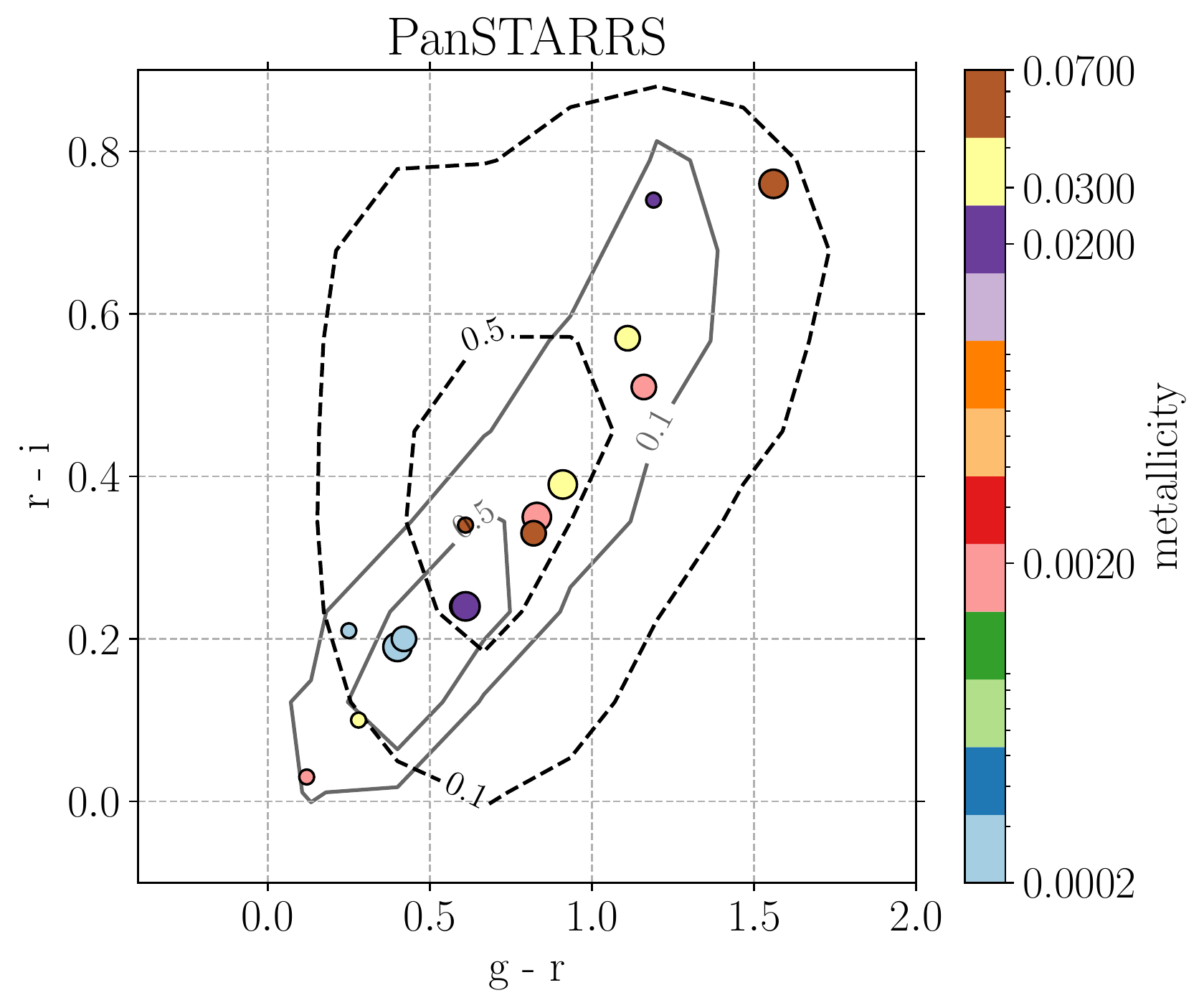}  & 
\includegraphics[trim= 0cm 0cm 3.5cm 0cm, clip=true, scale=0.38]{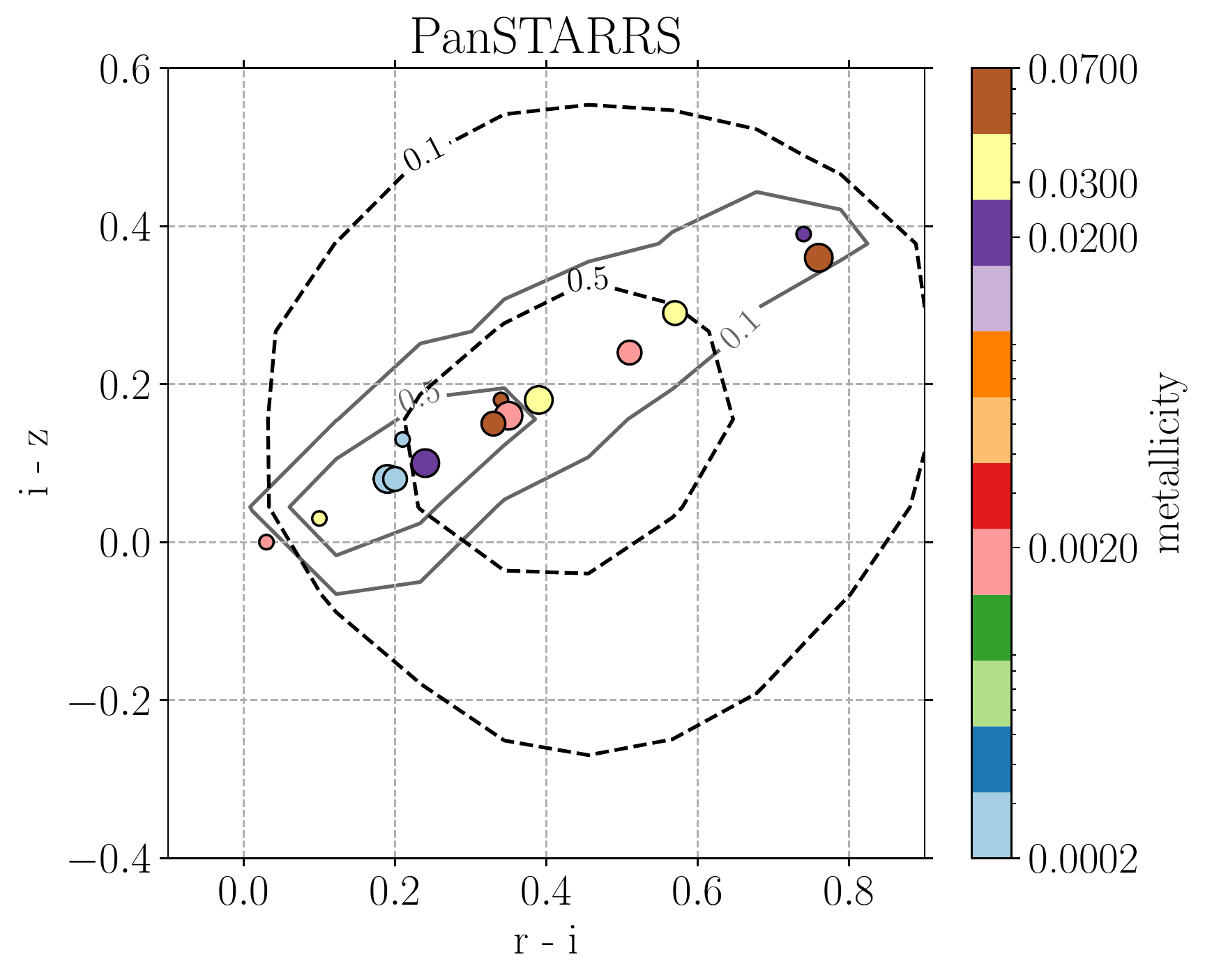} & 
\includegraphics[trim= 0cm 0cm 3.5cm 0cm, clip=true, scale=0.38]{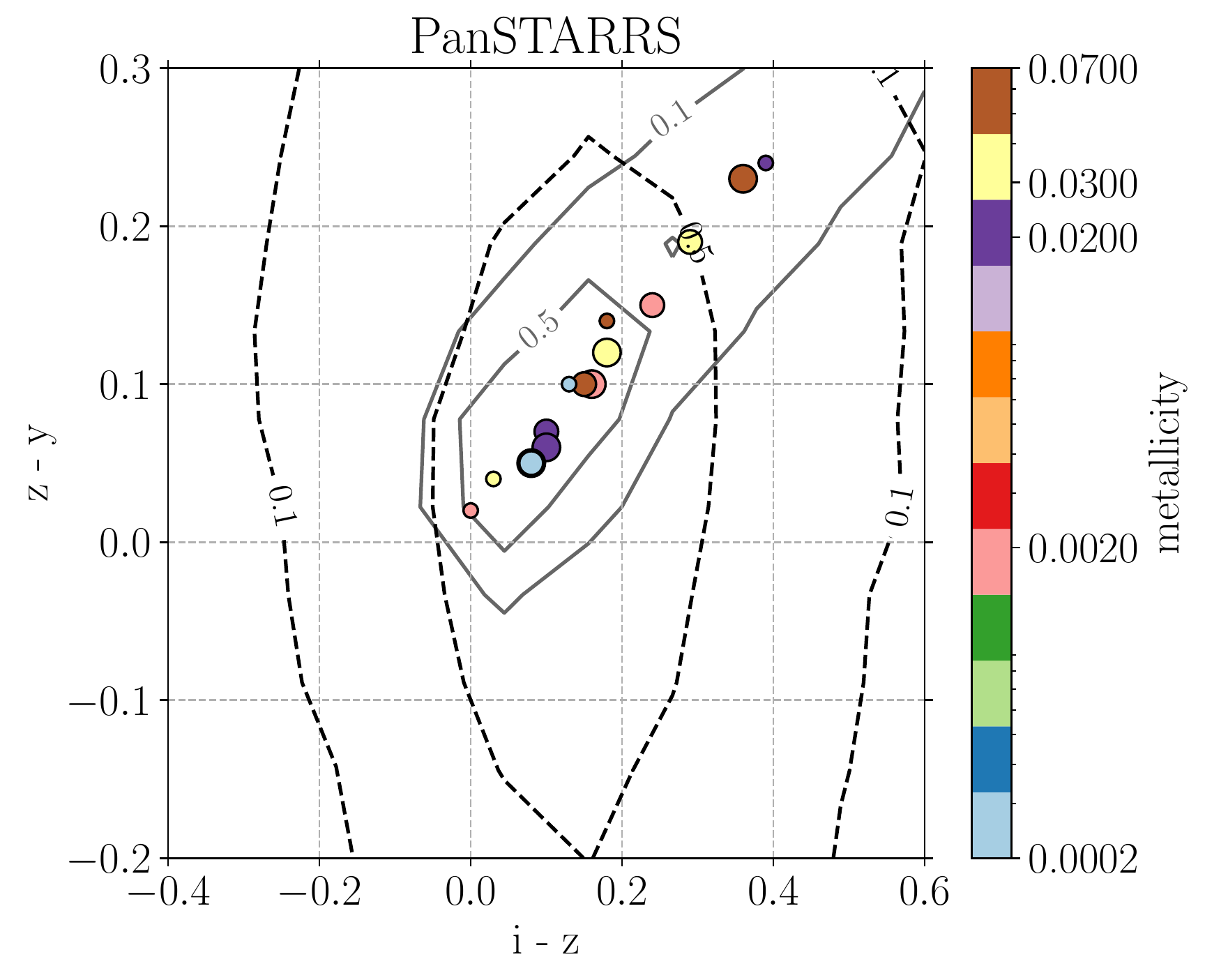} \\
\end{array}$
\end{center}
\caption[]{Colour-colour plots for different instruments and different bands. Circles: expected colours for the HCSCs with metallicity as indicated in the colour bar, and with circle size proportional to the age of the stellar population (1, 7, or 13 Gyr); dashed black contours: randomly selected galaxies; solid grey contours: randomly selected stars. Contour labels indicate the fraction of data outside the contour. Details on the origin and selection of the plotted data are given in Appendix \ref{sec: app_color_selection} (continues on next page).}
\label{fig: colors_sim}
\end{figure*} 

\begin{figure*}
\ContinuedFloat
\begin{center}$
\begin{array}{ccc}

\includegraphics[trim= 0cm 0cm 3.5cm 0cm, clip=true, scale=0.37]{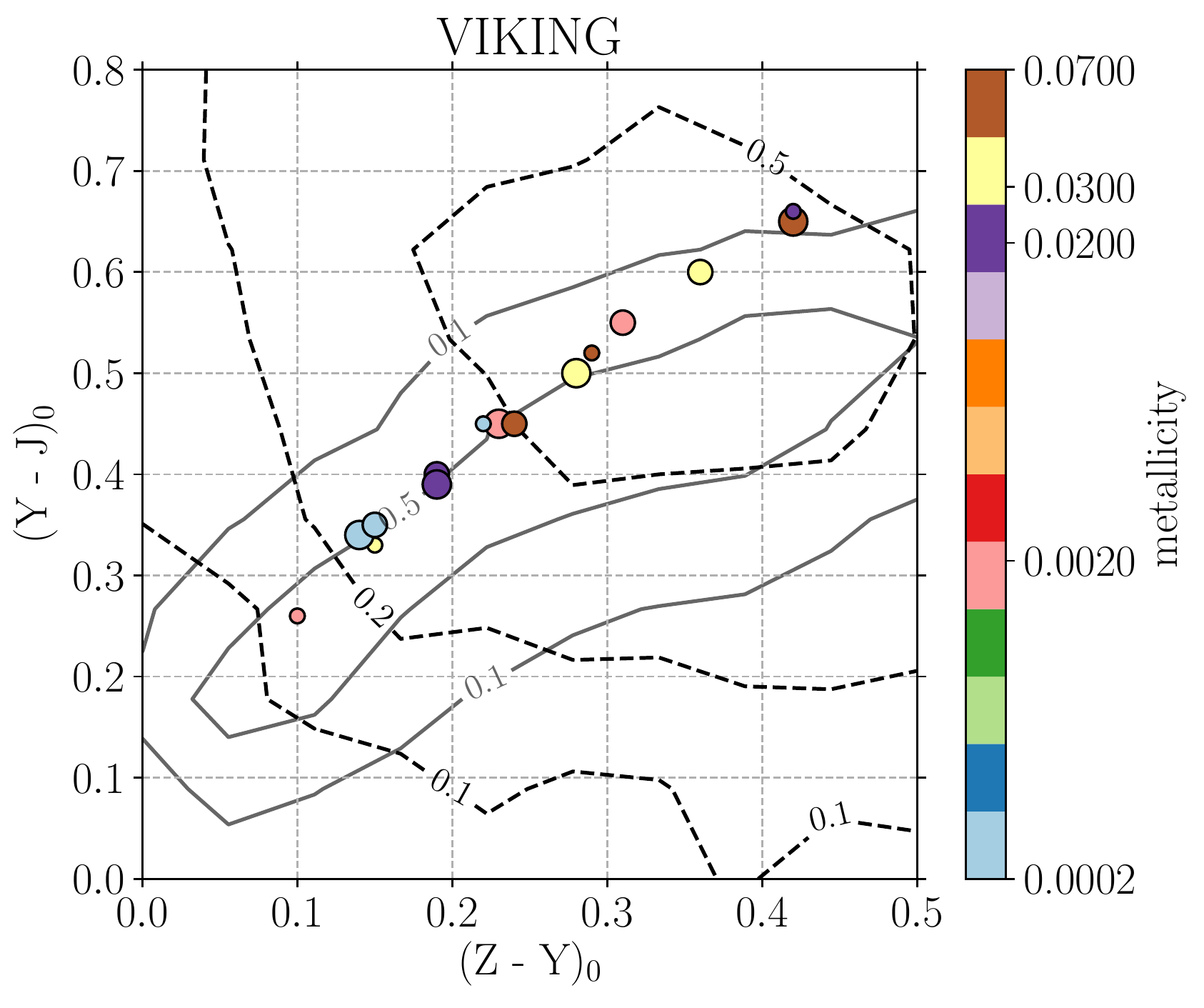} & \includegraphics[trim= 0cm 0cm 3.5cm 0cm, clip=true, scale=0.37]{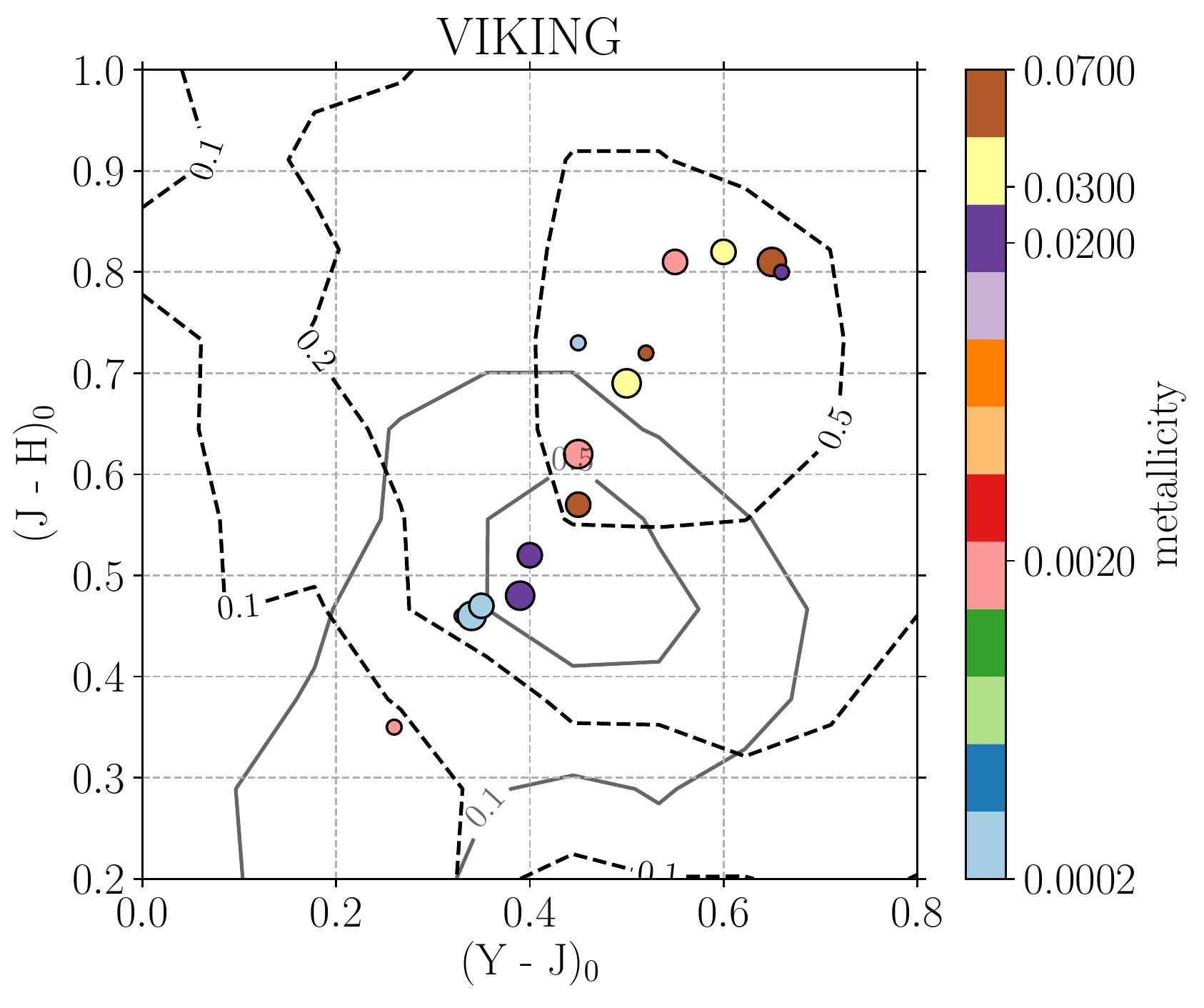} &
\includegraphics[trim= 0cm 0cm 0cm 0cm, clip=true, scale=0.37]{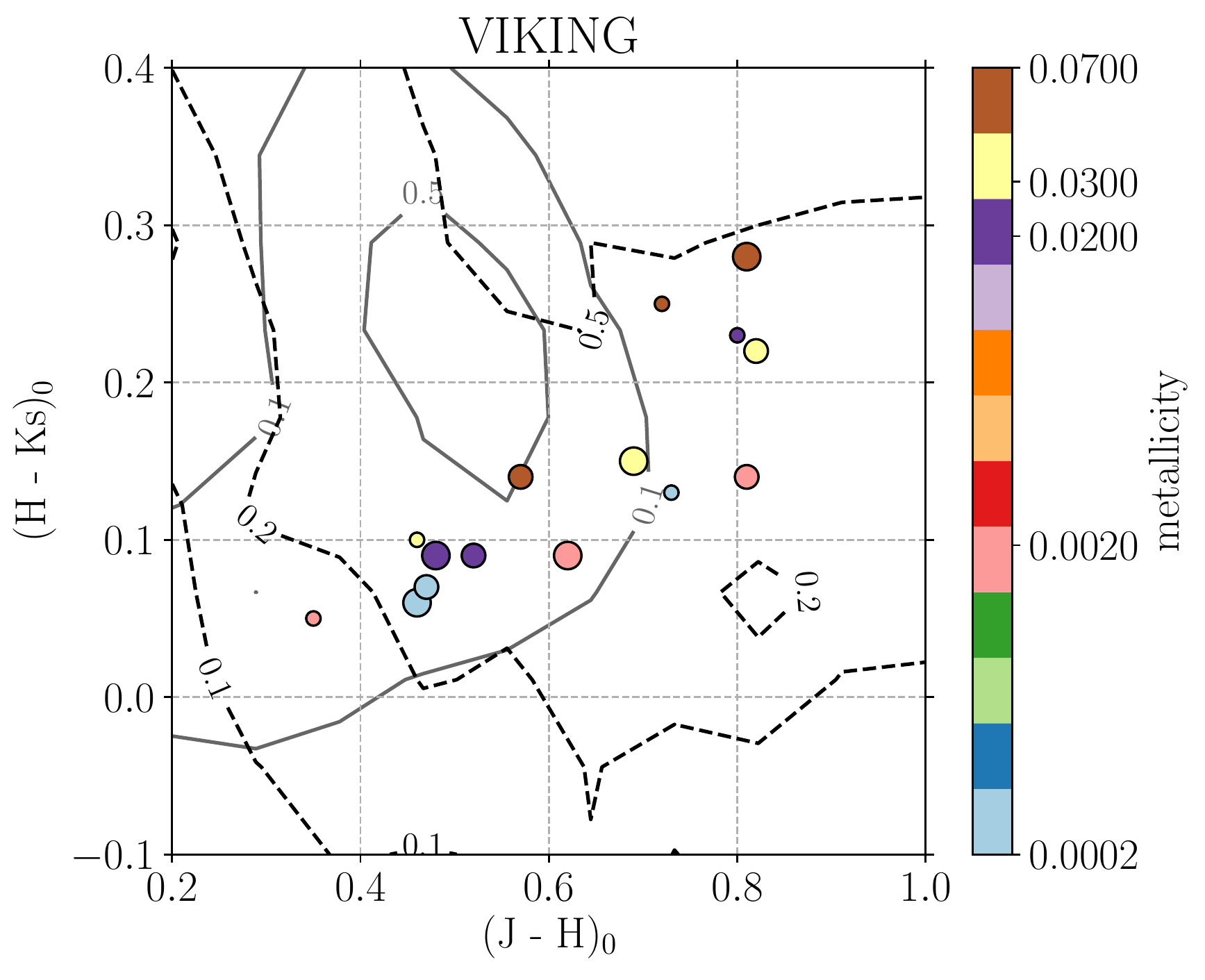} \\
\includegraphics[trim= 0cm 0cm 3.5cm 0cm, clip=true, scale=0.37]{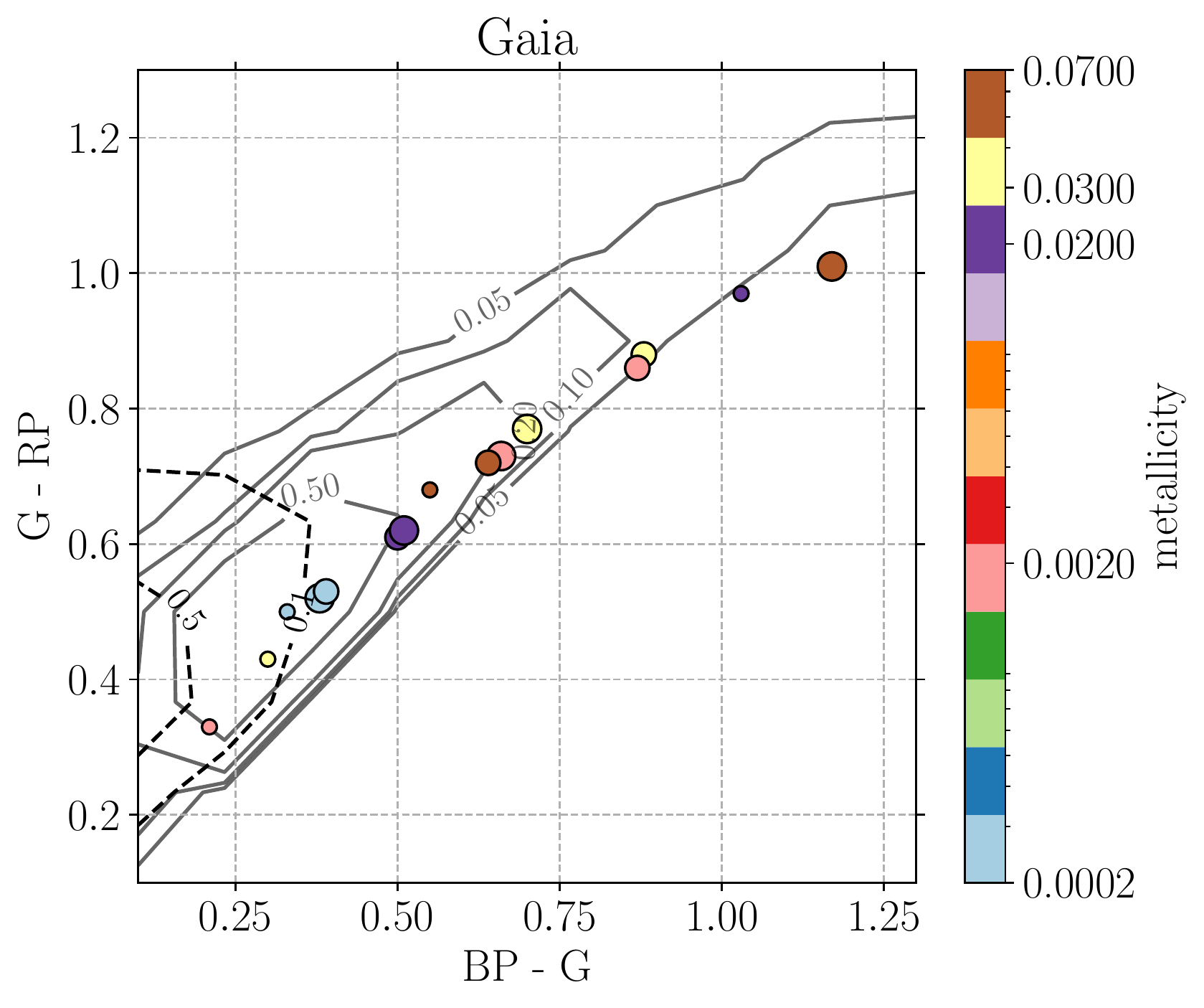}&\includegraphics[trim= 0cm 0cm 3.5cm 0cm, clip=true, scale=0.37]{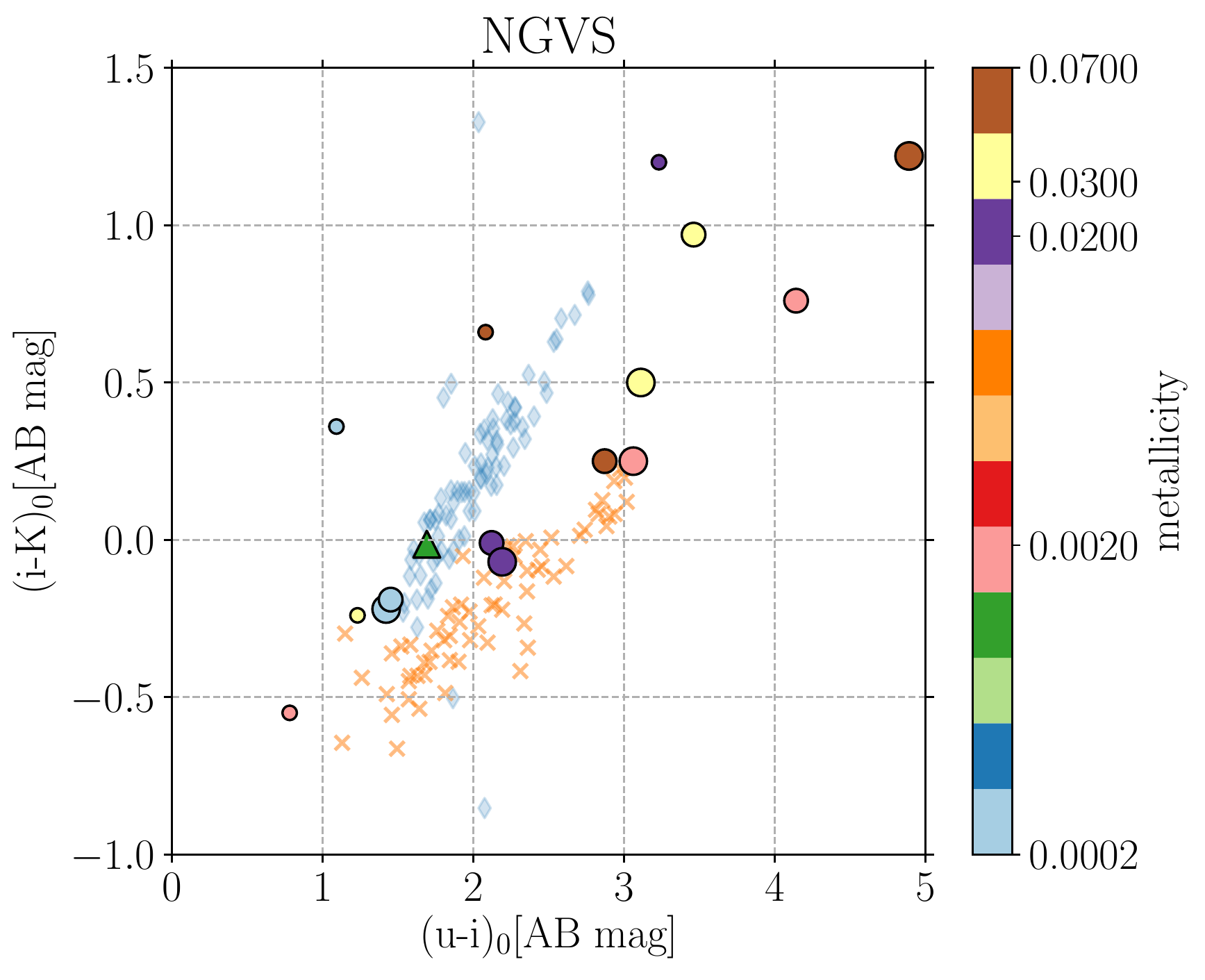}&\includegraphics[trim= 0cm 0cm 3.5cm 0cm, clip=true, scale=0.37]{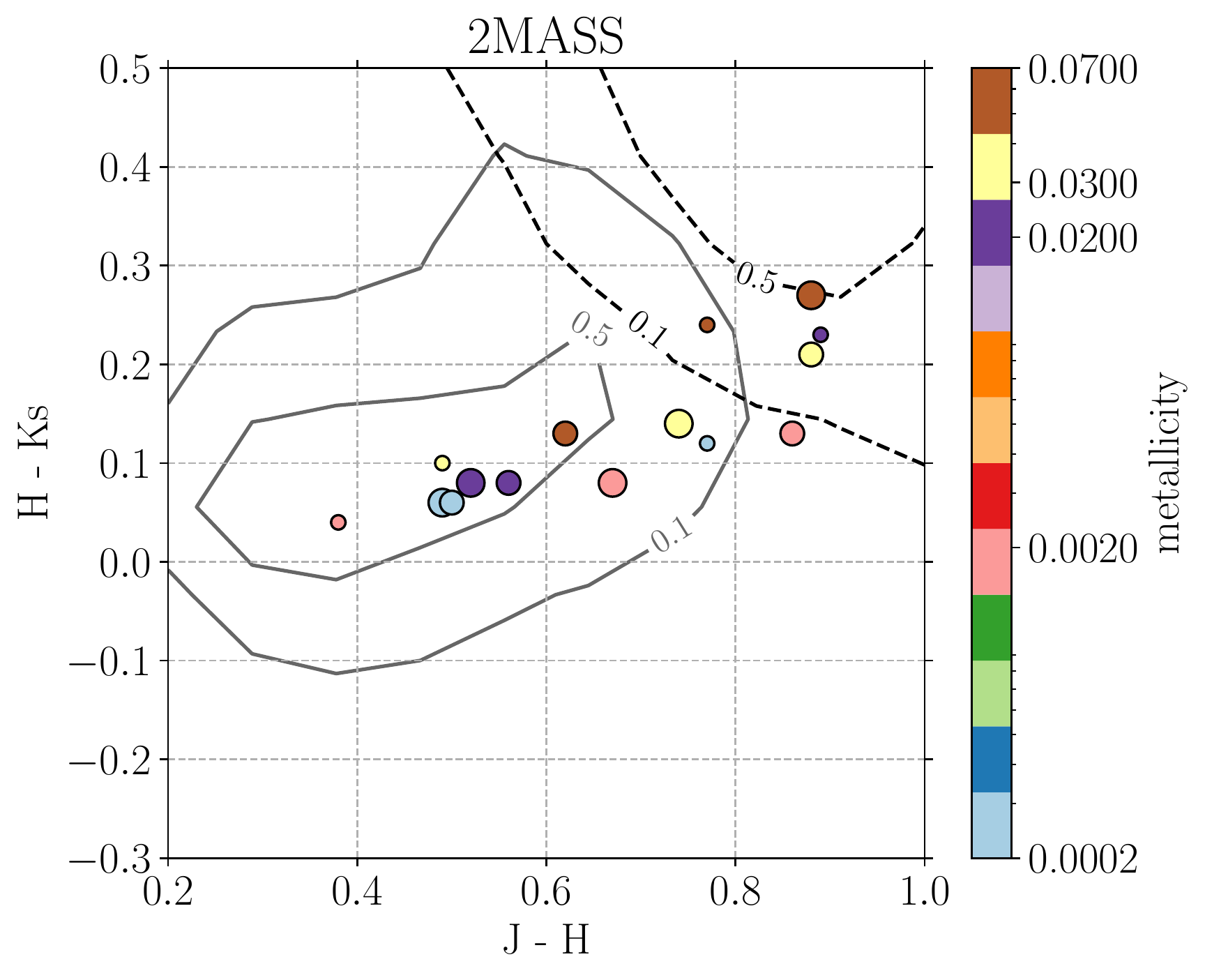} \\

\end{array}$
\end{center}
\caption[]{(Continued) \textit{Gaia}: black dashed contours are QSOs. NGVS: orange crosses are Galactic halo stars and blue diamonds are M31 globular clusters, the green triangle is the hypervelocity cluster HVGC-1 \citep[data from ][]{CaldwellSR14}; galaxies, which would be located above the globular clusters, have been omitted from this plot. 2MASS: solid contours are point sources; dashed contours are extended sources.}
\label{fig: colors_sim1}
\end{figure*}

\subsection{Morphology} \label{sec_morpho}
We used \textsc{skymaker} \citep{Bertin09} to produce mock images of HCSCs corresponding to Pan-STARRS \citep{ChambersMMF16} and \textit{Euclid}/NISP \citep{LaureijsAA11} observations. The result is shown in Fig.\ref{fig: hcss_mock_psnisp}. 

\textsc{skymaker} is a simulator of astronomical images; it takes as input a parameter file and a catalog. The former specifies a number of parameters, such as seeing, pixel scale, telescope details, and central wavelength, among many others. The latter is a catalog of coordinates, magnitudes, and a code to discern between stars and galaxies. If a point-spread function (PSF) is not provided, then the software derives one taking into account blurring due to the atmosphere, to the telescope motion, diffraction and aberration features, effects due to optical diffusion (i.e. scattered light), and intra-pixel response (i.e. the sensitivity variation within a pixel). Once the objects in the catalog are rendered, a sky background is added along with Poissonian and Gaussian noise, representing photon-noise and noise due to electronics, respectively.

To create the input catalog for \textsc{skymaker} we adopted the stellar positions derived via N-body simulations by \citet{MSK2009}, Fig.\ref{fig: hcss_nbody}. The simulation was realised with $N = 1.5 \times 10^4$ equal-mass particles and assuming an initial power-law density profile $\rho \propto r^{-7/4}$. An instantaneous kick of magnitude V$_{k}$ was delivered to the cluster along the $-$X direction at $t = 0$. We used the results obtained at $t=100$ (physical units can be obtained by multiplying the N-body time by GM$_{\bullet}$/V$_{k}^3$, with G the gravitational constant, M$_{\bullet}$ the black hole mass, and V$_{k}$ the kick velocity). Stellar locations were derived by inverting the cumulative mass profile derived from the N-body simulation. The process consisted of four steps: 1) we selected a random number uniformly distributed between zero and the total mass of the cluster; 2) by inverting the cumulative mass profile we selected a radial distance from the cluster centre; 3) we picked the N-body point located at that radial distance; 4) the star coordinate was extracted from a Gaussian distribution centred at the location of the N-body point, and characterised by a standard deviation given by 10 per cent the value of its 3D distance from the cluster centre.

The number of stars in the cluster, right after the kick, was determined so that the bound stellar mass would amount to:

\begin{align}\label{eq: mb}
M_{b} &= 11.6\gamma^{-1.75} \times M_{\bullet} \left(\frac{GM_{\bullet}}{r_{\bullet}V_{k}^2}\right)^{3-\gamma},
\end{align}
where $\gamma$, here assumed to be 7/4, as in the N-body simulation, is the slope of the stellar density profile before the kick ($\rho \propto r^{-\gamma}$), and $r_{\bullet}$ is the radius containing an integrated mass in stars equal to twice M$_{\bullet}$. Eq.\ref{eq: mb} was derived by combining eq. 5, 7, and 8 in \citet{MSK2009}. The scaling relation adopted for $r_{\bullet}$ is:

\begin{align}\label{eq: rb}
r_{\bullet} \approx 8\ \mathrm{pc} \left(\frac{M_{\bullet}}{10^7M_{\odot}}\right)^{0.46},
\end{align}
which we derived by fitting the data presented by \citeauthor{MSK2009} in their Fig.12 for ``power-law" galaxies, i.e. galaxies with a deprojected inner density profile steeper than $R^{-0.5}$ (e.g. \citealt{lau95}, but see also \citealt{Graham13}). Although ``core" galaxies (those with a central density profile shallower than $R^{-0.5}$), are more likely to result from a history of mergers \citep[e.g.][]{EbisuzakiMO91,MakinoE96,Merritt06_mergers}, and therefore more likely to be the progenitors of HCSCs, we prefer to use the scaling relation derived for ``power-law'' galaxies as it is better constrained at the low-mass end of the SMBH mass distribution, which is the focus of this work. The difference between the extrapolated values of r$_{\bullet}$ is, however, small: for M$_{\bullet} = 10^{5}$ M$_{\odot}$, we get r$_{\bullet,\mathrm{core}} \approx 0.8$ r$_{\bullet,\mathrm{power-law}}$; furthermore, if one would extend \citeauthor{MSK2009}'s Fig.12 to lower masses, such difference would turn out to be much smaller than the scatter in the data points.
\newline 

To derive magnitudes we inverted the evolved integrated initial mass functions produced via \textsc{parsec} along with the isochrones (Sec. \ref{sec_spectral}). Stellar masses and the corresponding magnitudes were then extracted and randomly assigned to the N-body points. In this process we ignored the effects of mass segregation as \citet{OLearyL12} found only moderate evidence that this process was taking place in their models, even though the dynamical simulations included stars spanning a factor 10 in mass. Similar results on mass segregation were obtained for simulations of globular clusters with a nuclear BH: the BH quenches mass segregation by scattering sinking particles out of the core \citep[e.g.][]{BaumgardtME04,GillTM08}.

We assumed the telescope and seeing parameters indicated in Appendix \ref{sec: con_files}, a distance to the cluster of 10 and 30 kpc, and a BH mass of 10$^5$ M$_{\odot}$. 
Finally, we point out that an accurate representation of the cluster morphology should take into account the cluster expansion. Dynamical studies conclude that the radius enclosing a fixed number of stars evolves as $r \propto t^{2/3}$, however the evolution is expected to deviate from this law when the number of stars in the cluster becomes small \citep[e.g.][and references therein]{OLearyL12}. Furthermore, \citet{MSK2009} warn that the true evolution of the cluster is likely affected by a number of phenomena which are still poorly understood and not fully implemented in dynamical simulations. For these reasons, as \citeauthor{MSK2009}, we ignored the cluster expansion.

\subsubsection{Effect of different kick velocities}
\begin{figure*}
\begin{center}$
\begin{array}{cccc}
\includegraphics[trim= 0cm 1cm 2cm 1cm, clip=true, scale=0.35]{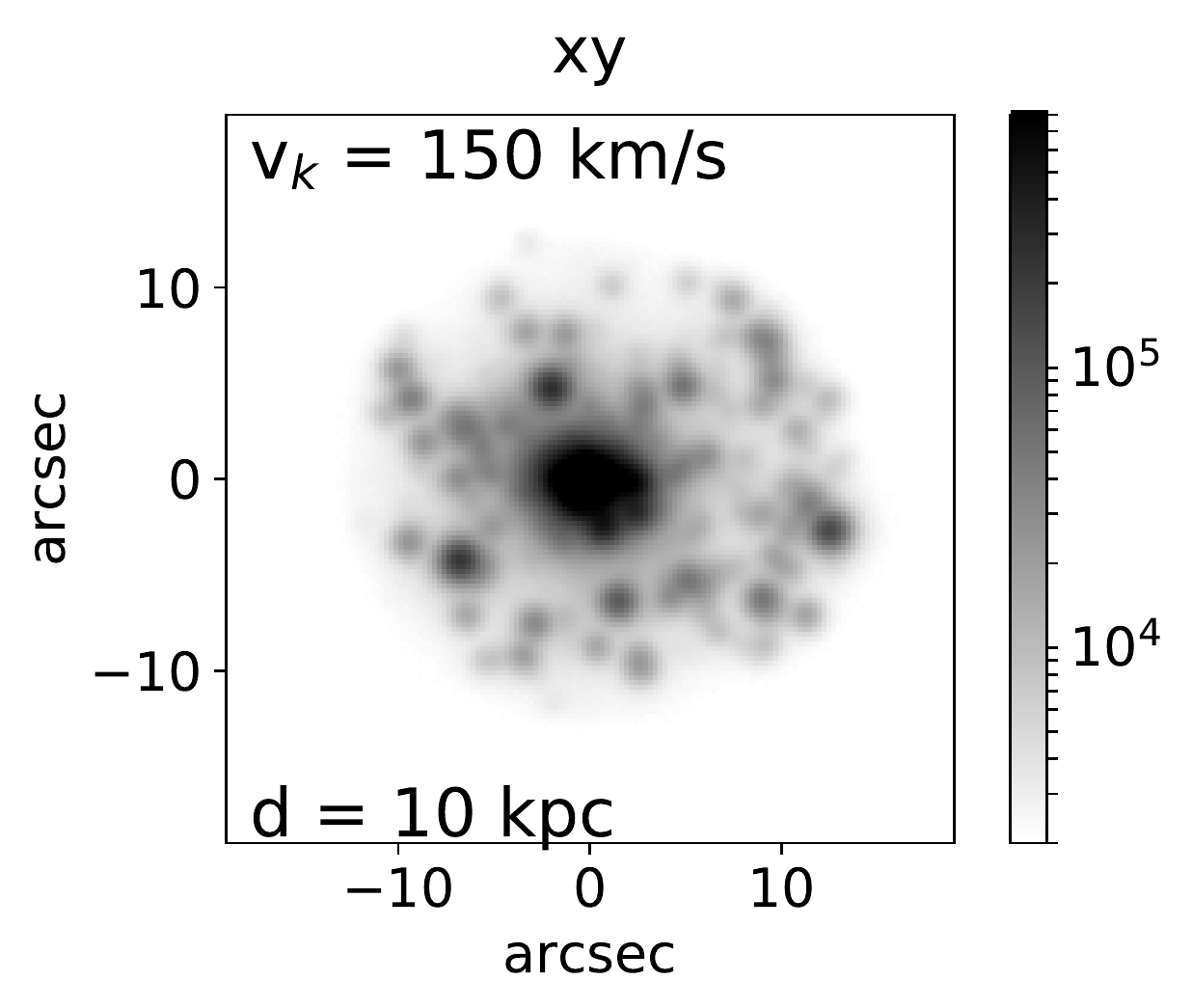}&
\includegraphics[trim= 2.2cm 1cm 0cm 1cm, clip=true, scale=0.35]{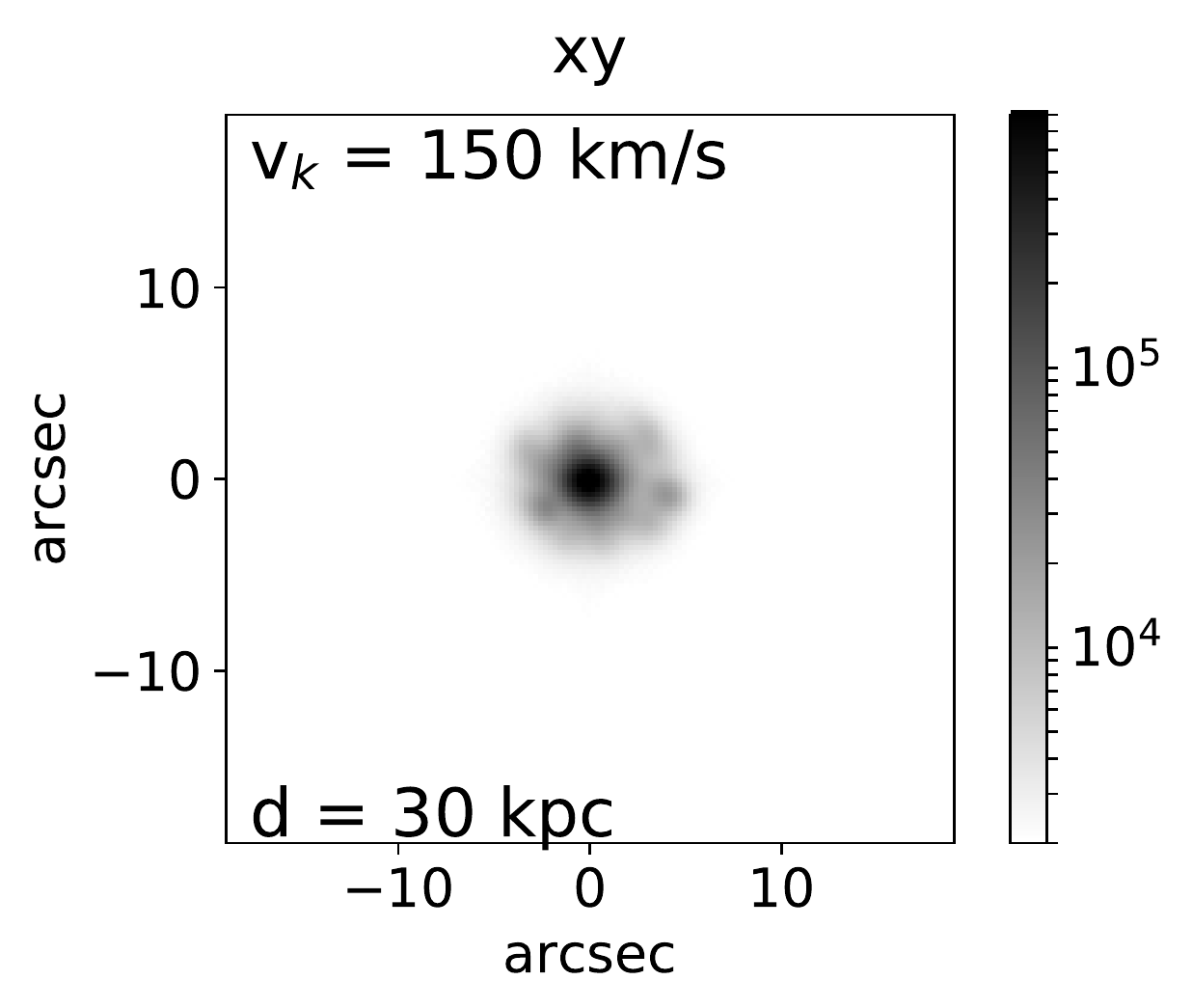}&
\includegraphics[trim= 2.2cm 1cm 2cm 1cm, clip=true, scale=0.35]{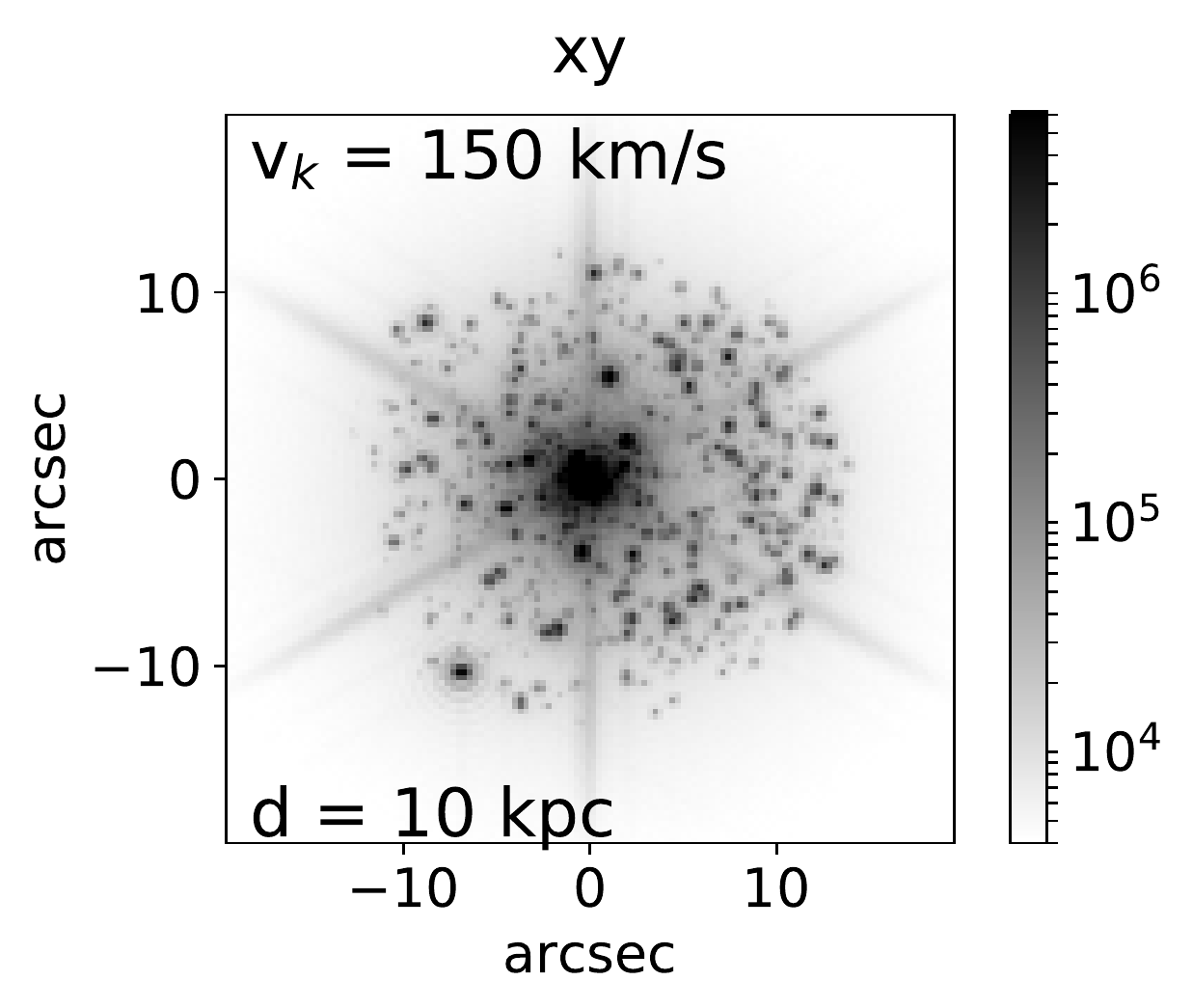}&
\includegraphics[trim= 2.2cm 1cm 0cm 1cm, clip=true, scale=0.35]{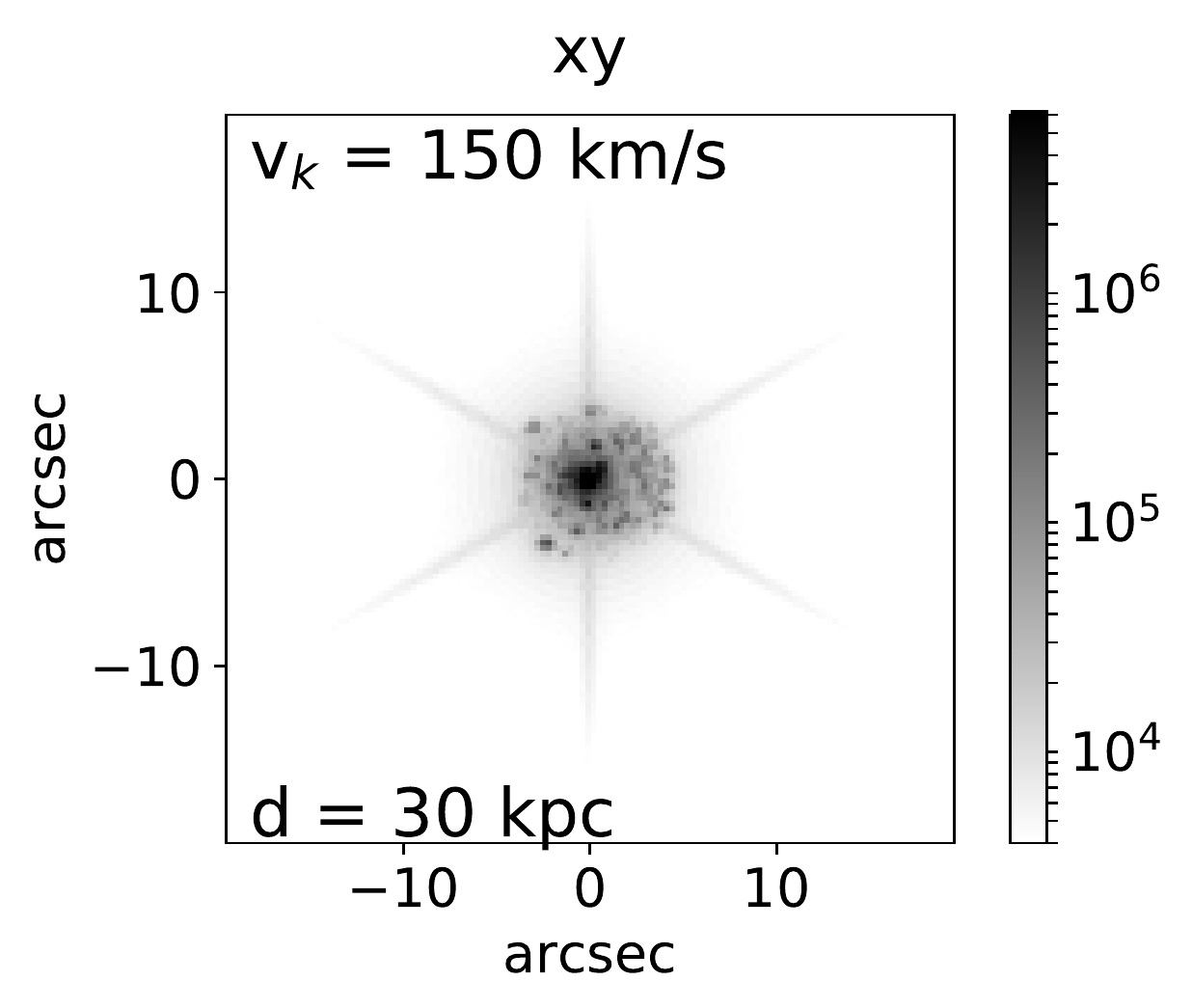}\\ 

\includegraphics[trim= 0cm 1cm 2cm 1cm, clip=true, scale=0.35]{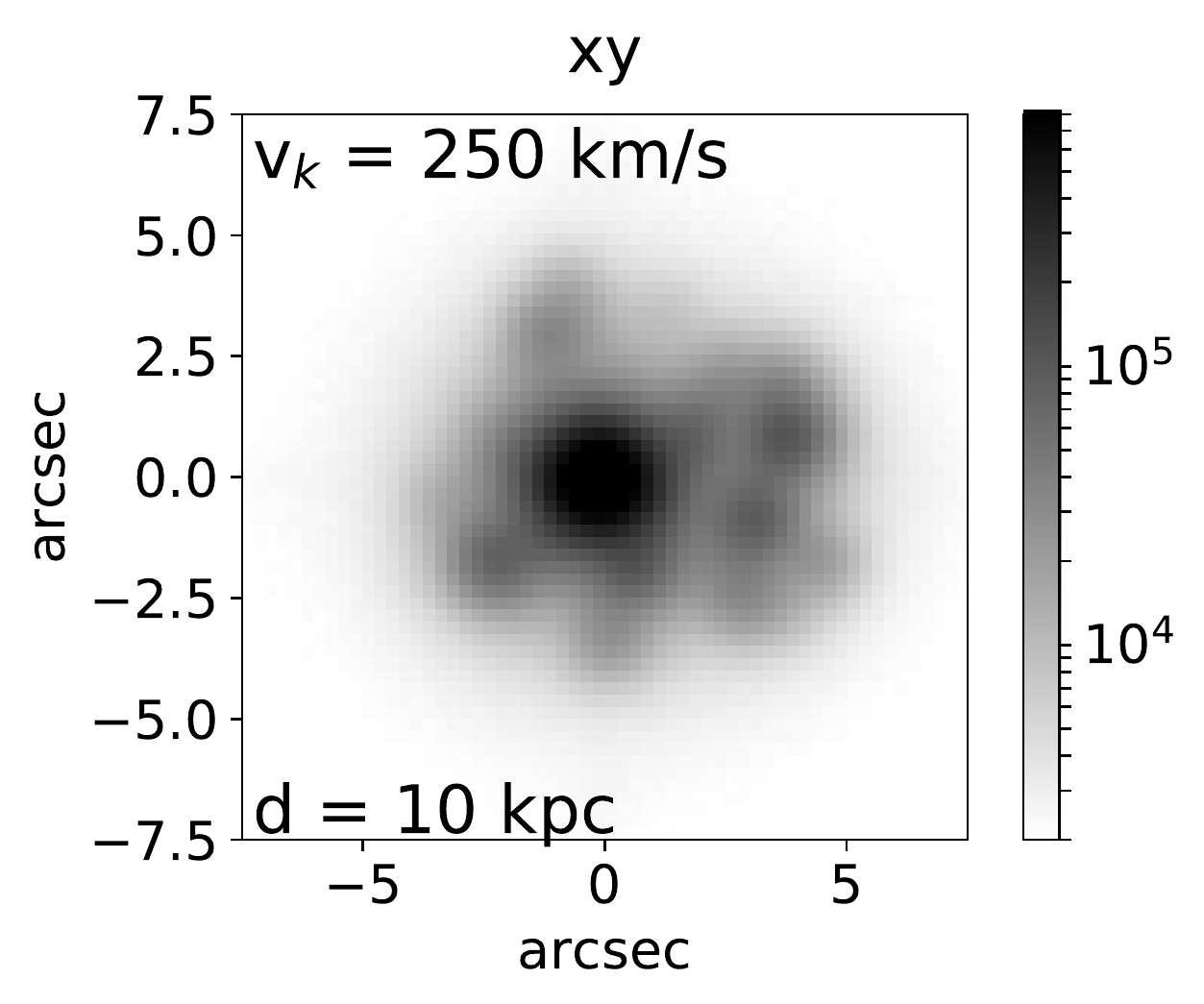}&
\includegraphics[trim= 2.4cm 1cm 0cm 1cm, clip=true, scale=0.35]{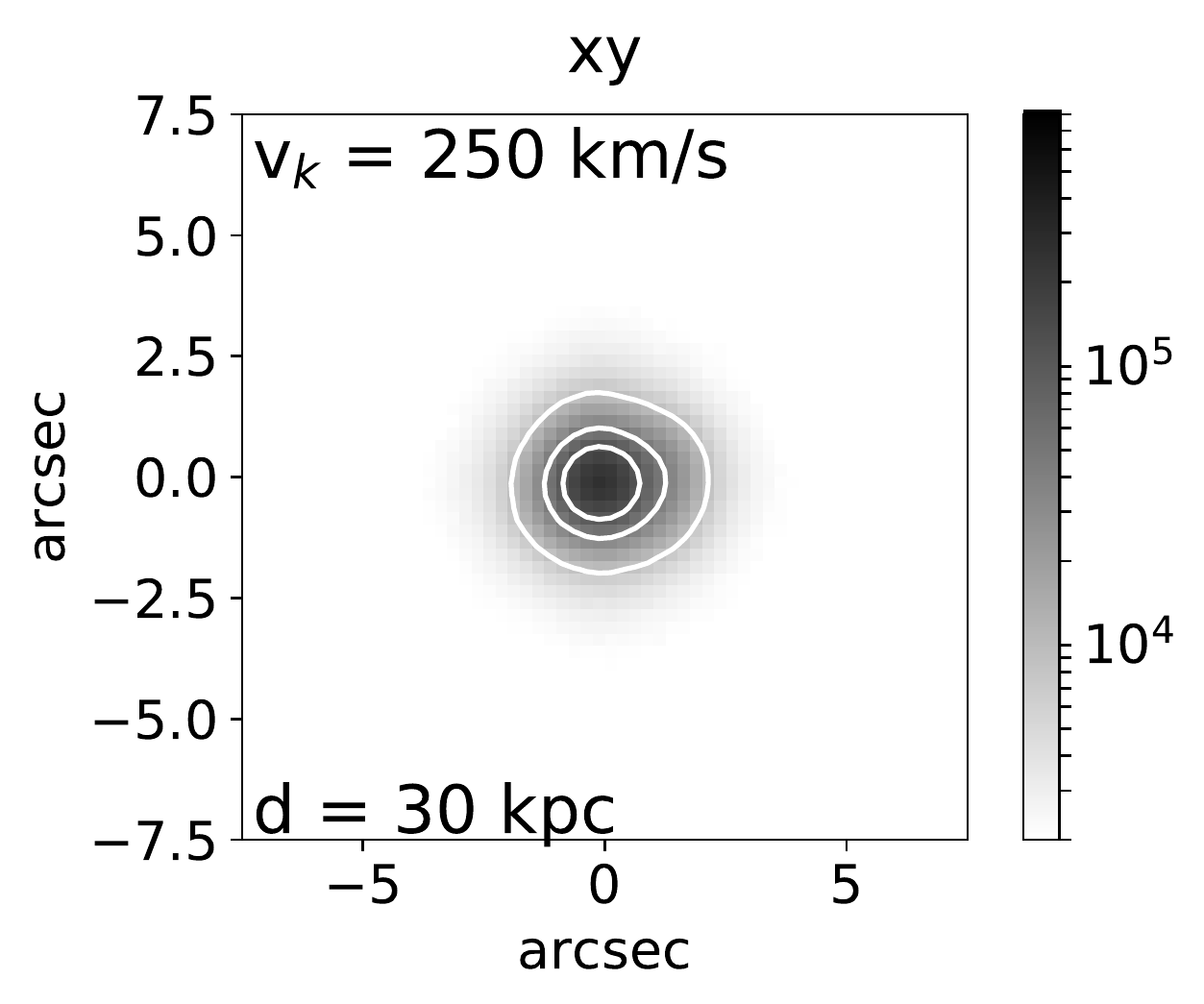}&
\includegraphics[trim= 2.4cm 1cm 2cm 1cm, clip=true, scale=0.35]{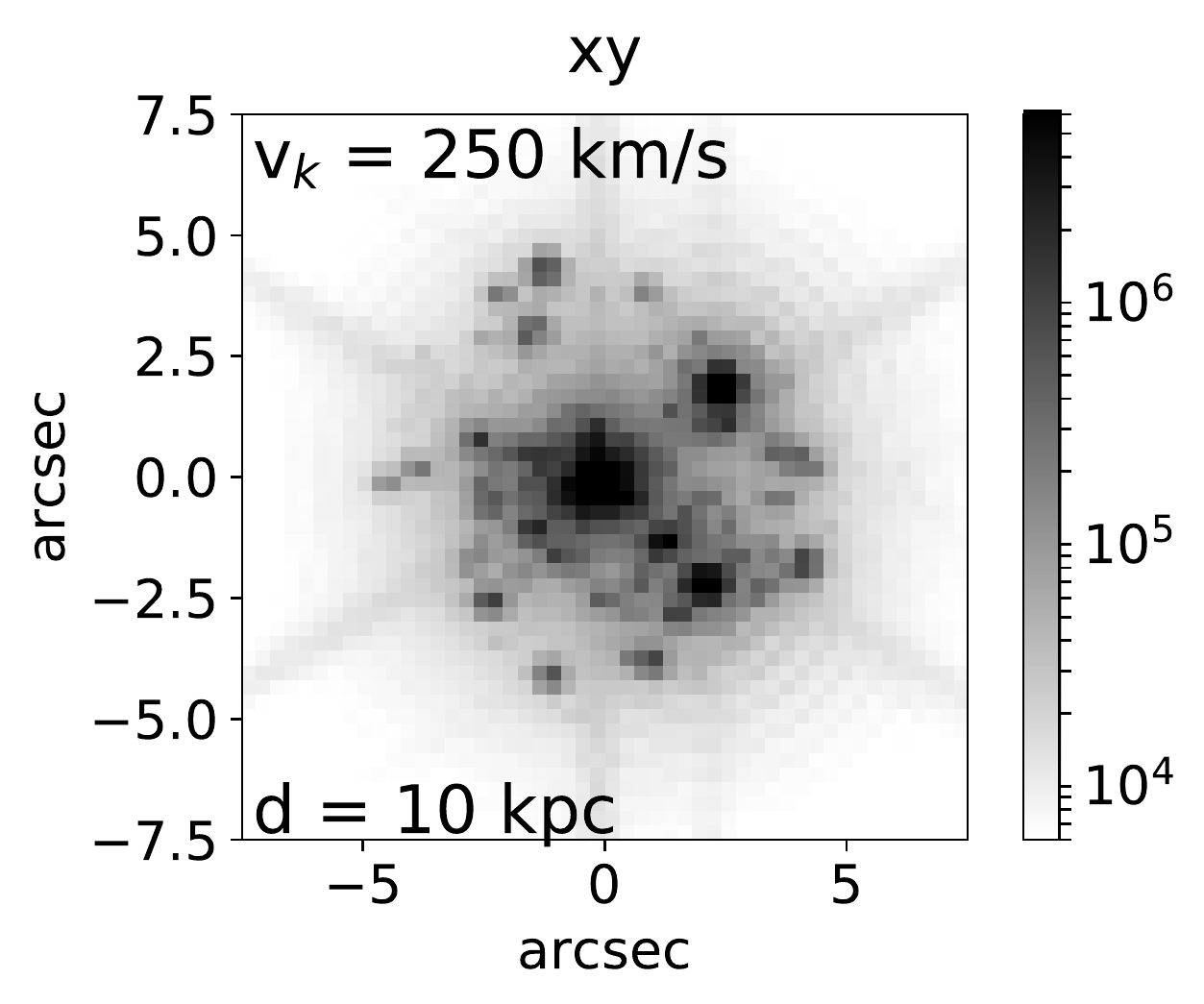}&
\includegraphics[trim= 2.4cm 1cm 0cm 1cm, clip=true, scale=0.35]{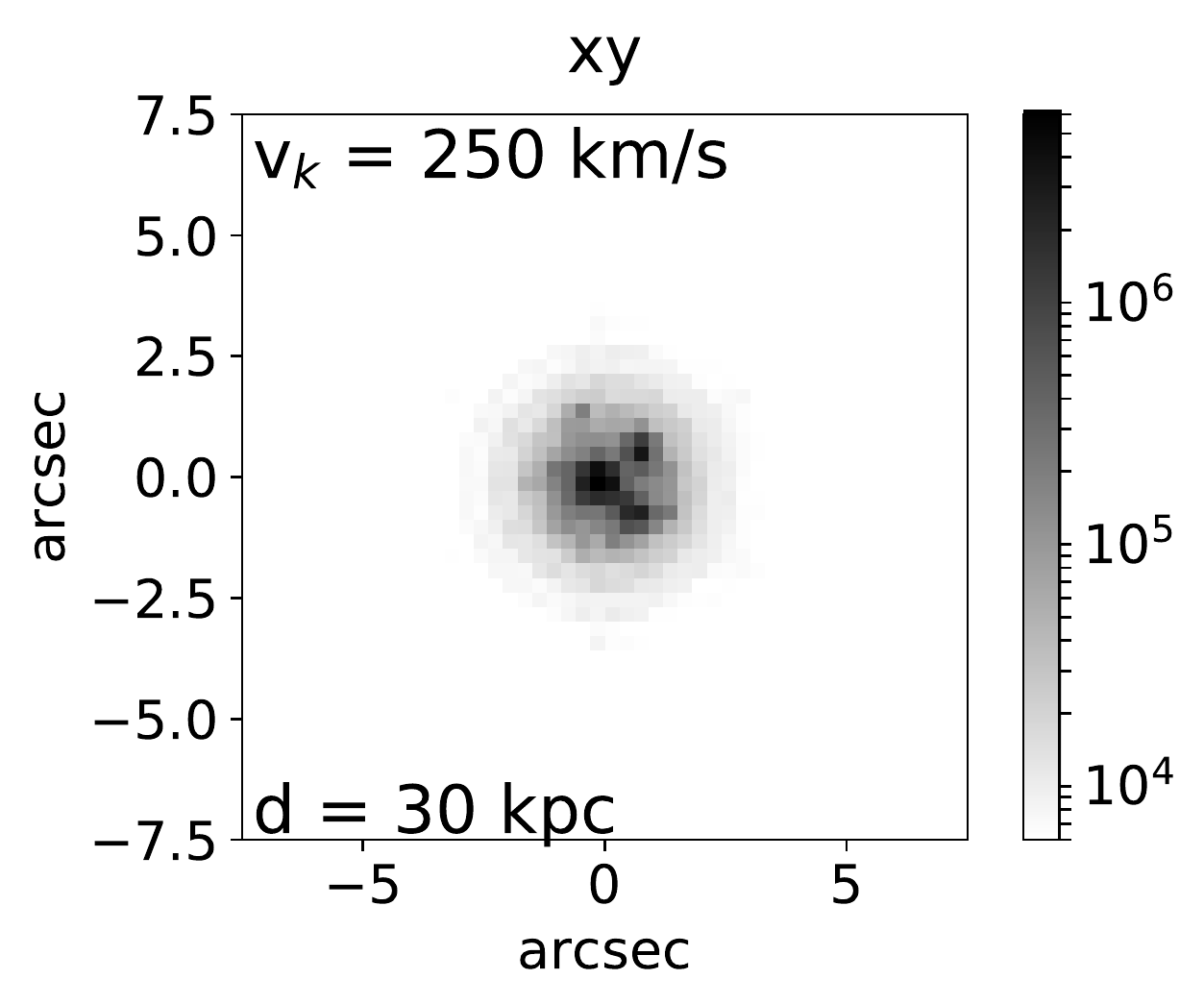}\\ 

\includegraphics[trim= 0cm 0cm 2cm 1cm, clip=true, scale=0.35]{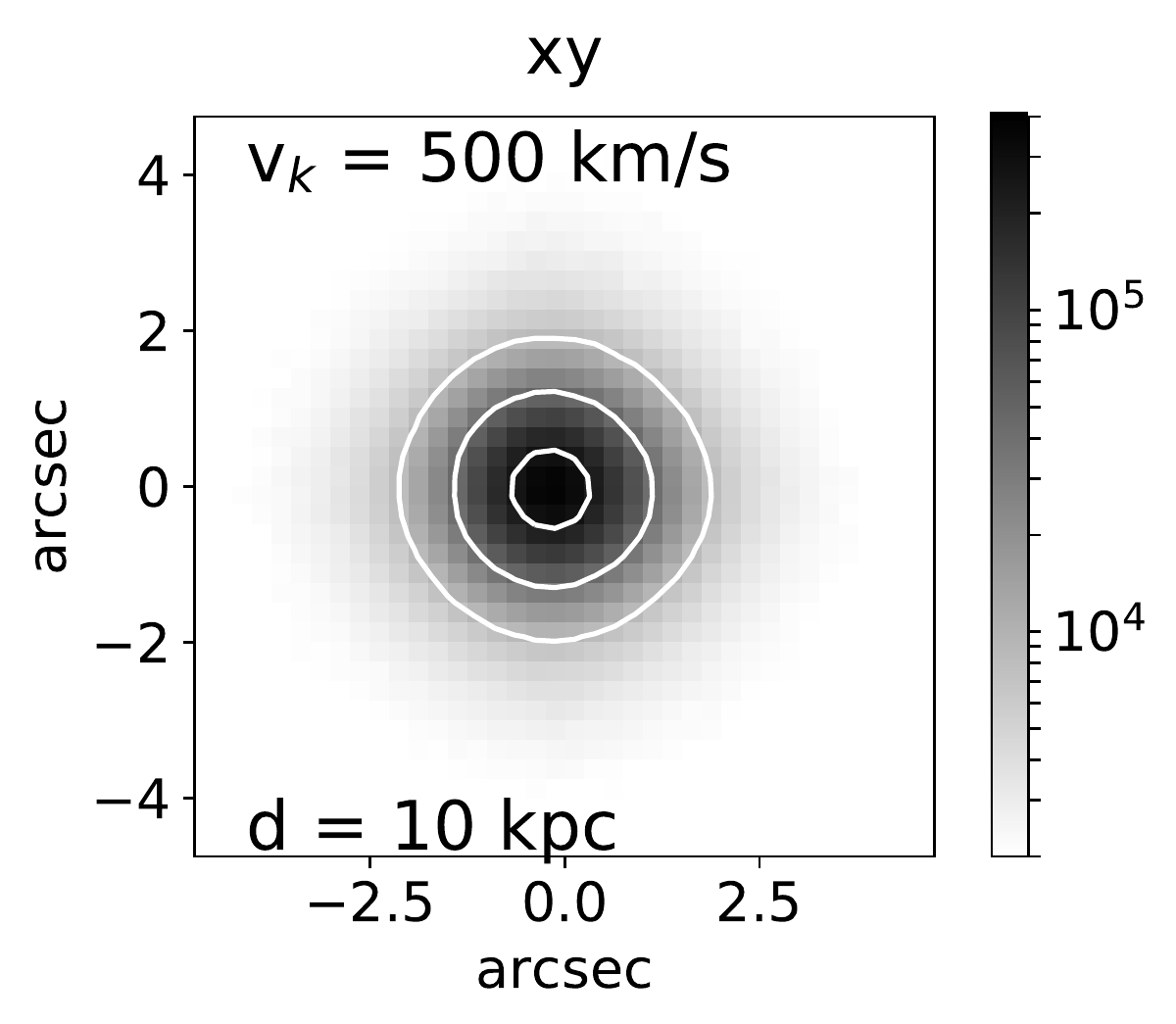}&
\includegraphics[trim= 1.8cm 0cm 0cm 1cm, clip=true, scale=0.35]{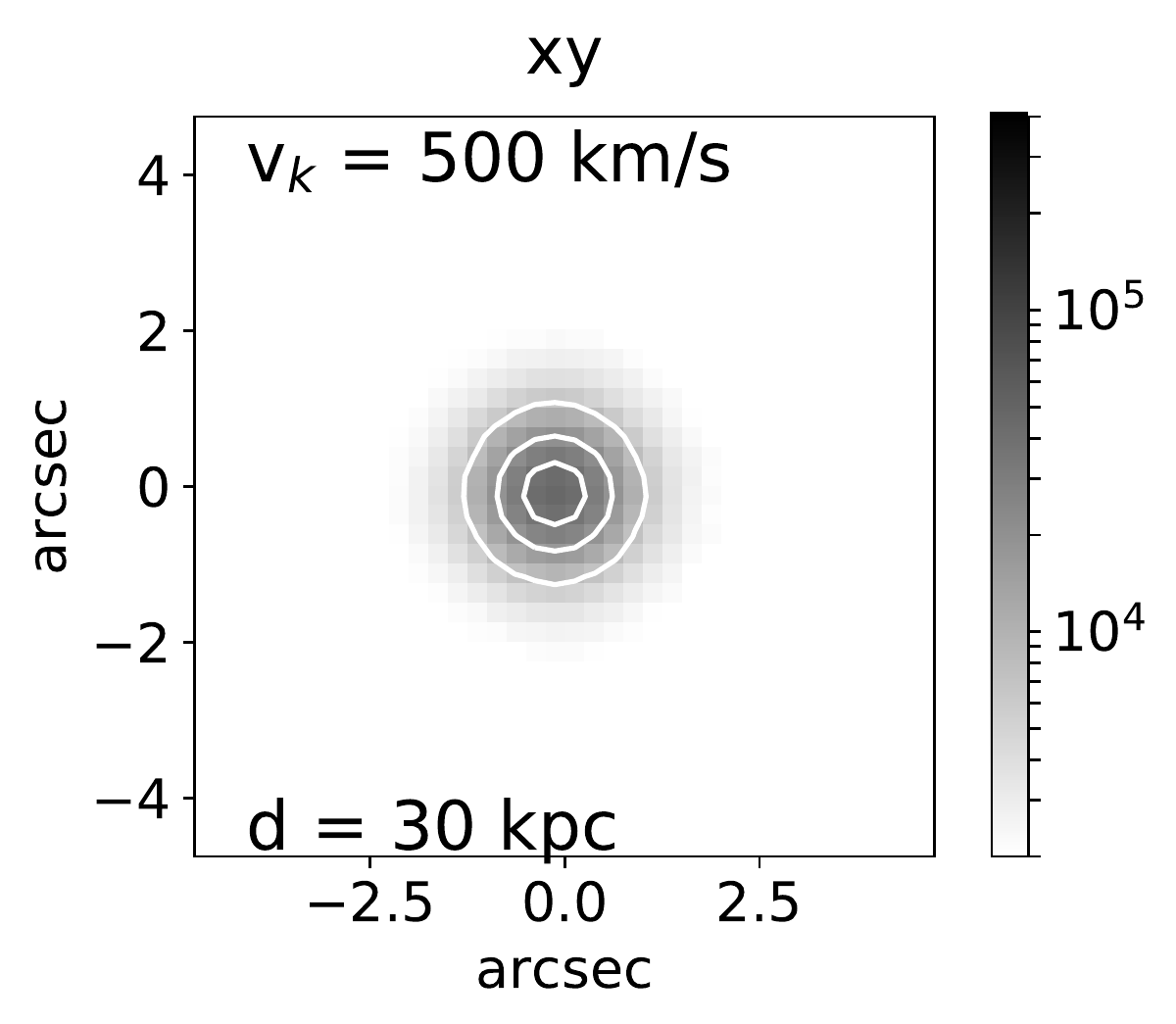}&
\includegraphics[trim= 1.8cm 0cm 2cm 1cm, clip=true, scale=0.35]{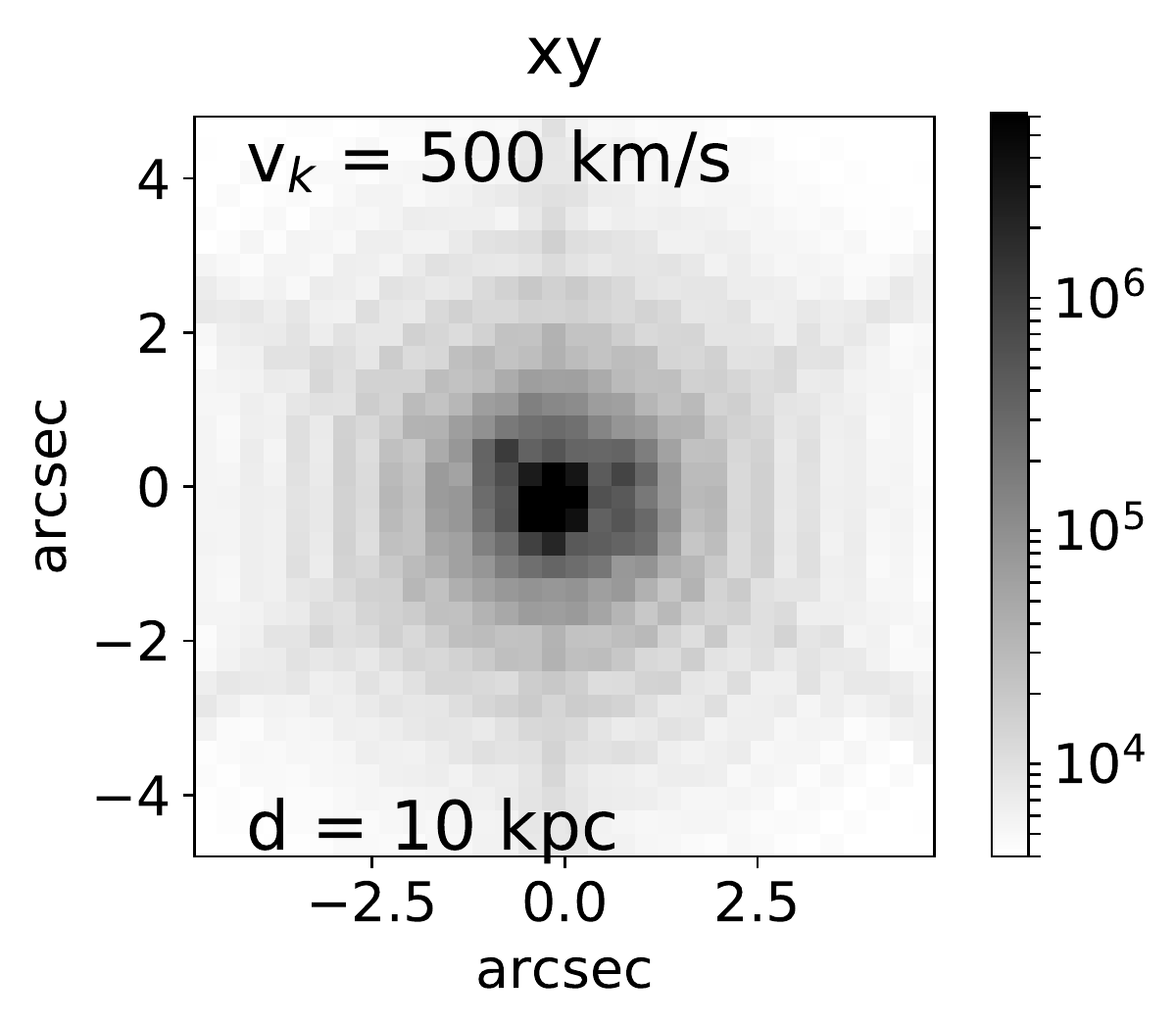}&
\includegraphics[trim= 1.8cm 0cm 0cm 1cm, clip=true, scale=0.35]{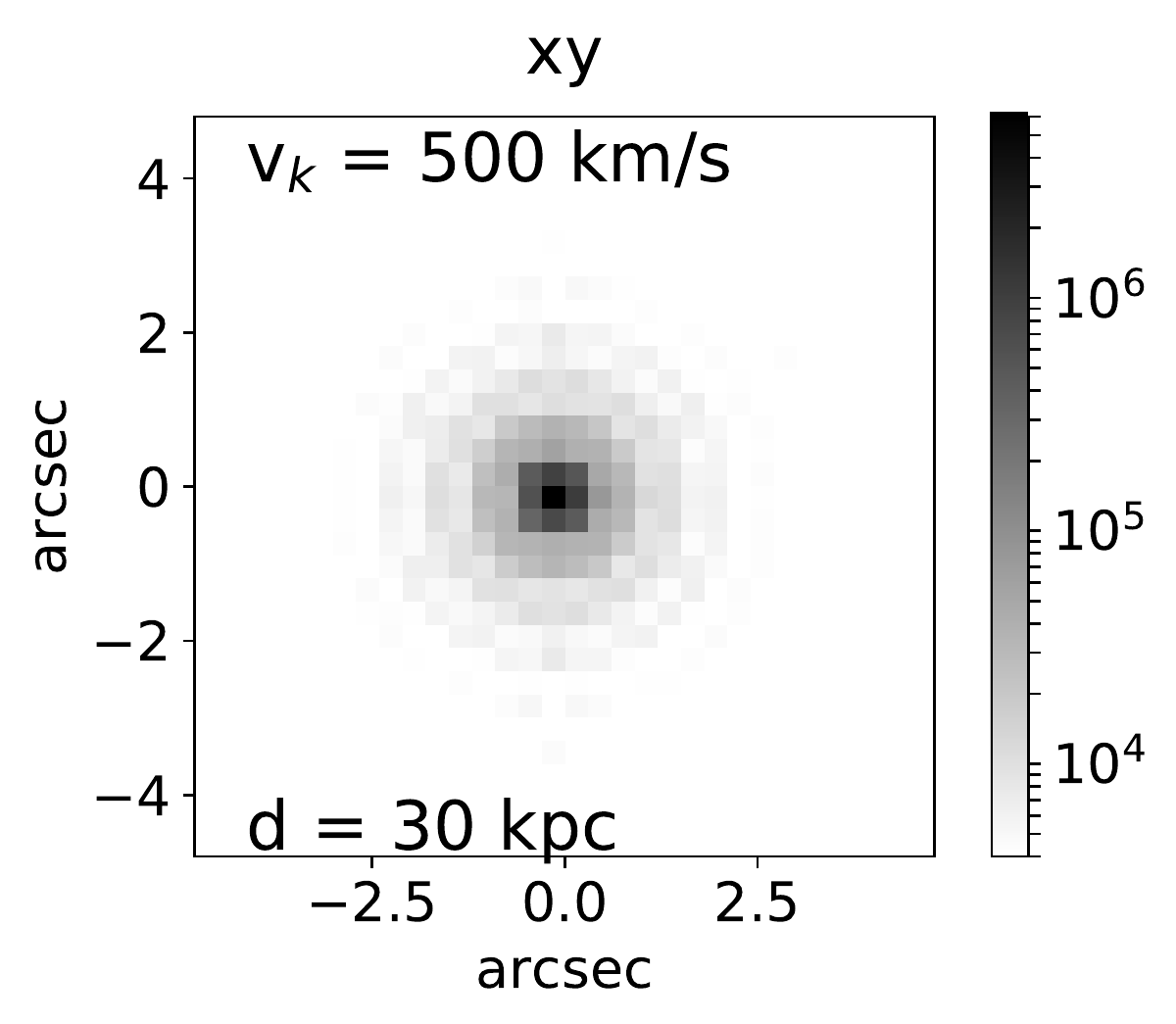}\\ 

\end{array}$
\end{center}
\caption{Renderings of HCSCs for Pan-STARRS (first and second column) and for \textit{Euclid}/NISP (third and fourth column). The cluster is bound to a $10^{5}$ M$_{\odot}$ black hole at distances of 10 and 30 kpc from the observer, as indicated in the bottom right of each panel. The kick velocity increases from 150 to 500 km s$^{-1}$ as indicated in the top right corner of the panels. Simulated exposure times are $t=40$ seconds for Pan-STARRS and $t=116$ seconds for NISP. The age of the stellar population is $\tau_{*} = 7$ Gyr, and the metallicity is Z$=0.02$ (solar). The time since the kick is $\tau_{k} = 100$ in N-body units, or $\tau_{k} = 1.25, 0.27, 0.03$ $\times 10^{4}$ yr; we note, however, that this dynamical state of the cluster should persist for a relaxation time (10$^{6}$ - 10$^{7}$ yr). The number of stars in each cluster is approximately 10000, 3000, and 500 (decreasing at higher values of V$_{k}$). The renderings correspond to the projection on the \textit{xy} plane of Fig.\ref{fig: hcss_nbody}, where the kick-induced asymmetry is maximised. Flux units: counts. Note the spatial-scale change in each row. Arbitrary contours are added to highlight the clusters morphology.}
\label{fig: hcss_mock_psnisp}
\end{figure*}

As eq.\ref{eq: mb} shows, the stellar mass bound to the recoiling BH right after the kick is a function of the kick velocity. The effects of different kick velocities on a HCSC bound to a 10$^{5}$ M$_{\odot}$ black hole are shown in Fig. \ref{fig: hcss_mock_psnisp}, which simulates a 40 second Pan-STARRS-DR1 exposure in the $r$-band, and a 116 second \textit{Euclid}/NISP exposure in the \textit{J}-band (these are the nominal durations of single exposures for the $3\pi$ Pan-STARRS survey, e.g. \citealt{SchlaflyFJ12}, and for the wide field survey planned for \textit{Euclid}, e.g. \citealt{Carry18}).

\subsubsection{Cluster evolution after the kick}
\label{subs: evafter_kick}
After receiving the gravitational-wave kick, the cluster is believed to evolve via resonant relaxation \citep{MSK2009,OLearyL12}. More specifically, a fraction of the stars is lost via tidal disruption events, and a fraction is ejected because of large-angle scattering \citep{Henon69,LinT80}. \citet{MSK2009} and \citet{OLearyL12} performed dynamical simulations to quantify the effects due to these processes, finding that the rate of stellar loss depends on the mass of the BH, with clusters around more massive BHs evolving more slowly than clusters around lower mass BHs.

\begin{figure*}
\begin{subfigures}
\begin{center}$
\begin{array}{cccc}
\includegraphics[trim= 0cm 1cm 2cm 1cm, clip=true, scale=0.38]{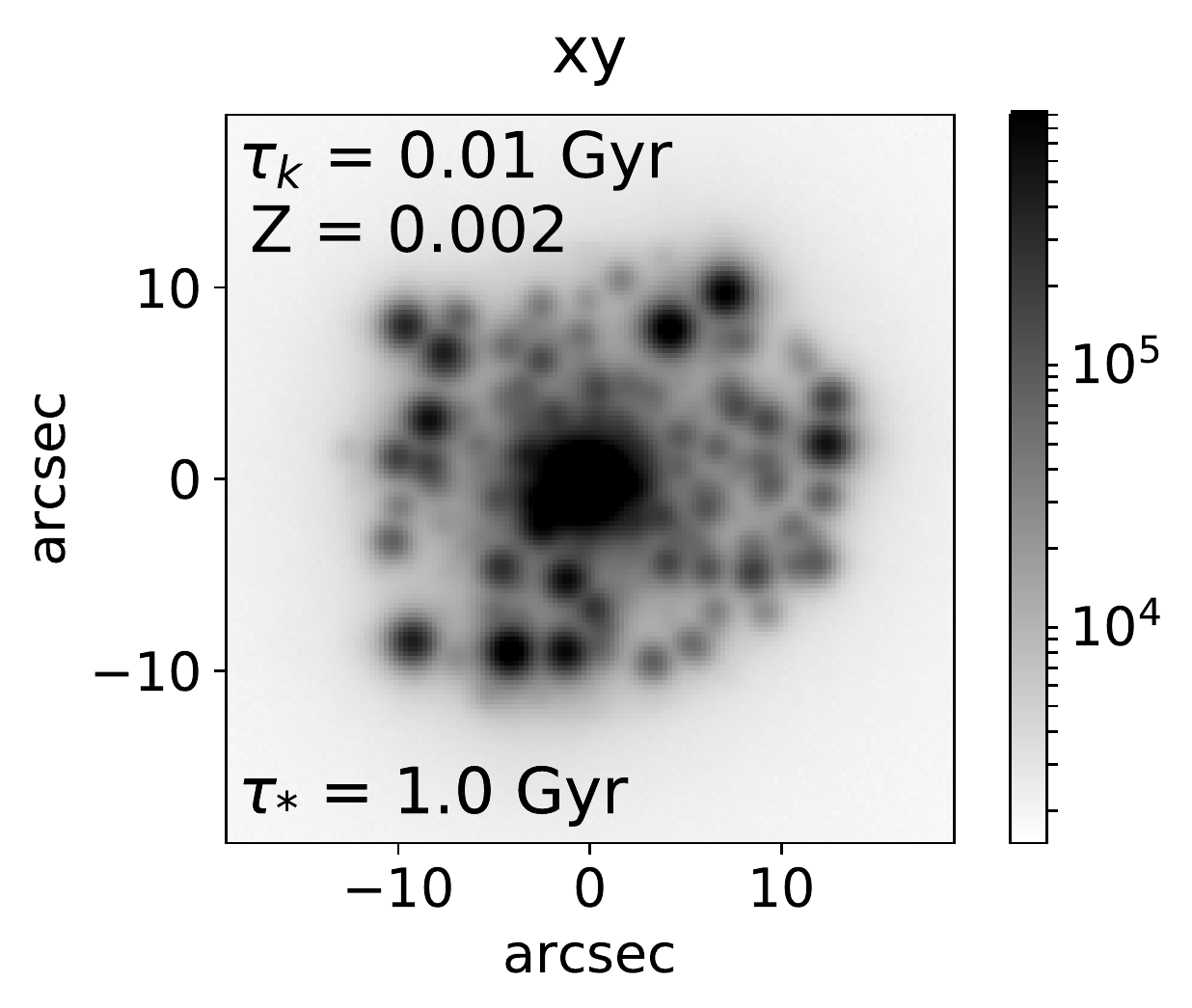}&
\includegraphics[trim= 1cm 1cm 2cm 1cm, clip=true, scale=0.38]{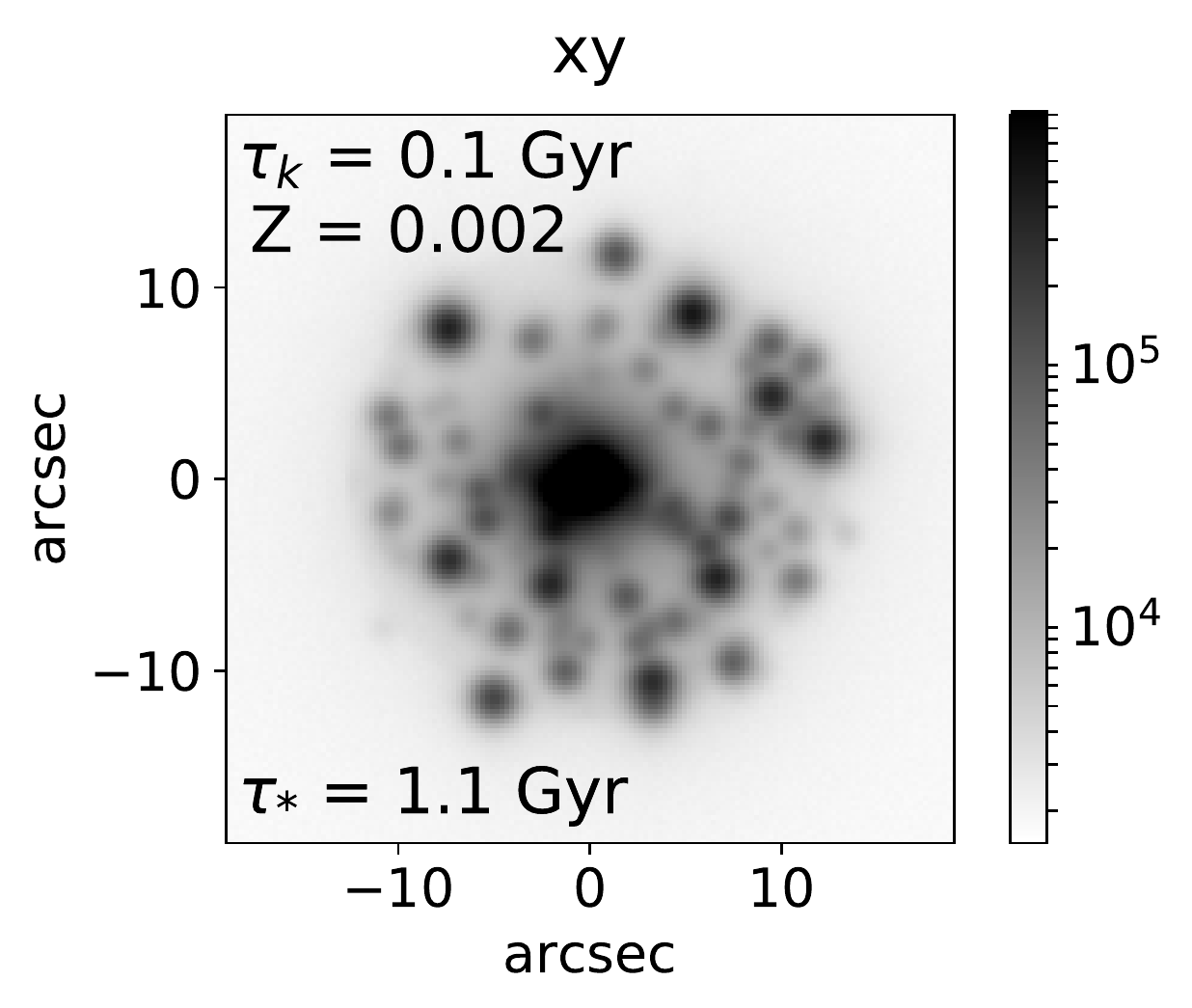}&
\includegraphics[trim= 1cm 1cm 2cm 1cm, clip=true, scale=0.38]{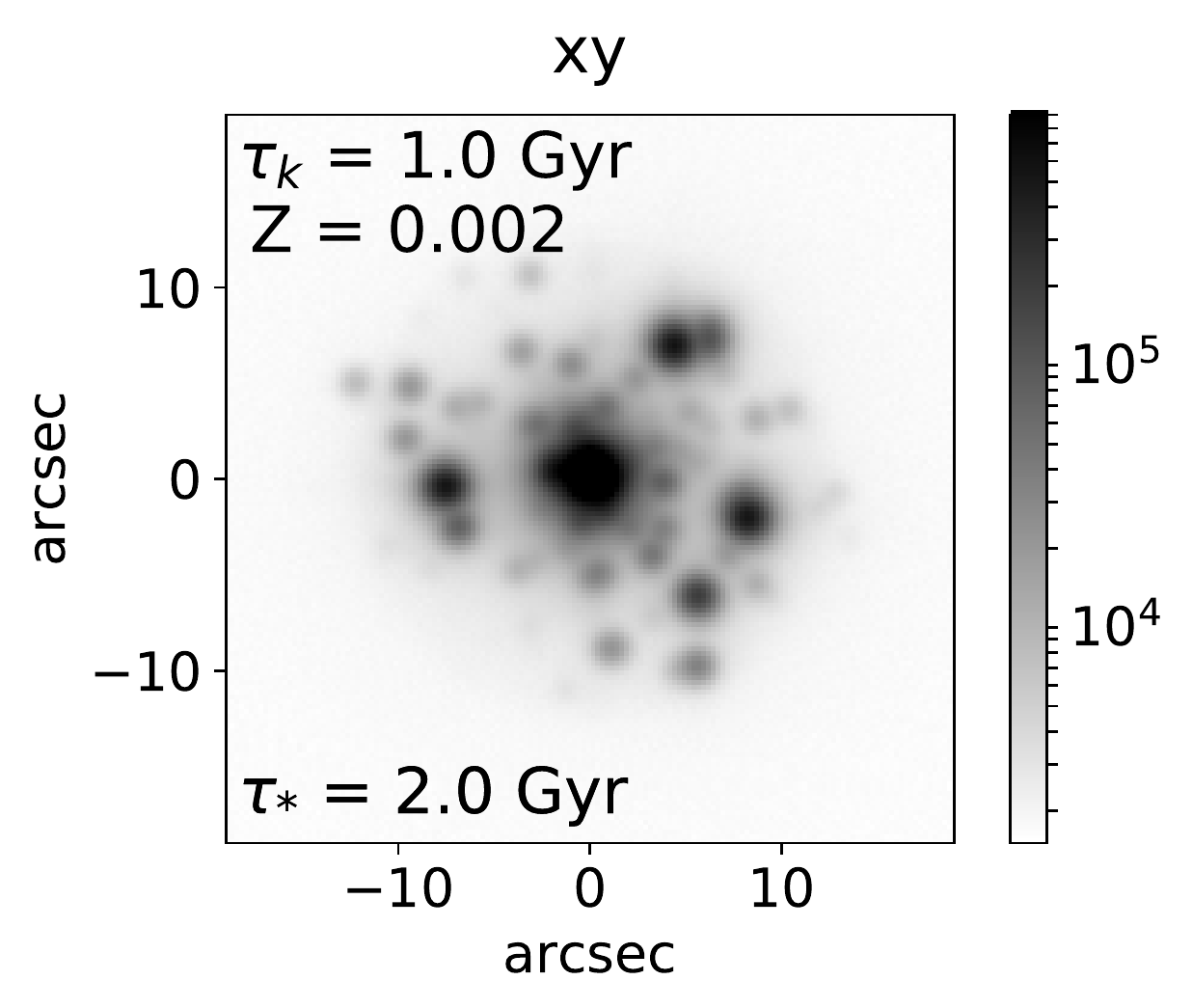}&
\includegraphics[trim= 1cm 1cm 0cm 0.9cm, clip=true, scale=0.38]{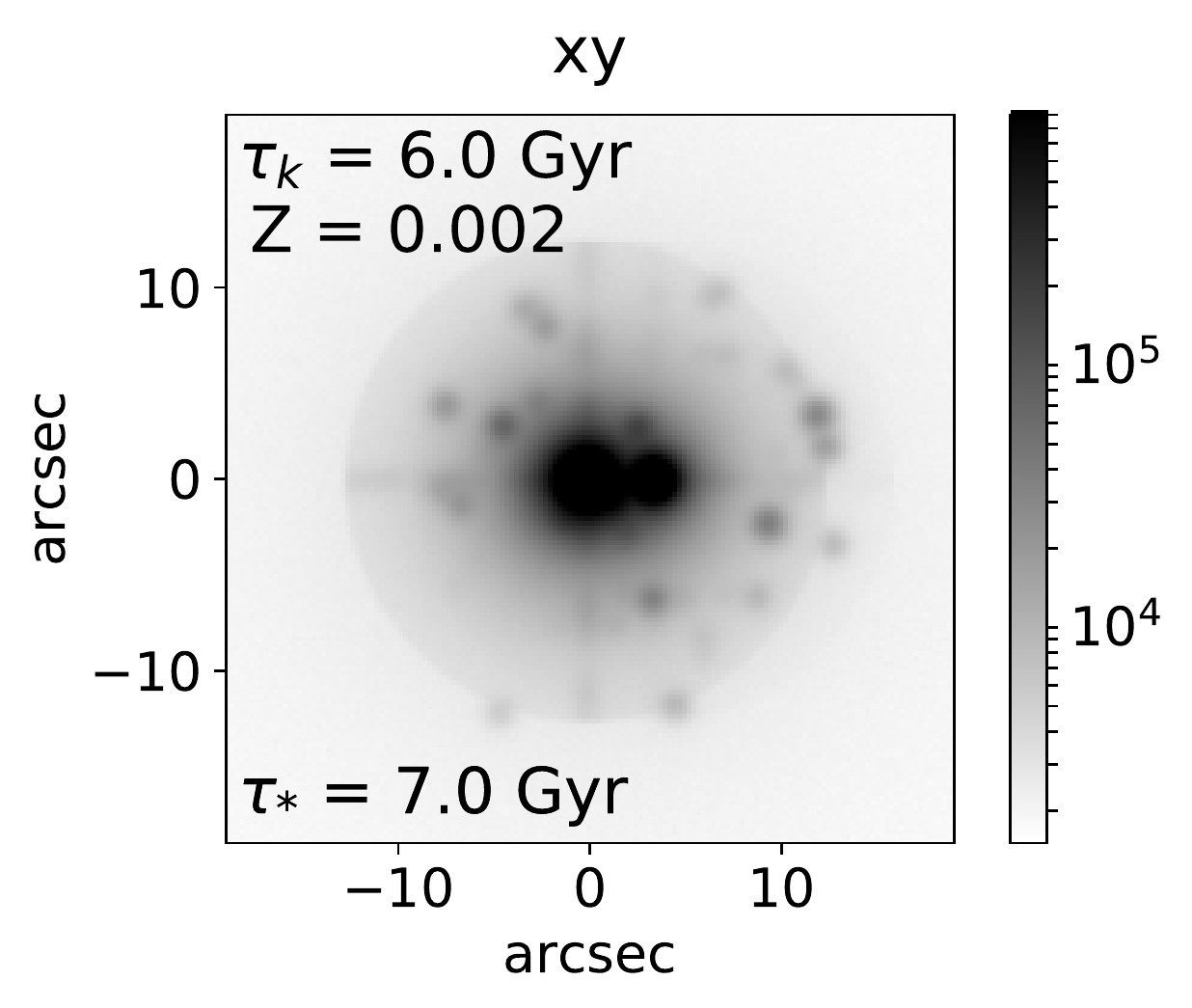}\\

\includegraphics[trim= 0cm 0cm 2cm 1cm, clip=true, scale=0.38]{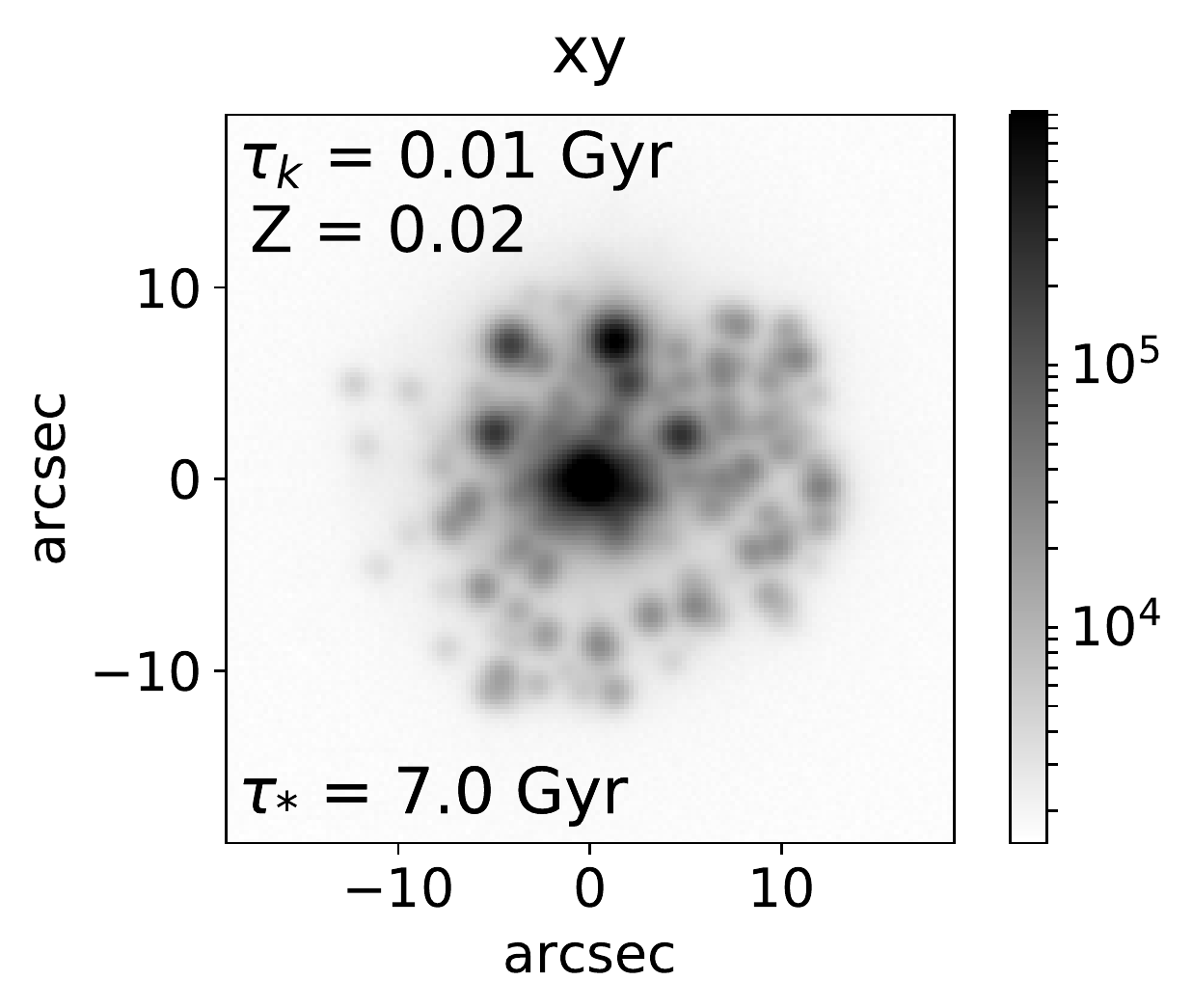}&
\includegraphics[trim= 1cm 0cm 2cm 1cm, clip=true, scale=0.38]{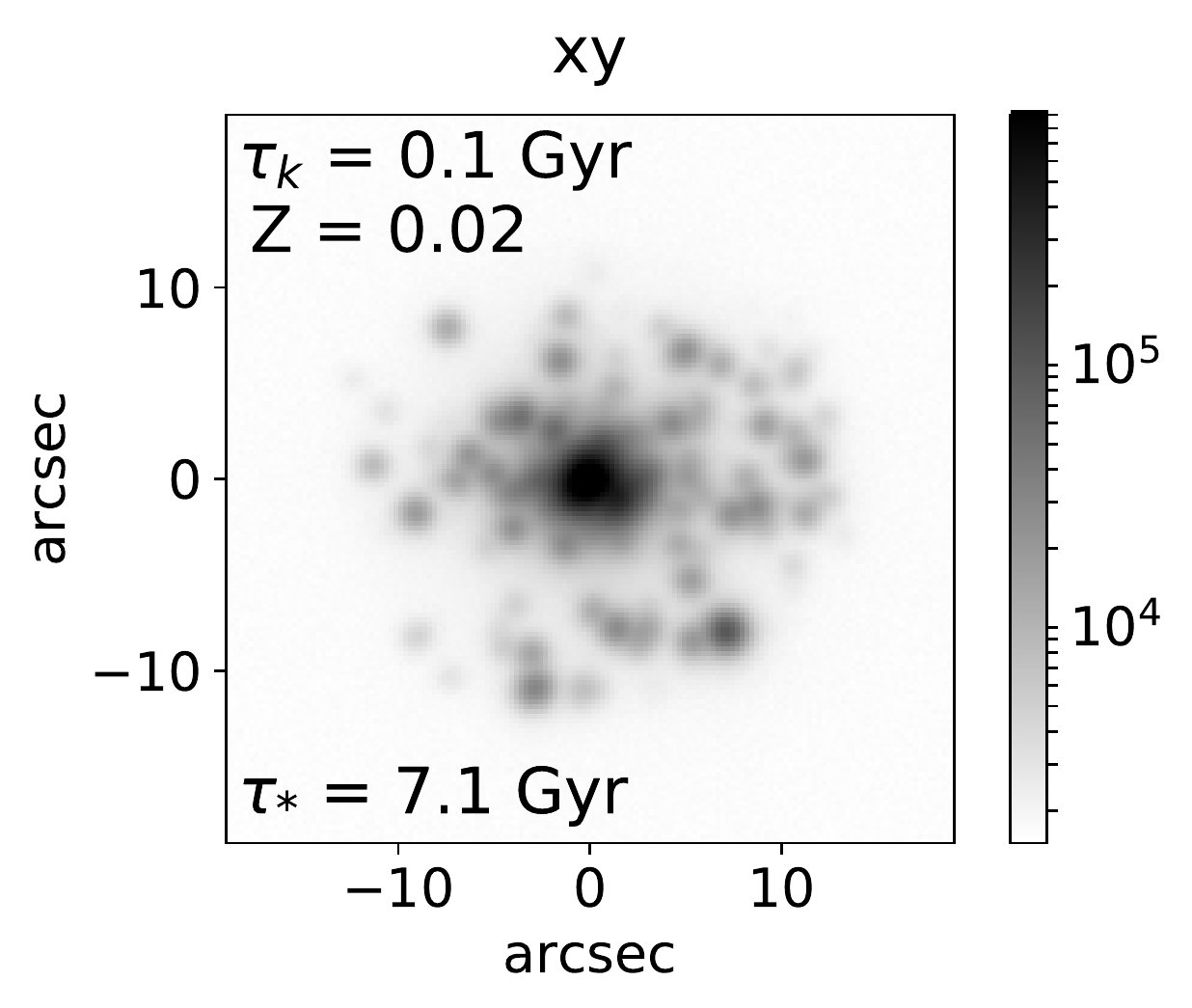}&
\includegraphics[trim= 1cm 0cm 2cm 0.9cm, clip=true, scale=0.38]{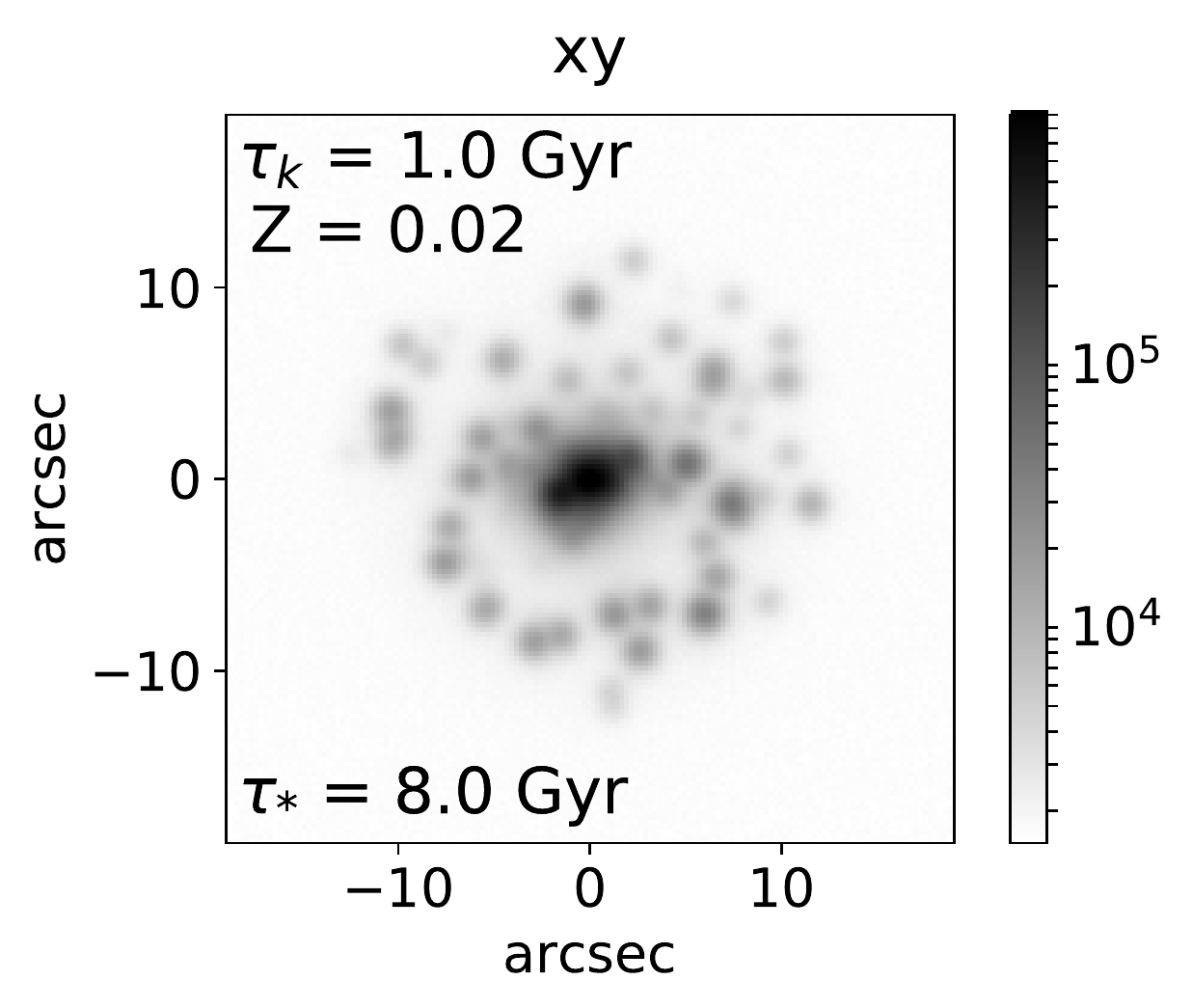}&
\includegraphics[trim= 1cm 0cm 0cm 0.9cm, clip=true, scale=0.38]{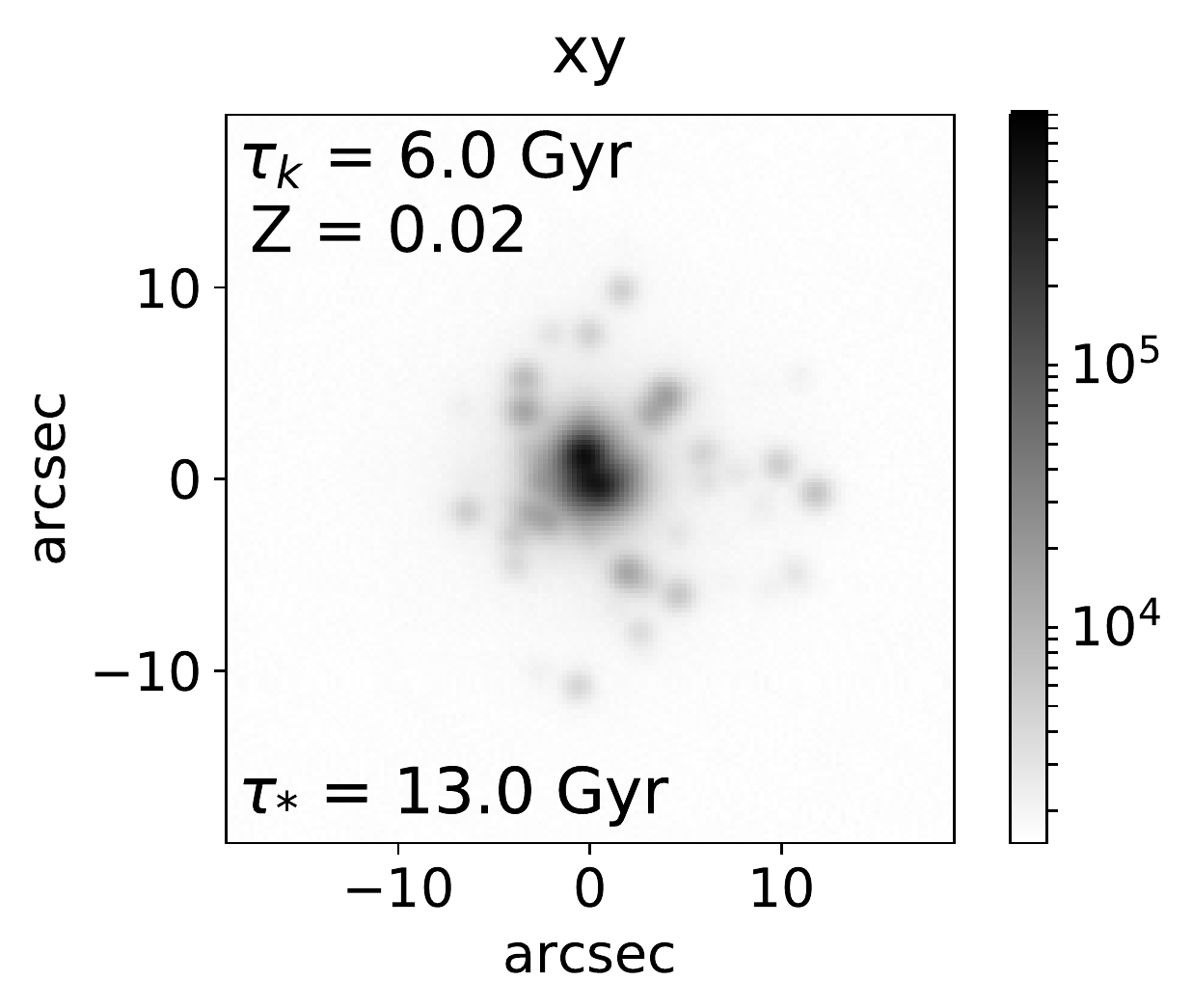}\\

\end{array}$
\end{center}
\caption{Mock Pan-STARRS-DR1 images of an HCSC as a function of the time since the kick, $\tau_{k}$, and as a function of the age of the stellar population, $\tau_{\star}$. Assumed parameters are: kick velocity V$_{k} = 150$ km s$^{-1}$, distance = 10 kpc, black hole mass M$_{\bullet}= 10^{5}$ M$_{\odot}$. The number of stars in each panel (left to right) is approximately 8000, 5000, 2000, and 1000.}
\label{fig: hcss_mock_ps1_tk}

\begin{center}$
\begin{array}{cccc}

\includegraphics[trim= 0cm 1cm 2cm 1cm, clip=true, scale=0.37]{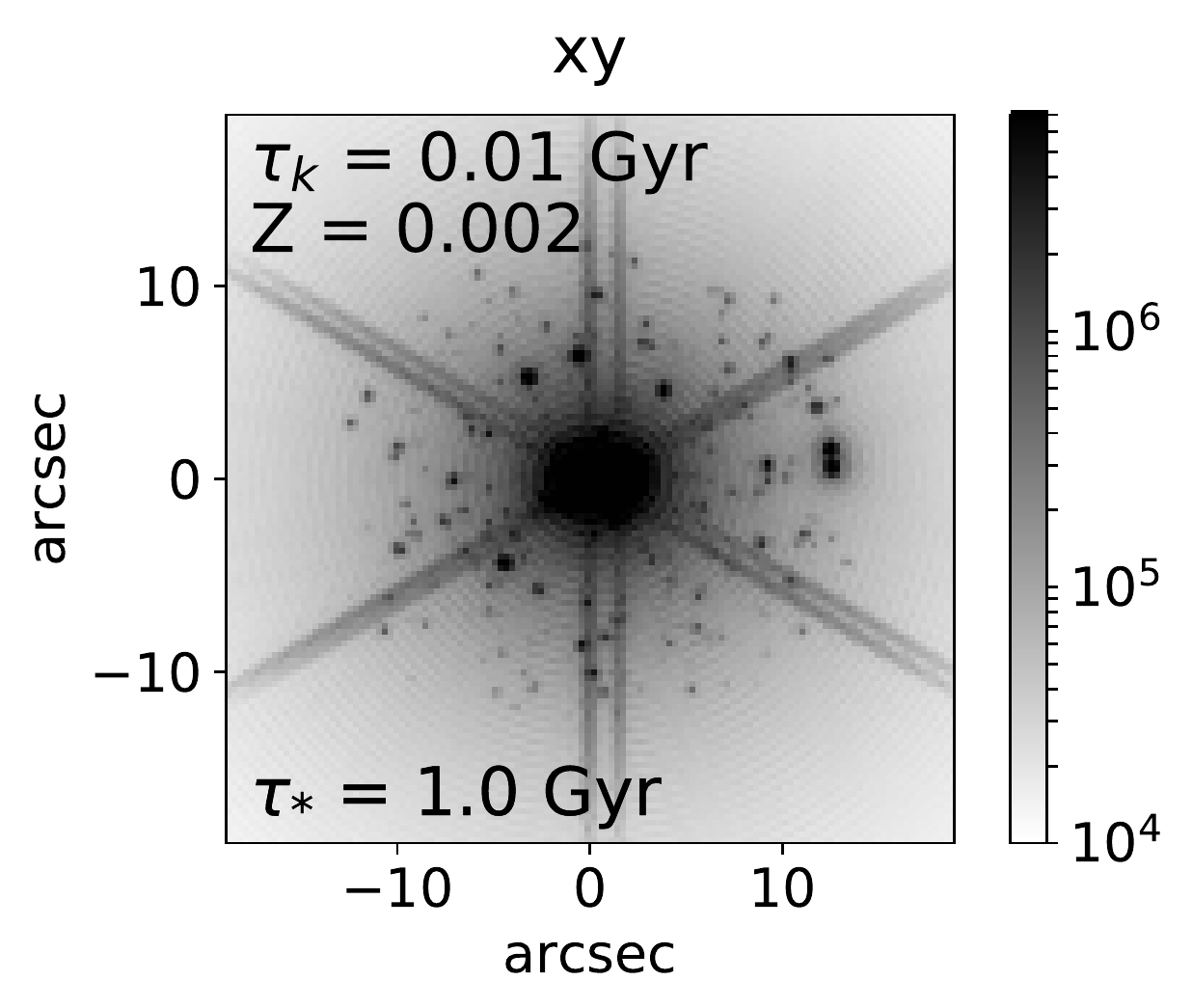}&
\includegraphics[trim= 1cm 1cm 2cm 1cm, clip=true, scale=0.37]{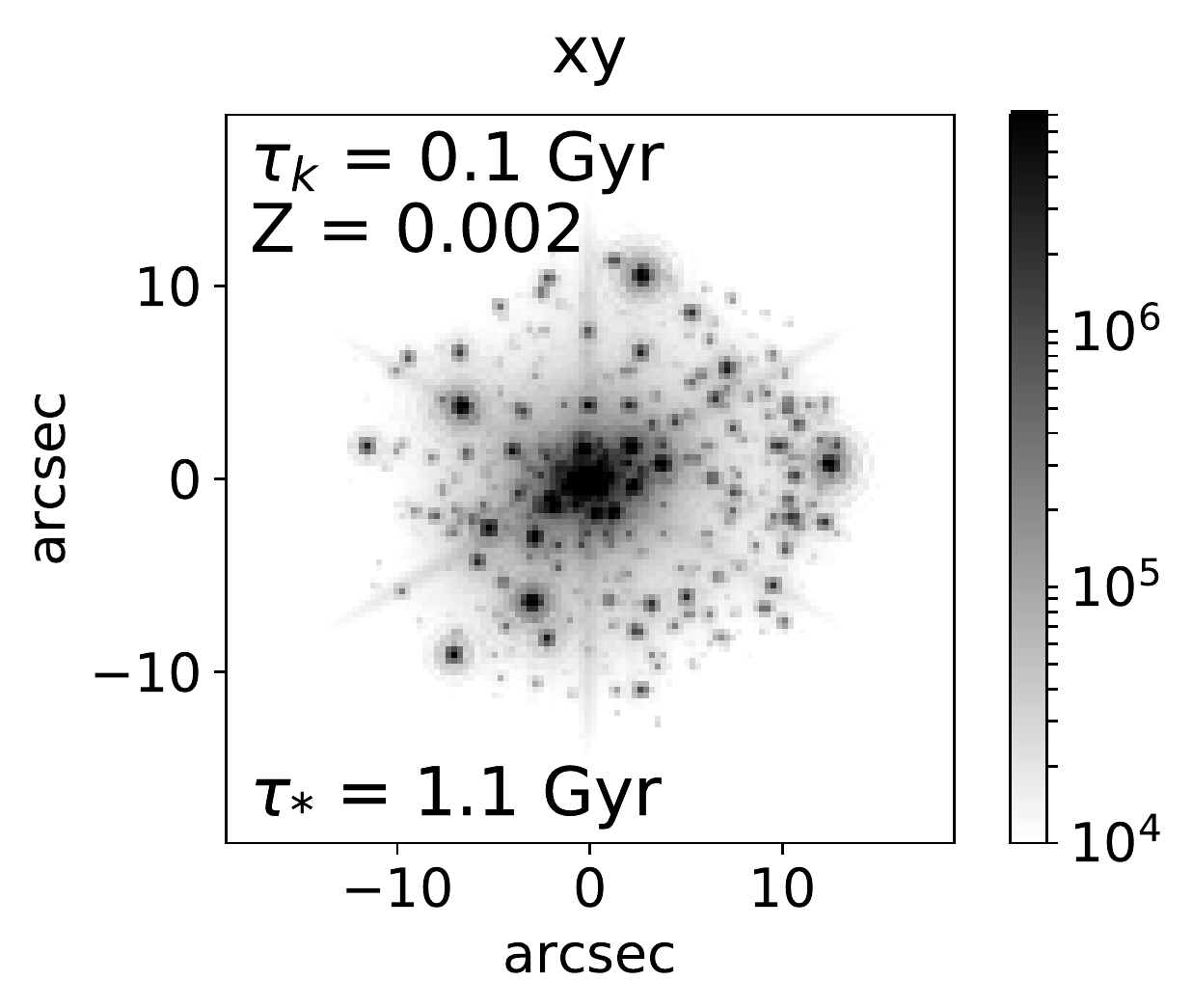}&
\includegraphics[trim= 1cm 1cm 2cm 1cm, clip=true, scale=0.37]{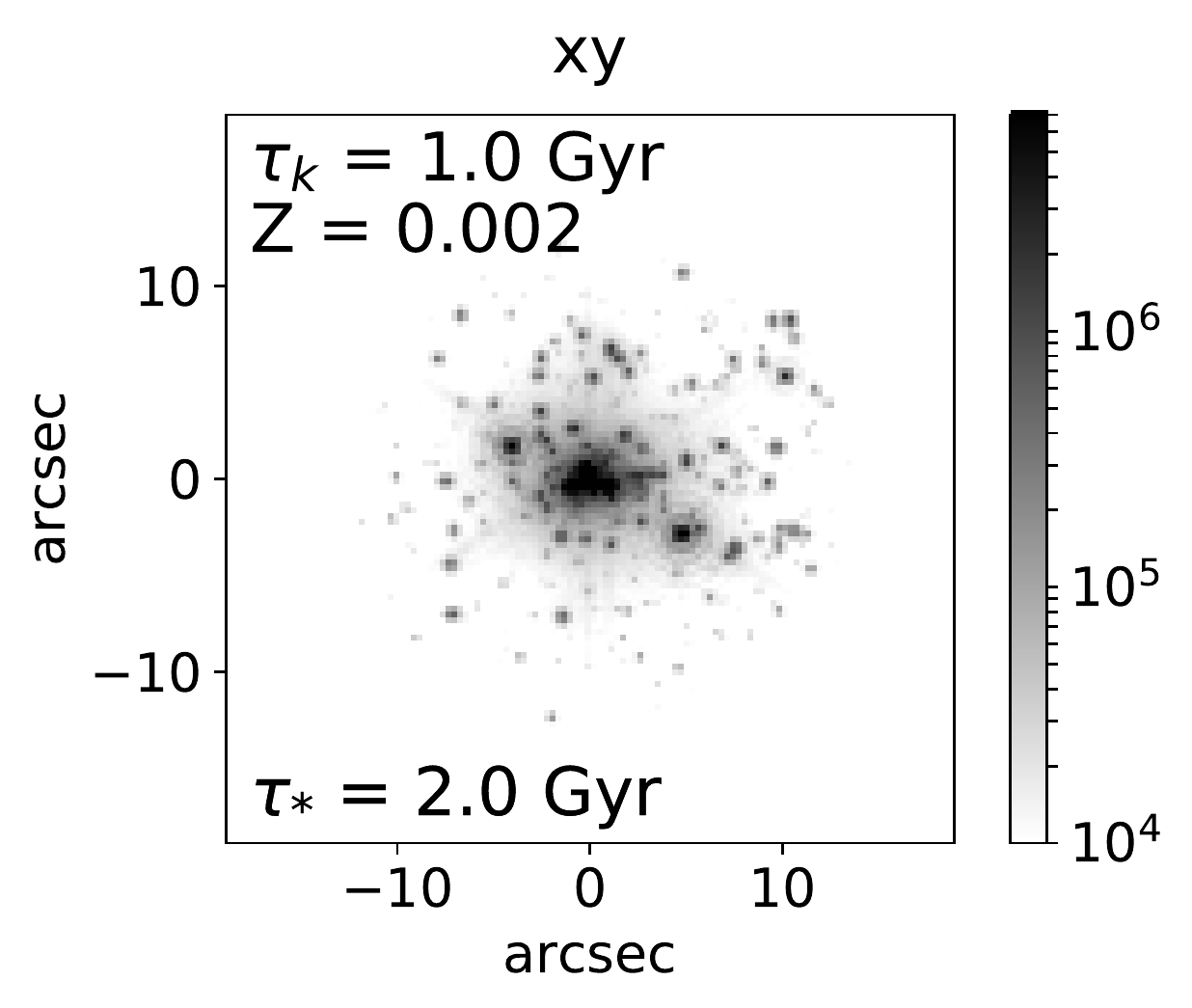}&
\includegraphics[trim= 1cm 1cm 0cm 0.9cm, clip=true, scale=0.37]{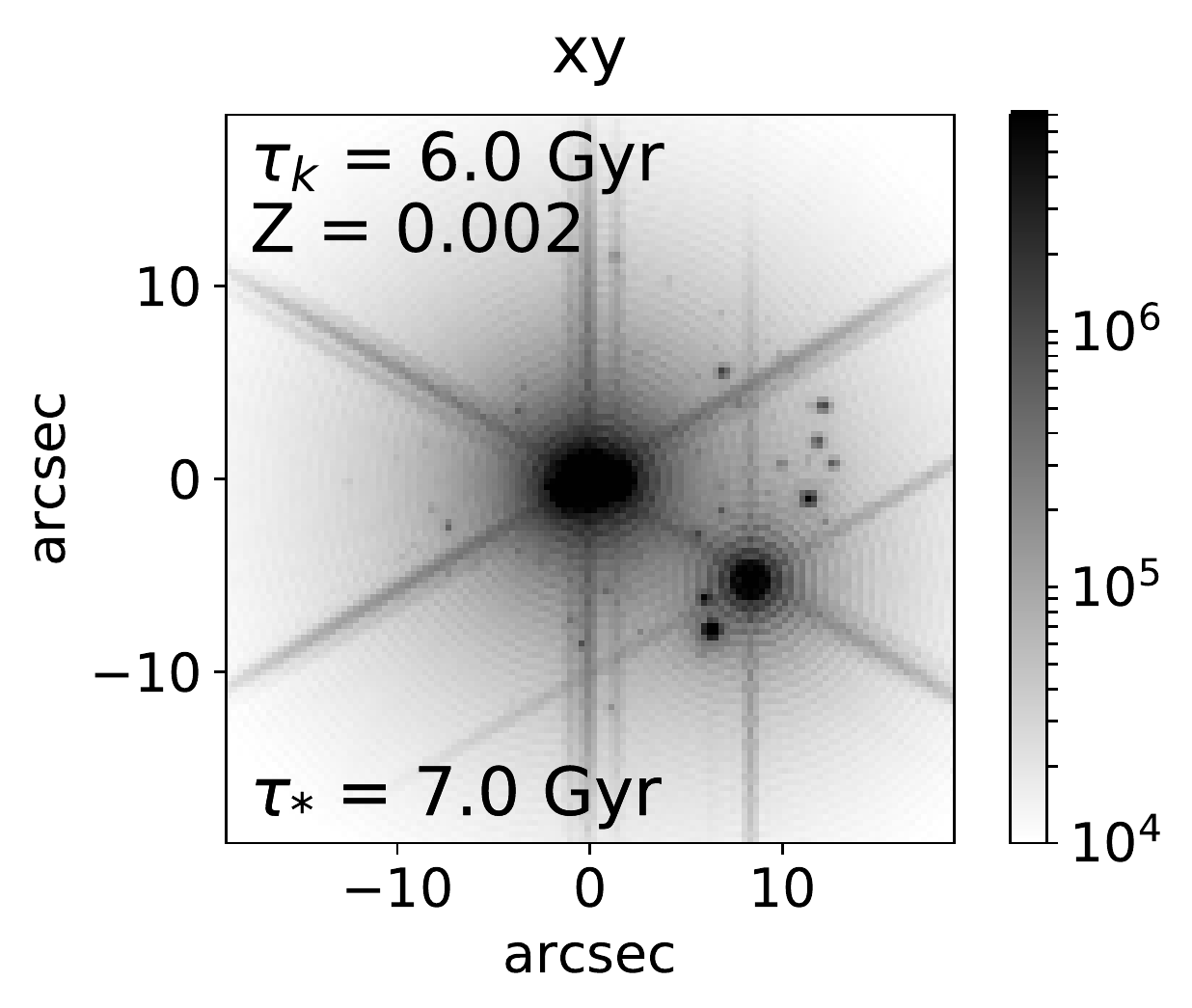}\\

\includegraphics[trim= 0cm 0cm 2cm 1cm, clip=true, scale=0.38]{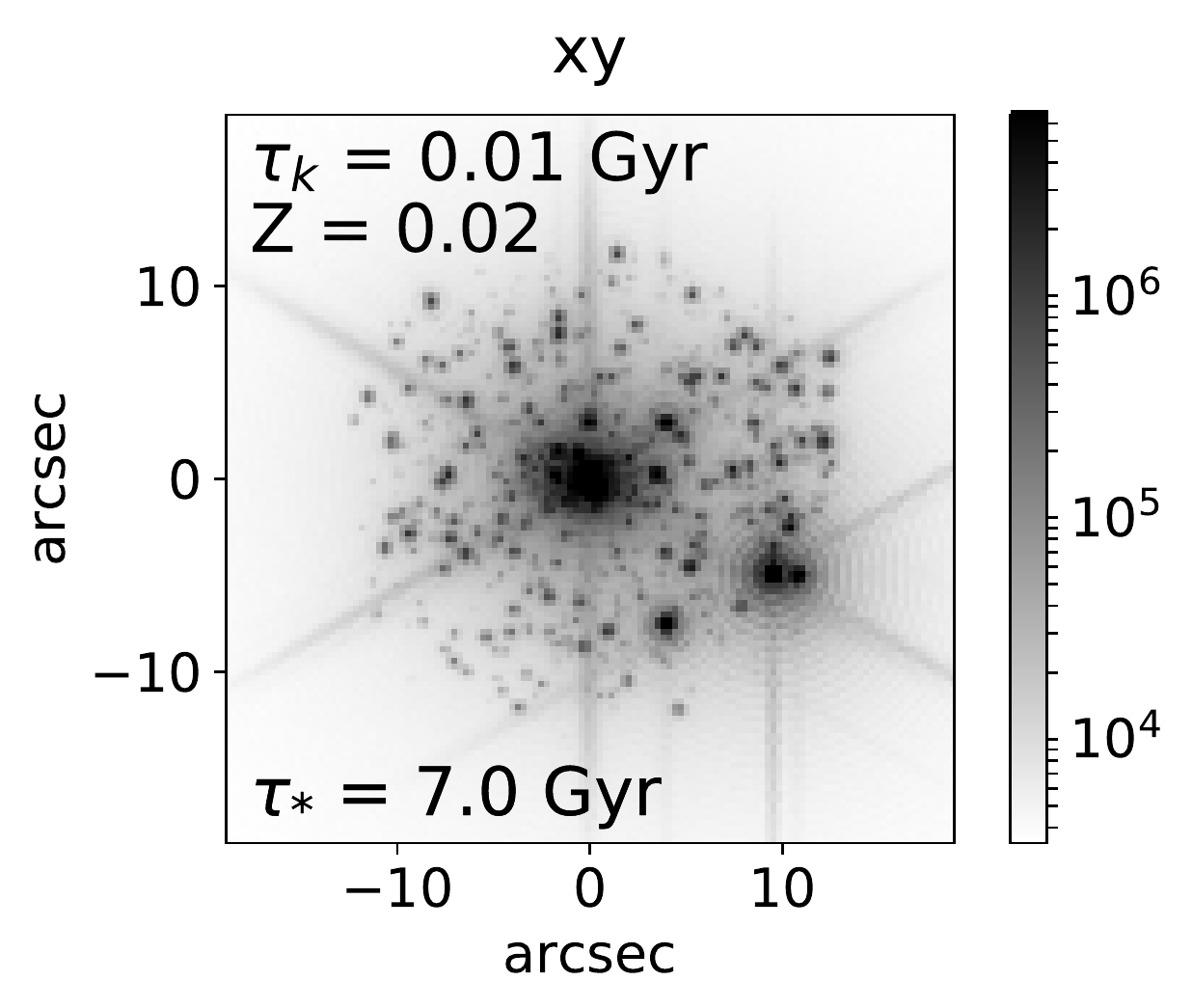}&
\includegraphics[trim= 1cm 0cm 2cm 1cm, clip=true, scale=0.38]{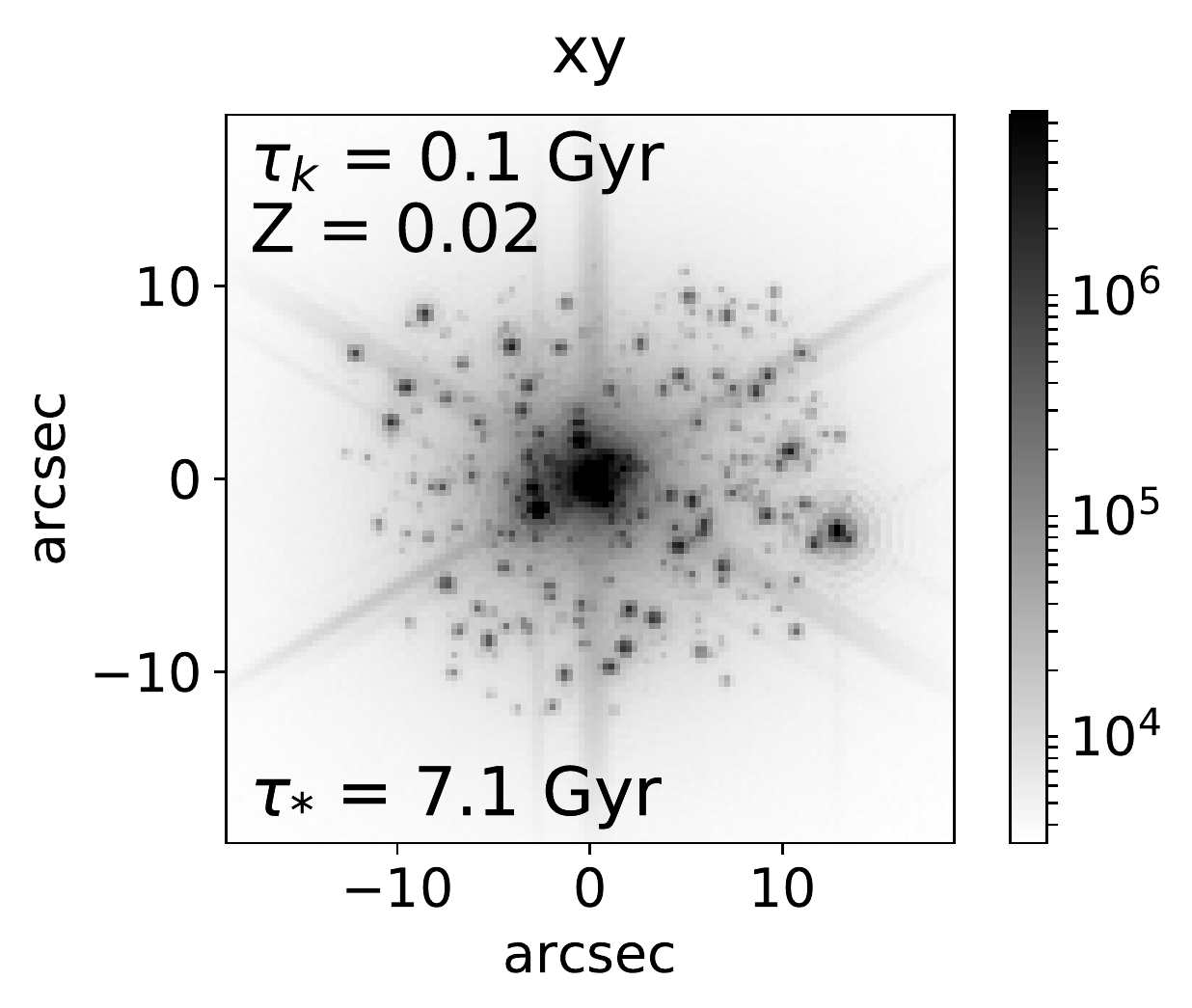}&
\includegraphics[trim= 1cm 0cm 2cm 0.9cm, clip=true, scale=0.38]{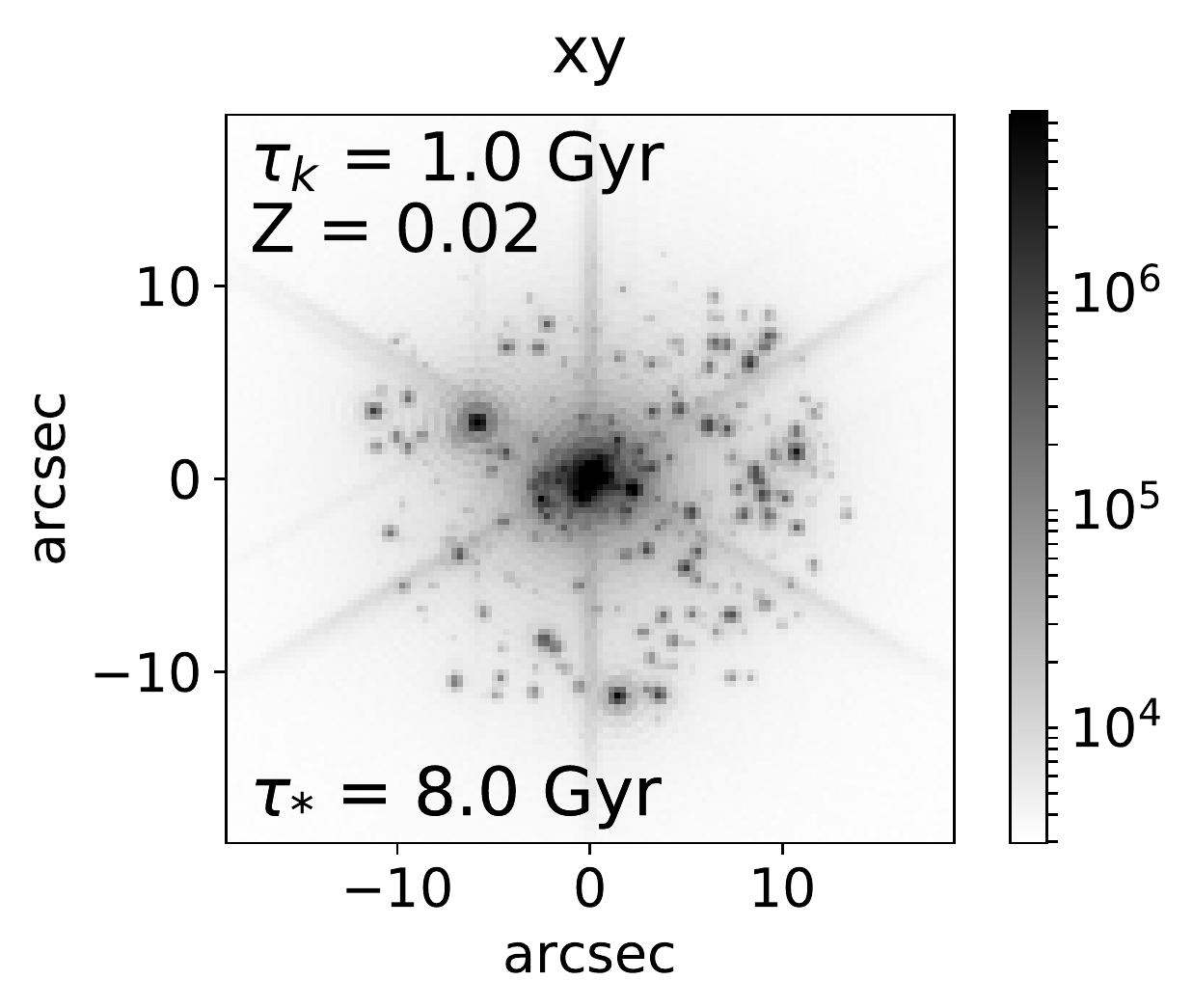}&
\includegraphics[trim= 1cm 0cm 0cm 0.9cm, clip=true, scale=0.38]{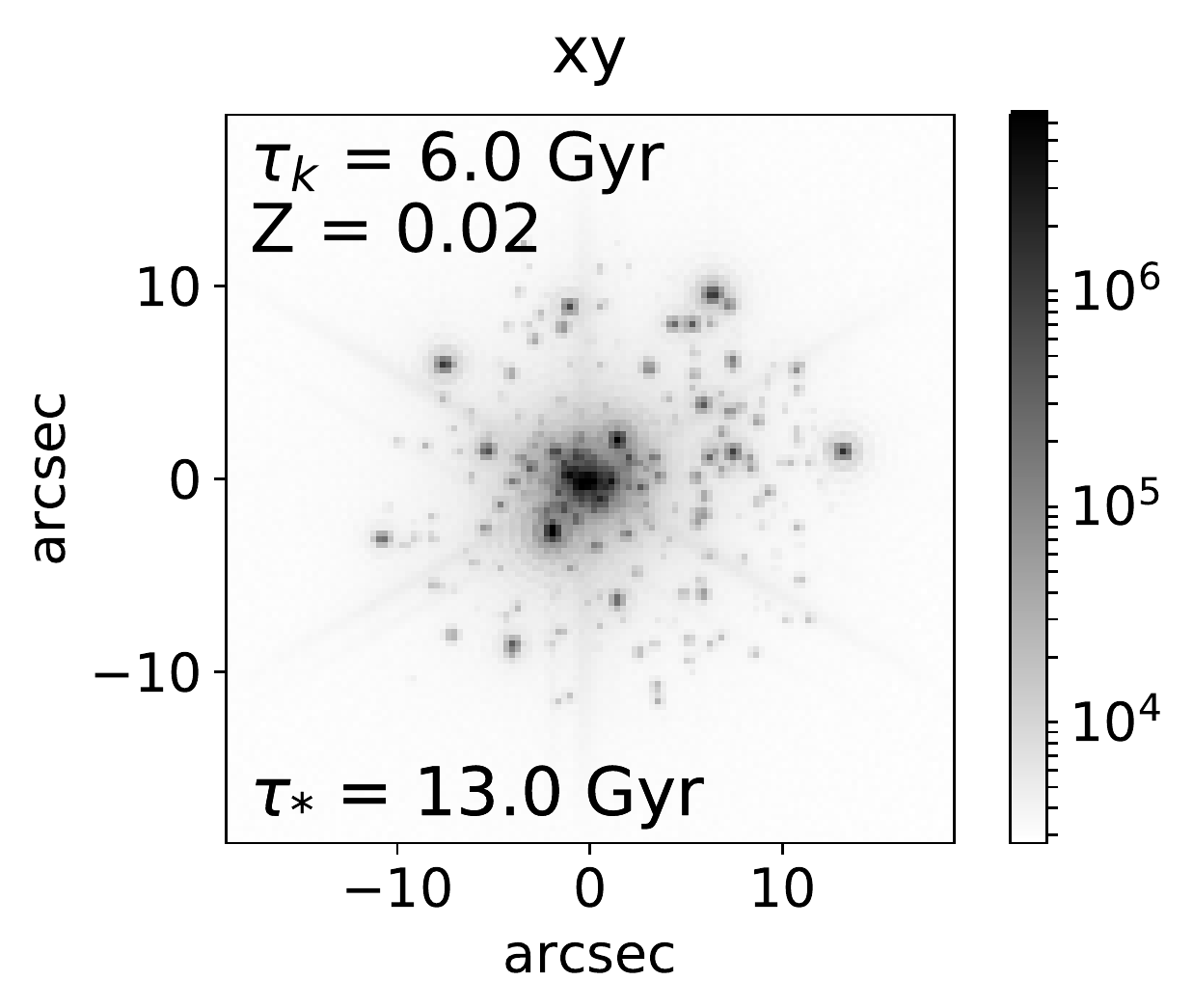}\\

\end{array}$
\end{center}
\caption{As in Fig.\ref{fig: hcss_mock_ps1_tk} for \textit{Euclid}/NISP.}
\label{fig: hcss_mock_nisp_tk}
\end{subfigures}
\end{figure*}

To implement the time evolution we used the results of the Fokker-Planck simulation of \citet{OLearyL12} for a cluster bound to a $10^{5}$ M$_{\odot}$ black hole. We reproduced their function using a smoothly-broken power law:

\begin{align}
f(t) = A\left(\frac{t}{t_{b}}\right)^{-\alpha_1}\left\{\frac{1}{2}\left[1 + \left(\frac{t}{t_{b}}\right)^{1/\Delta}\right]\right\}^{(\alpha_1 - \alpha_2)\Delta},
\end{align}

\noindent where the amplitude $A$ was matched to the number of stars bound to the BH at the time of the kick ($\tau_{k} \leq 10^{6}$ yr), and the remaining parameters being $t_{b} = 6.79 \times 10^{6}$ yr, $\alpha_{1} = -0.12$, $\alpha_{2} = 0.34$, and $\Delta = 1.76$.
The resulting images, derived from the \textit{xy} panel of Fig.\ref{fig: hcss_nbody}, are presented in Fig.\ref{fig: hcss_mock_ps1_tk} and \ref{fig: hcss_mock_nisp_tk}, for Pan-STARRS and \textit{Euclid}/NISP, respectively. 

\section{Discussion}  \label{sec_discussion}
To narrow down the number of HCSC candidates from the \textit{mare magnum} of galaxies and stars sampled in a survey one can start with a selection based on colour. For unresolved candidates, this might be the only parameter available, followed by an estimate of the absolute magnitude, kinematics, and by the stellar velocity-dispersion, if spectroscopy is feasible. Resolved candidates will offer the possibility of further cuts based on the morphology. We discuss these aspects in the following sections. 


\subsection{Colours}
\label{discussion_colours}
Fig.\ref{fig: colors_sim} shows that the predicted optical and NIR colours of the clusters are not peculiar with respect to the general population of stars and galaxies. This is not surprising, as we have not implemented (and we are not aware of) any peculiar process which could be at play in these clusters, and which could leave a distinctive imprint on their stellar population. 

In the optical, the predicted colours follow the locus of stars and galaxies. It is only in the \textit{JHK}$_{s}$ colour diagram that HCSCs show an offset from the peaks of the distributions for stars and galaxies.  
It is for this reason that adding a NIR colour-cut to a set of candidates selected on the basis of their optical colours decreases the sample size by a factor ten. 
Still, the number of candidates obtained via a search based solely on optical and NIR colours is too large to be constraining.

\citet{MunozPL14} showed that a \textit{uiK$_{s}$} colour-colour diagram allows for a clean separation between stars, globular clusters, and galaxies. Such colour-colour diagram was used by \citet{CaldwellSR14} to investigate the nature of the hypervelocity cluster HVGC-1, finding that it falls in the region defined by the globular clusters. We computed the expected \textit{uiK$_{s}$} colours for our sample of simulated HCSCs (Fig.\ref{fig: colors_sim1}, central panel, bottom row); unfortunately, we found a large scatter which hampers the diagnostic power of this diagram in the search for HCSCs.

It is clear that a colour selection alone will not yield a suitable sample of candidates, unless it turns out that some peculiar process takes place into the exotic environment of such clusters, leaving a strong signature on its spectro-photometric properties, for example an enhanced rate of stellar mergers leading to an excess of blue stragglers. 
Preliminary simulations show that increasing the number of blue stragglers affects mostly the UV colours of sub-solar metallicity clusters; instead, adopting the GALEX \citep[][]{Bianchi99} transmission curves, HCSCs with metallicity Z $\geq 0.02$ (solar and above) have always colours in the range $6 \leq$ (FUV - NUV) $\leq 8$, independently of the presence of blue stragglers. 
Evolved stragglers falling on the red-giant branch affect optical and NIR colours introducing a higher variance in the result of the simulation (e.g. making a young cluster with solar-metallicity appear similar to a super-solar metallicity cluster with an old stellar population).

We draw the reader's attention to the following three points. First, sub-solar metallicity clusters might be unusual: HCSCs, being dislodged galactic nuclei, will likely show super-solar metallicity, and, being residuals of the galaxy-assembly process, their stellar population might also be old (e.g. if a HCSC was ejected at redshift z=1, and no star-formation took place in it, then its stellar population should be at least 7 Gyr old). While an old stellar population with super-solar metallicity might seem an odd combination, we note that several-times super-solar solar metallicity has been inferred for the broad-line regions of high-redshift quasars, e.g. \citealt{JuarezMM09}. The second point to note is that, to date, the only instrument performing sky-surveys and sensitive in the near UV is \textit{Gaia} (the \textit{G}$_{BP}$ passband covers the wavelength range 330-680 nm, e.g. \citealt{Weiler18, MaizApellnizW18}), with no dedicated facilities planned for the foreseeable future. Unfortunately, also our predicted \textit{Gaia} colours mostly follow the stellar locus. Finally, a pilot search through public databases shows that only a tiny fraction (below 1 per cent) of candidate HCSCs with optical and infrared coverage posses also good quality UV data. 

We note that our models assume a single stellar population (including blue stragglers) and a single metallicity. On the other hand, HCSCs might share some properties with ancient nuclear star clusters (NSCs): if a NSC was present at the time of the recoil, the HCSC would consist of a subset of stars from the inner region of the NSC. The study of these compact objects is still in its infancy, and it is limited to the local Universe, however, it is becoming clear that they are characterised by complex stellar populations \citep[e.g.][]{MastrobuonoBPG19, NeumayerST20}. While our individual models for HCSCs might be simplistic, the overall set explores a range of ages and metallicities, including some extreme cases. Clusters with more complex star-formation histories fall within the colour-colour loci defined by the simple models presented here, with the scatter in the distribution being inversely proportional to the number of stars bound to the BH. By comparison, \citet{MSK2009} assumed a single stellar population and a metallicity following a Gaussian distribution; \citet{OLearyL12} proposed two models, the first with constant star-formation rate for the 5 Gyr preceding the GW-kick, the second with a single stellar population formed at the time of the kick; furthermore, they explored three metallicity histories: a constant solar metallicity, a constant subsolar metallicity (Z = $2\times10^{-4}$), and the estimated time-evolution for the Galactic Centre. The \textit{ugr} colours derived here for SDSS can be directly compared with those derived by \citeauthor{OLearyL12}. The models produce consistent results, with the larger scatter recovered here in the red-side of the colour diagram being ascribed to bright evolved blue stragglers. 

To summarise, the combination of parameters producing models which deviate from the bulk of the population of stars and galaxies corresponds to super-solar metallicities and old stellar ages. In light of the observations of super-solar metallicities in the nuclei of high-redshift quasars, this combination, although unusual, might not be unrealistic, allowing to select ancient HCSCs.

In the following section we show that an HCSC with an old stellar population resembles K and M giant stars, and we highlight the spectro-photometric differences between the two classes of objects.

\subsection{Spectra}
In Fig.\ref{fig: hcss_stellar_library} we presented the model spectra derived as explained in Sec.\ref{sec_spectral}. Our derivation of the spectra did not take into account the internal dynamics of the cluster and the resulting shape of the absorption line profiles; for this we refer the reader to \citet{MSK2009}, where the authors dealt with this aspect in great detail. Rather, we will compare the model spectra with the observed spectra of single stars to identify the most probable interlopers, and to identify any difference - besides a high velocity dispersion (i.e. strongly broadened absorption lines) - that would allow to distinguish stars from clusters.

We compared the models with the stellar spectra of the Pickles Atlas \citep{Pickles98}, a library of 131 stellar spectra at solar abundance, including all normal spectral types and luminosity classes. We matched the binning of our models and the Pickles spectra, and we selected those resembling the most to the HCSC models (i.e. those producing a smaller residual when subtracted from the model). In Table \ref{tab: match_spec} we indicate, for each metallicity and stellar population age assumed for the cluster, the best-matching spectrum (multiple spectra are indicated whenever the residuals differ by less than 10 per cent). In Fig.\ref{fig: hcss_stellar_library} we show the simulated spectra, the best-matching spectra from the Pickles Atlas, and the residuals.

From Table \ref{tab: match_spec} we can see that HCSCs with a young stellar population ($\tau_{\star} \approx 1$ Gyr) and sub-solar metallicity resemble G and F stars, either on the main sequence, or on the giant and sub-giant phase. At the other end of the spectrum, clusters with an old stellar population ($\tau_{\star} \approx 13$ Gyr) and super-solar metallicity resemble giant K and M stars. 
Overall, most of the times, the simulated spectra resemble those of giant K stars. It is, therefore, likely that a colour-colour selection alone would be heavily contaminated by such objects. 

How could we distinguish an unresolved HCSC from a giant star? Any distance information would help constraining the absolute magnitude of the candidates, applying a further selection cut on the sample.
Moreover, the plots presented in Fig.\ref{fig: hcss_stellar_library} show often, but not systematically, a blue excess in the spectra of HCSCs. This is likely due to the contribution of the hottest main sequence stars and the blue stragglers in the cluster. 

We note, again, that the addition of a population of blue/yellow/red-stragglers acts as a confounding ingredient, adding stochasticity to the appearance of the cluster: simulations which do not include stragglers produce spectra which age as expected, i.e. with the peak shifting towards longer wavelengths and emission in the red portion of the spectrum becoming more and more prominent as the age of the stellar population increases. The addition of stragglers, instead, does not simply contribute with a population of blue-stragglers, adding emission to the blue spectral component. Rather, this population also contributes with a few bright giant stars giving a major contribution to the integrated spectrum, ageing, instead of rejuvenating, the appearance of the cluster. We also note that the adopted number of blue stragglers (0.3, 1, and 2 per cent for $\tau_{\star} = 1, 7, 13$ Gyr) could be a lower limit: as explained in Sec.\ref{sec_spectral}, this fraction was derived for Galactic open clusters, where the stellar density is much lower than the one expected for HCSCs. The high density in the nuclei of HCSCs might enhance the production rate of blue stragglers.

To summarise, the simulated spectra of HCSCs often resemble those of K-type giant stars, however, the presence of a blue excess could be used to distinguish an unresolved HCSC from a single star.

\begin{table}
\begin{center}
\caption{Spectral type of single-star spectra from the Pickles Atlas library resembling the most a HCSC simulated with the metallicity and stellar population age indicated in the table. Spectral class indicated in upper-case letter, luminosity class indicated with roman lower-case letters. The prefixes \textit{r} and \textit{w} indicate metal-rich and metal-weak stars, respectively.} 
\scalebox{1}{
\begin{tabular}{l lll}
\hline
\hline
Metallicity	 					&  	\multicolumn{3}{c}{Stellar population age (Gyr)}\\
& &    \\
	 					&  	1					& 7					& 13	\\
\hline
&&\\
0.0002					& G0iii					& K0v				& G8v, K0v, G0iii	\\
0.002					& F02iv, F0v				& \textit{r}K5iii				& \textit{r}K2iii		\\
0.02						& M2iii, M1iii				& (\textit{w})K0iii				& \textit{w}K0iii, K2v	\\
0.03						& F8iv, G0iv				& M1iii, M0iii			& \textit{w}K4iii		\\
0.07						& (\textit{w})K3iii					& K2iii				& M2iii, M3iii	\\

&&\\
\hline
\hline
\end{tabular}}
\label{tab: match_spec}
\end{center}
\end{table}

\subsection{Morphology of a resolved HCSC}
Fig.\ref{fig: hcss_mock_psnisp}, \ref{fig: hcss_mock_ps1_tk}, and \ref{fig: hcss_mock_nisp_tk} show rendered versions of the N-body model presented in Fig.\ref{fig: hcss_nbody}. For this model the time since the kick is t = 100 $\times$ GM$_{\bullet}$/V$^{3}_{k}$; although this quantity amounts to 2700 yr for M$_{\bullet} = 10^{5}$ M$_{\odot}$ and V$_{k} = 250$ km s$^{-1}$, this state of the cluster should be representative for a full relaxation time, i.e. approximately $10^{6}$ yr. After that time, we assume that the cluster starts loosing stars as described in Sec.\ref{subs: evafter_kick}. 

Fig.\ref{fig: hcss_mock_psnisp} shows the effect of increasing kick velocities for a cluster with solar metallicity (Z=0.02), a stellar population of intermediate age ($\tau_{*} = 7$ Gyr), and located at distances of 10 and 30 kpc.
The first two columns show the rendering of a Pan-STARRS $r$-band image. At the resolution of Pan-STARRS (median seeing $\theta_{r} \approx 1\farcs2$ and pixel size $p = 0\farcs25$) the cluster corresponding to V$_k = 150$ and 250 km s$^{-1}$ is (barely) resolved to a distance of 30 kpc, showing evidence for a low-density envelope and a denser, compact core. For V$_{k} = 500$ km s$^{-1}$ the cluster is featureless, at visual inspection, and barely resolved at best. With the adopted saturation level and exposure time (reflecting the duration of a single exposure in the 3$\pi$ Pan-STARRS survey, \citealt{SchlaflyFJ12}), the core is saturated when V$_{k}$ = 150 km s$^{-1}$ and for V$_{k}$ = 250 km s$^{-1}$ at d = 10 kpc.

The last two columns of Fig.\ref{fig: hcss_mock_psnisp} show the rendering of a \textit{Euclid}/NISP \textit{J}-band image. As NISP will be diffraction-limited, with a pixel size of $p\approx 0\farcs3$, it is not surprising to find a much sharper image, with the cluster being resolved also for V$_k = 500$ km s$^{-1}$ at d = 10 kpc. The core is saturated in all six renderings. 
It must be stressed, however, that the rendering depends on the background and the instrument PSF, two parameters which, at the time of writing, are not well-known.

An observable quantity that one could derive for a HCSC and use it to mine catalogs, similarly to \citet{OLearyL12}, is the Petrosian radius, i.e. the radius where the local surface brightness equals the average surface brightness within that radius \citep{Petrosian76}. However, this parameter is not ideal for clusters which are resolved and which include bright stars in their halo. Such clusters will be treated as a collection of individual objects in a survey catalog. Moreover, for the cluster as a whole, the condition defining the Petrosian radius might be satisfied at multiple radii because the local surface brightness would not be monotonic. Therefore, this parameter can not be used for most of the scenarios explored here.

\subsubsection{Kick signatures}

From Fig.\ref{fig: hcss_nbody} it is clear that at t = 100 the cluster core is still flattened, with the major-axis aligned with the kick direction. However, this important morphological feature (it carries information on the kick direction) is lost in the rendered images presented in Fig.\ref{fig: hcss_mock_psnisp}, \ref{fig: hcss_mock_ps1_tk}, and \ref{fig: hcss_mock_nisp_tk}. The reason resides in the combination of stellar density profile and resolution effects: first, the cluster is characterised by a steep density profile; i.e. already at $r\lesssim 1\ \times $ GM$_{\bullet}$/V$^{2}_{k}$ (in N-body units), or $r \lesssim 0.02$ pc (for M$_{\bullet} = 10^5$ M$\odot$ and V$_{k} = 150$ km s$^{-1}$) the stellar density is $10^{2} - 10^{5}$ times lower than it is at the very nucleus (the exact value depends on the adopted density distribution before the kick, see Fig.3 in \citeauthor{MSK2009}; we used simulations produced from an initial density $\rho \propto r^{-7/4}$). 
Second, when magnitudes and stellar evolutionary stages are assigned to the N-body points, a few of them correspond to the evolutionary stages ``red giant branch'' (RGB), ``core helium burning'' (CHEB), or ``early asymptotic giant branch'' (EAGB). These are all bright stars (-5 $<$ M$_{\mathrm{J}} <$ 3, -1 $<$ M$_{\mathrm{r}} <$ 3), and the great majority of them are packed in the central density cusp. As a result, such bright stars are not resolved in the mock observations, and the convolution with the PSF erases the photometric asymmetry which, naively, would be expected from the spatial distribution alone. 

To constraint the kick direction one could, therefore, rely on the asymmetry of the extended envelope, which evaporates as time goes by. As it is evident in Fig.\ref{fig: hcss_nbody}, at t = 100 GM$_{\bullet}$/V$^{3}_{k}$ (or after one relaxation time) stars are still distributed asymmetrically in the envelope: 
defining the envelope as the region outside the elliptical bulge with semi-major axes 13 and 10 (in N-body units), about 40 per cent of the envelope stars are located within the quadrant defined by position angles in the range $-45:45$ deg, where PA = 0 corresponds to the positive side of the X-axis, or the counter-kick direction. However, Fig.\ref{fig: hcss_mock_ps1_tk}, and \ref{fig: hcss_mock_nisp_tk}, which are renderings of the \textit{xy} panel of Fig.\ref{fig: hcss_nbody}, show that already at $t \geq 10^{7}$ yr such asymmetry looses its prominence. Projection effects, and higher recoil velocities (implying a lower number of bound stars), would only weaken the asymmetry and the conspicuity of the halo as a whole.

\subsubsection{Light profile}
\begin{figure*}
\begin{center}$
\begin{array}{ccc}
\includegraphics[trim= 1.1cm 0cm 0cm 0cm, clip=true, scale=0.35]{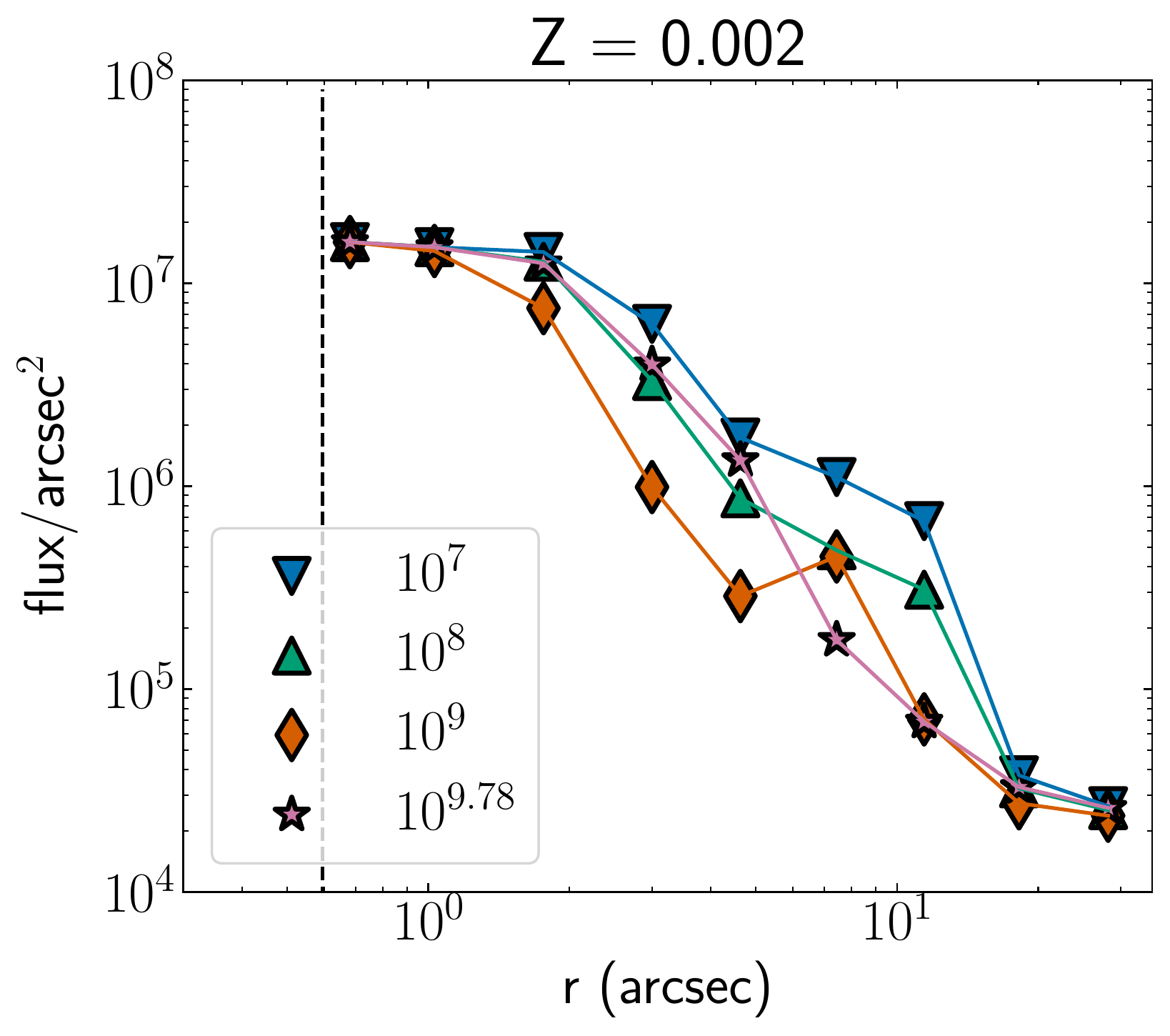} &
\includegraphics[trim= 0.2cm 0cm 0cm 0cm, clip=true, scale=0.35]{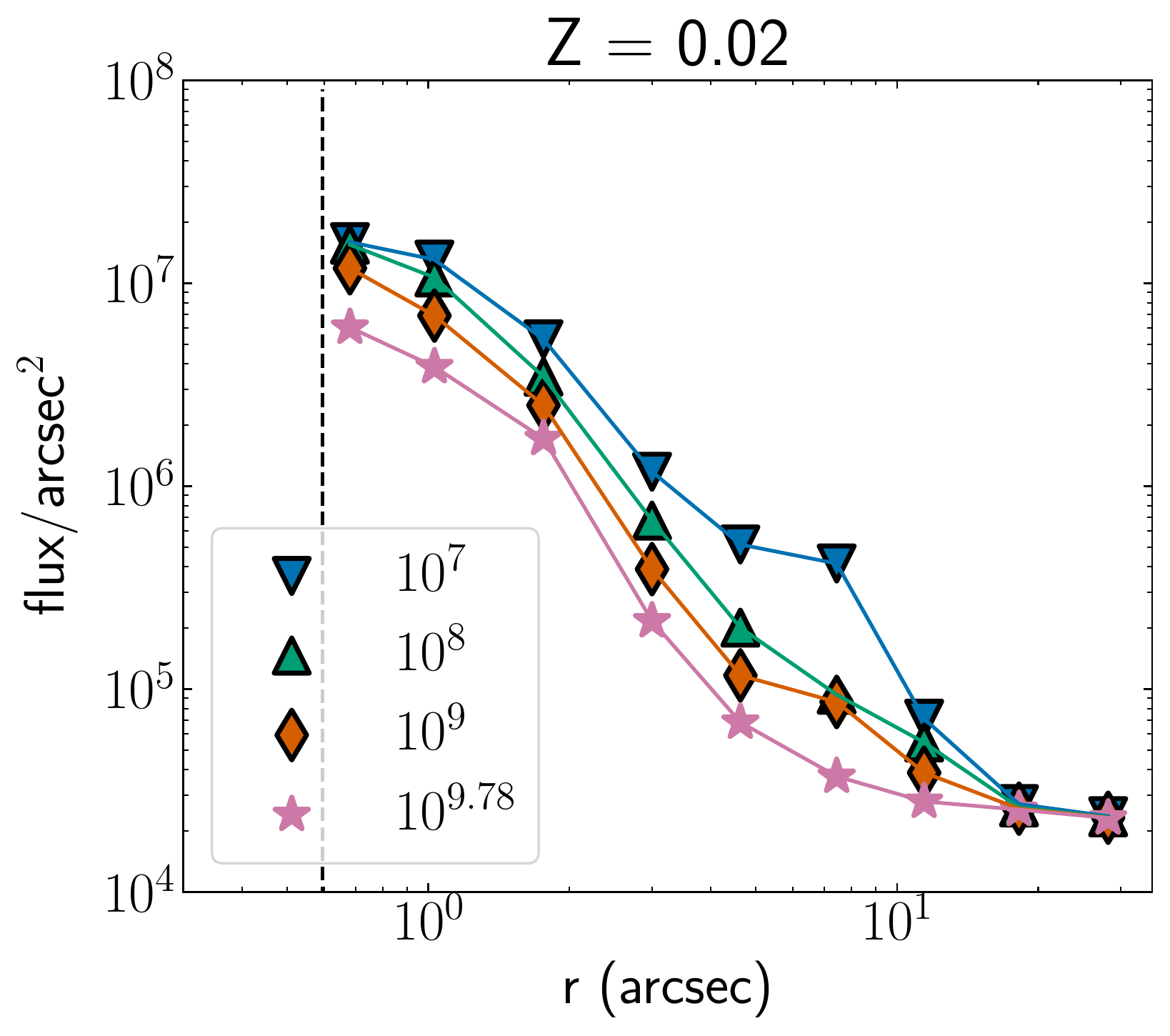} &
\includegraphics[trim= 0.2cm 0cm 0cm 0cm, clip=true, scale=0.35]{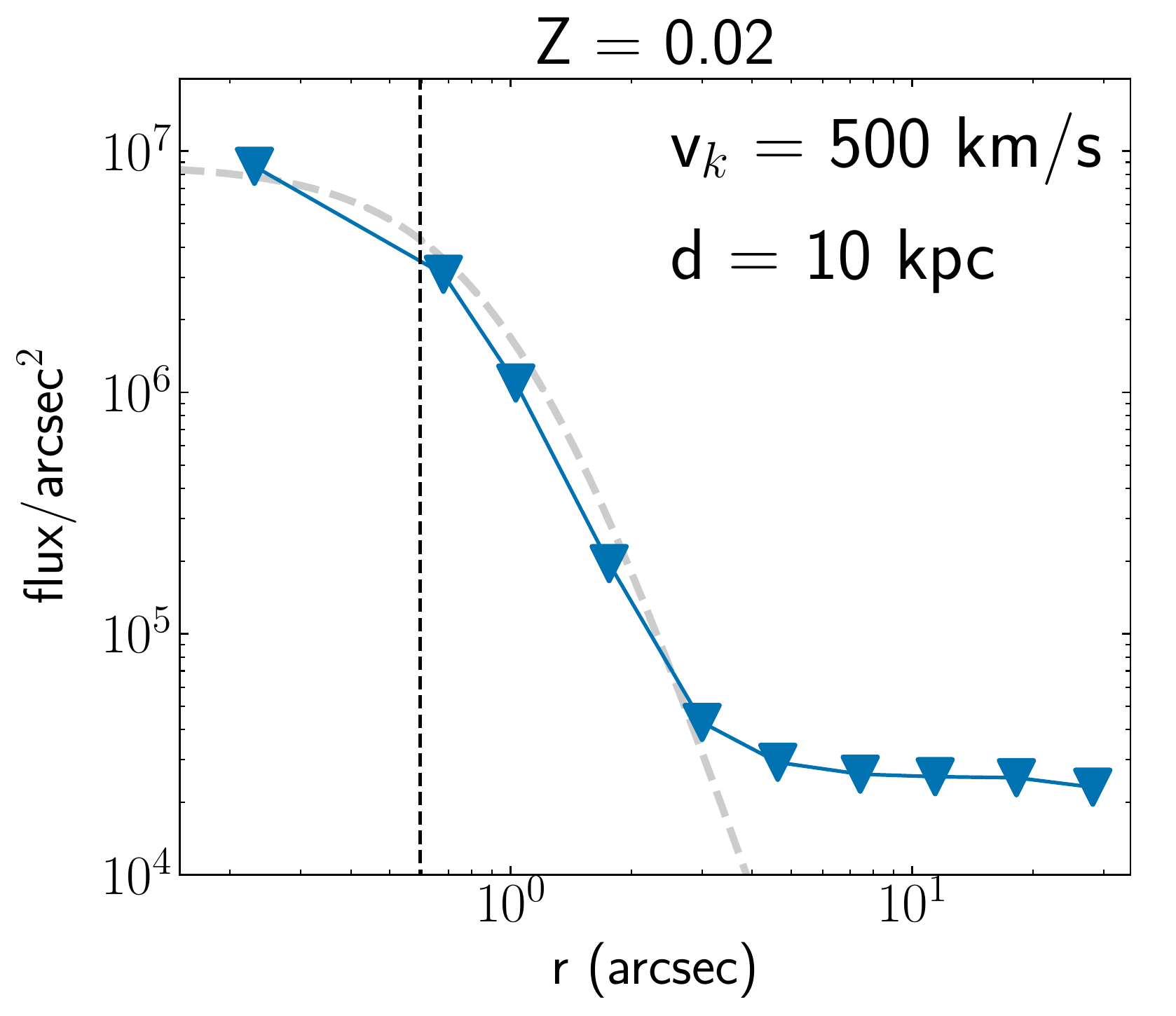}\\

\includegraphics[trim= 1.1cm 0cm 0cm 0cm, clip=true, scale=0.35]{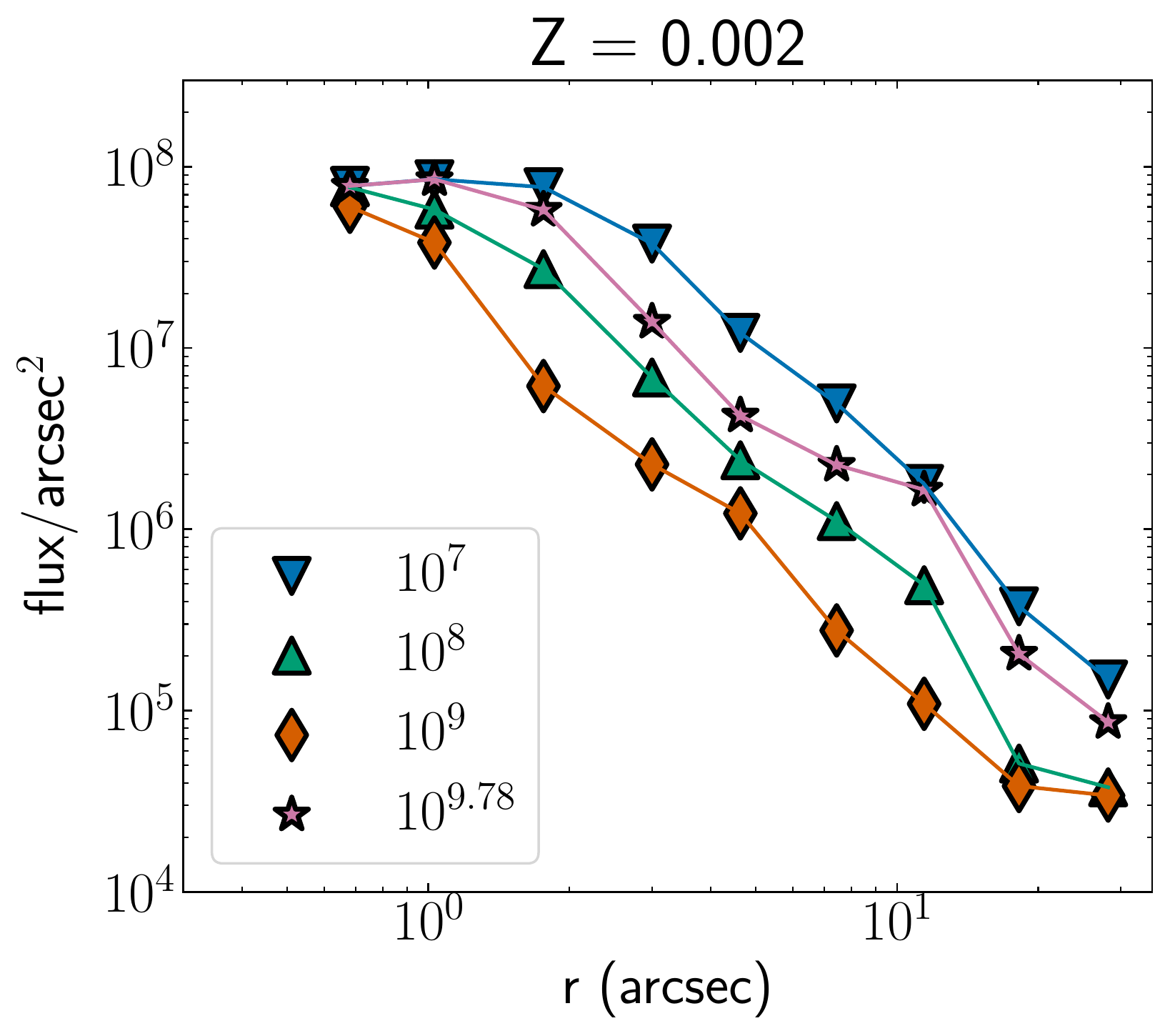} &
\includegraphics[trim= 0.2cm 0cm 0cm 0cm, clip=true, scale=0.35]{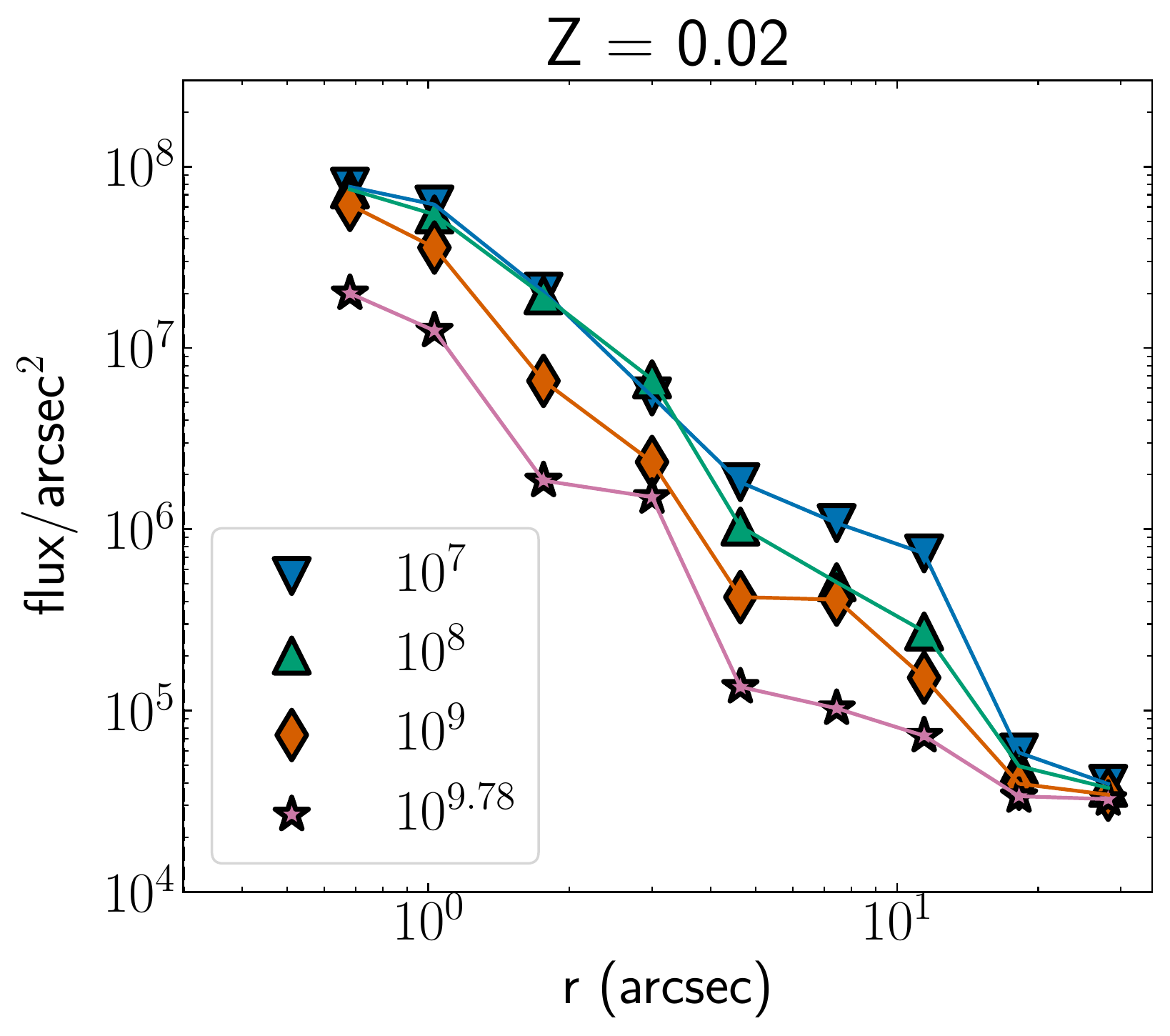} &
\includegraphics[trim= 0.2cm 0cm 0cm 0cm, clip=true, scale=0.35]{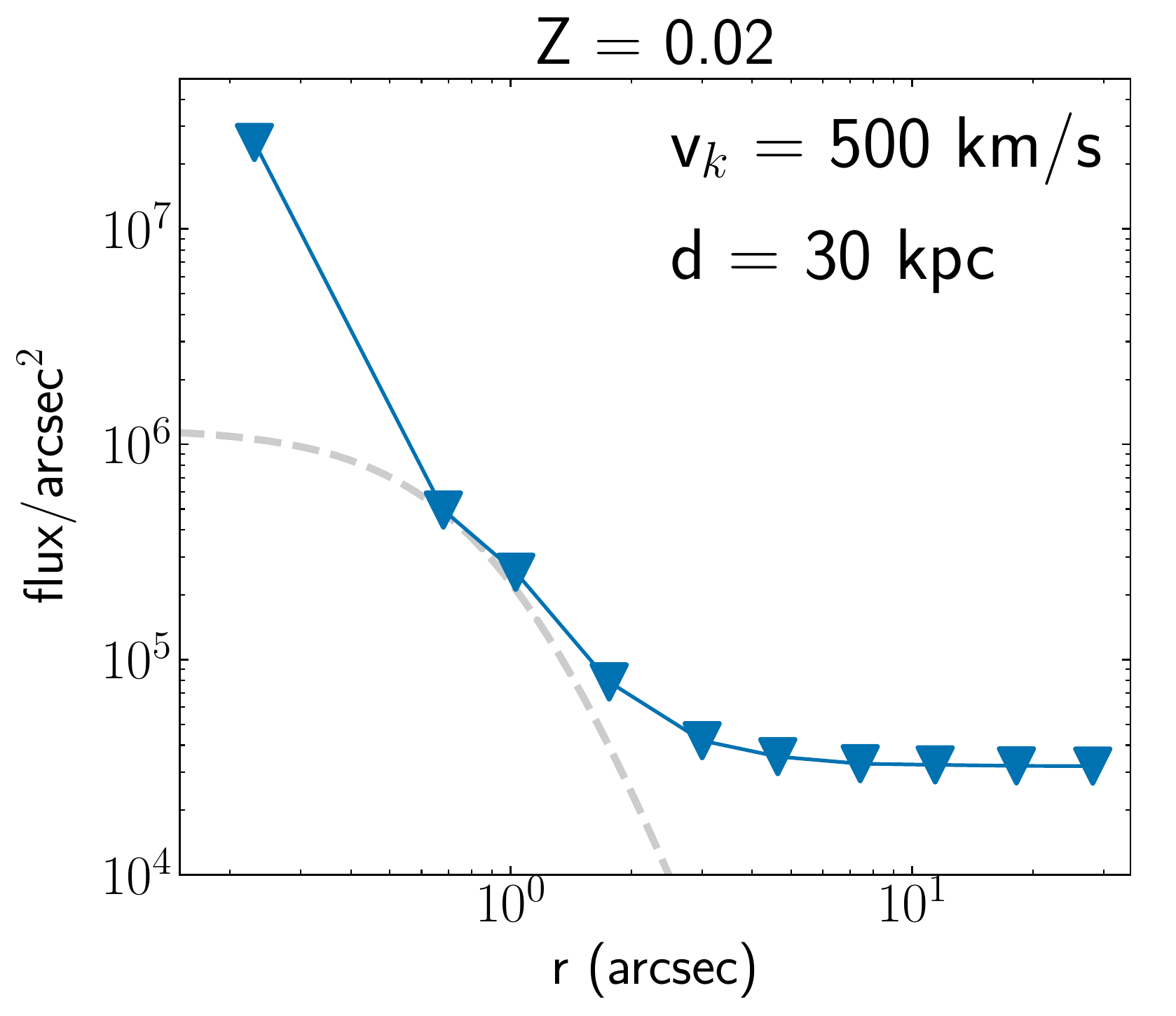}\\
\end{array}$
\end{center}
\caption{Light profiles derived from the rendered HCSCs presented in Fig.\ref{fig: hcss_mock_ps1_tk} and \ref{fig: hcss_mock_nisp_tk} (first and second column) and for two clusters from Fig.\ref{fig: hcss_mock_psnisp} (third column) for Pan-STARRS (top row) and \textit{Euclid}/NISP (bottom row). The plot shows the average intensity of light within annuli. The adopted metallicity is indicated on top of the panels. The time since the kick ($\tau_{k}$) is indicated in the legend in years for the clusters from Fig.\ref{fig: hcss_mock_ps1_tk} and \ref{fig: hcss_mock_nisp_tk}. Symbols with a solid black edge indicate the presence of saturated pixels in the nucleus. The vertical dashed line, in the top row, marks the typical FWHM/2 of the Pan-STARRS seeing in the $r$-band. The luminosity of the older cluster with metallicity Z = 0.002 increases because of a few stars on the early asymptotic giant branch. For comparison, a \citet{Plummer11} profile, representative of globular clusters, is plotted as a dashed grey curve in the third column. We stress that the inner flattening visible in the first and second column is due to the presence of saturated pixels.}
\label{fig: hcss_lightprof}
\end{figure*}

What is the light profile that one would recover from an image of a resolved HCSC? Similarly to \citet{OLearyL12}, we derived, for the clusters rendered in Fig.\ref{fig: hcss_mock_ps1_tk} and \ref{fig: hcss_mock_nisp_tk}, the logarithmic slope of the cumulative light profile: $\Gamma_{i} = \mathrm{d\ ln}(I_{i})/\mathrm{d\ ln}(r) \approx \mathrm{ln}(I_{i+1}/I_{i})/\mathrm{ln}(r_{i+1}/r_{i})$ and the average intensity within annuli of radius r$_{i}$ = 0.23, 0.68, 1.03, 1.76, 3.00, 4.63, 7.43, 11.42, 18.20, 28.20 arcsec (the same adopted to produce the SDSS and Pan-STARRS catalogs). The width of the annuli is $\delta = 1.25r_{i} - 0.8r_{i}$. The derived slopes are given in Table \ref{tab: logslope}; plots of the intensity profiles are shown in Fig.\ref{fig: hcss_lightprof}.

The clusters presented in Fig.\ref{fig: hcss_mock_ps1_tk} and \ref{fig: hcss_mock_nisp_tk} are nearby and ejected with low velocity. Therefore, they are well-resolved and individual stars in the halo can be discerned. However, Fig.\ref{fig: hcss_mock_psnisp} shows that clusters ejected at higher velocities and located further away would be almost featureless and with a smooth profile. In Fig.\ref{fig: hcss_lightprof} we show the light profiles for two of those clusters,  making a comparison with a \citet{Plummer11} profile, which is typical for globular clusters. The comparison shows that the profile of a HCSC observed with a resolution of about 1$^{\prime\prime}$ (as for PanSTARRS and other ground-based surveys) is similar to a Plummer profile. From the light profile alone, the HCSC can, therefore, be misclassified as a globular cluster. On the contrary, observations at higher resolution (such as those expected for \textit{Euclid}) reveal the presence of a nuclear cusp, which deviates from the Plummer profile.

\begin{figure*}
\begin{center}$
\begin{array}{ccc}
\includegraphics[trim= 0cm 0cm 3cm 0cm, clip=true, scale=0.625]{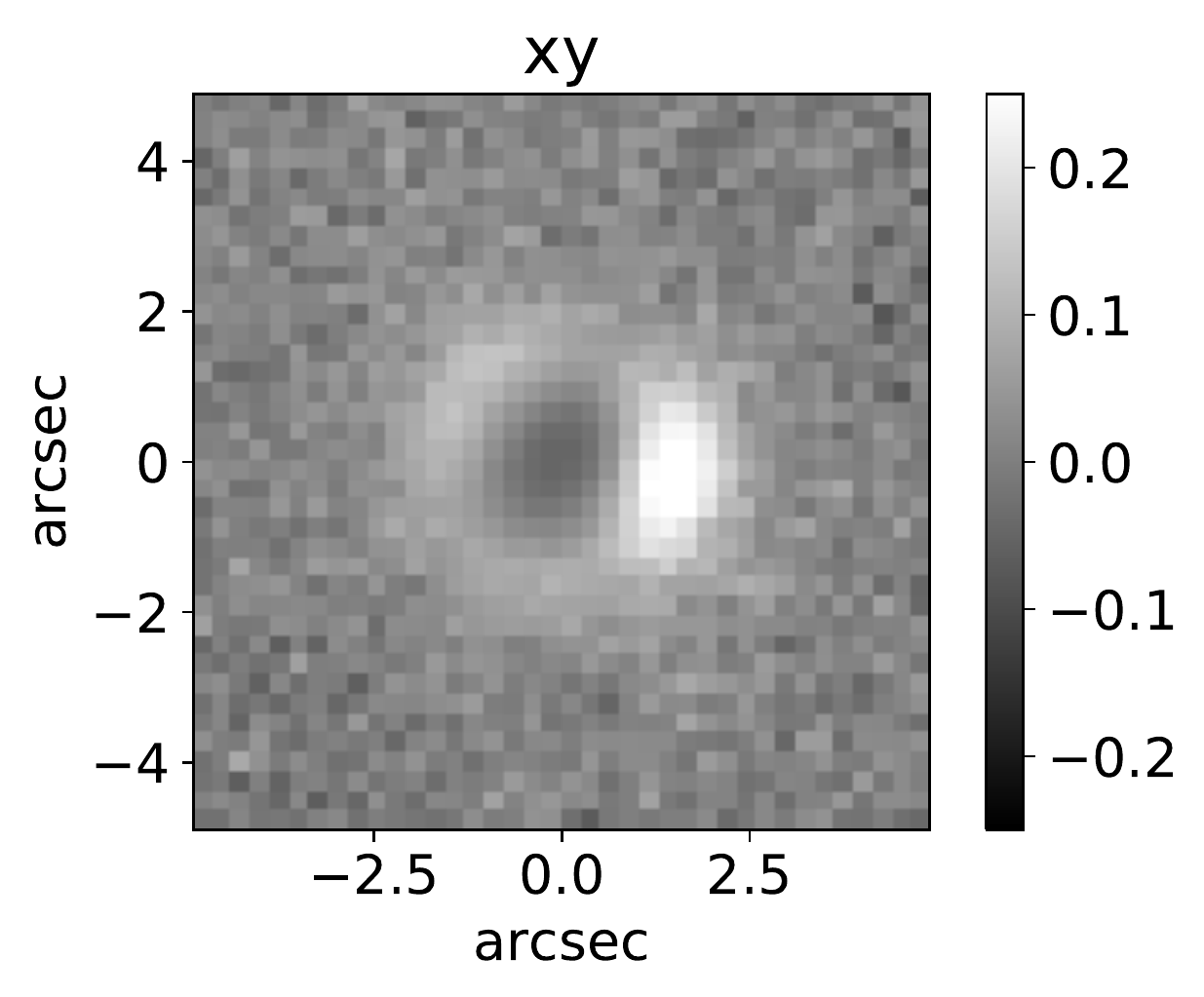} 
\includegraphics[trim= 1.85cm 0cm 3cm 0cm, clip=true, scale=0.625]{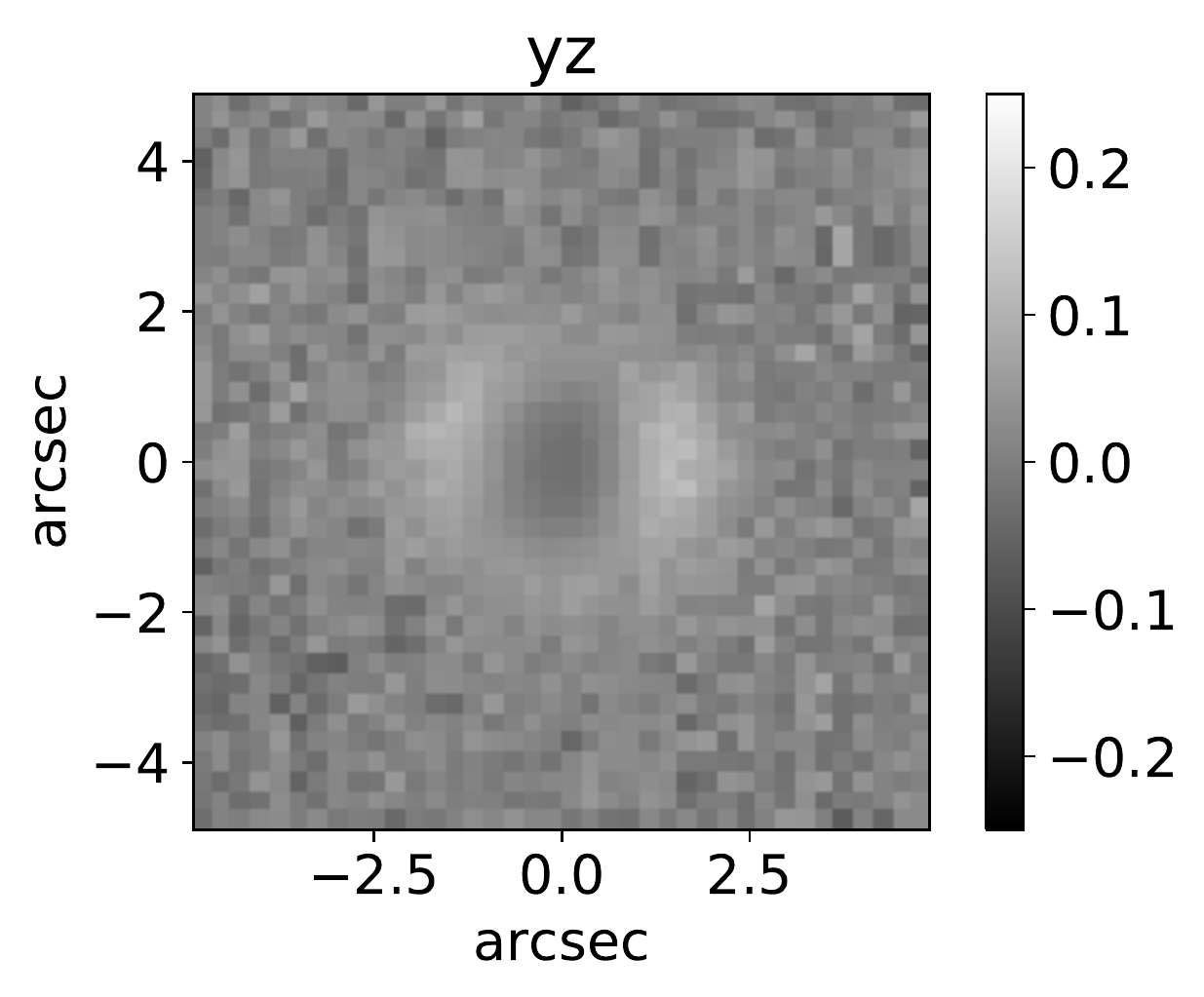} 
\includegraphics[trim= 1.85cm 0cm 0cm 0cm, clip=true, scale=0.625]{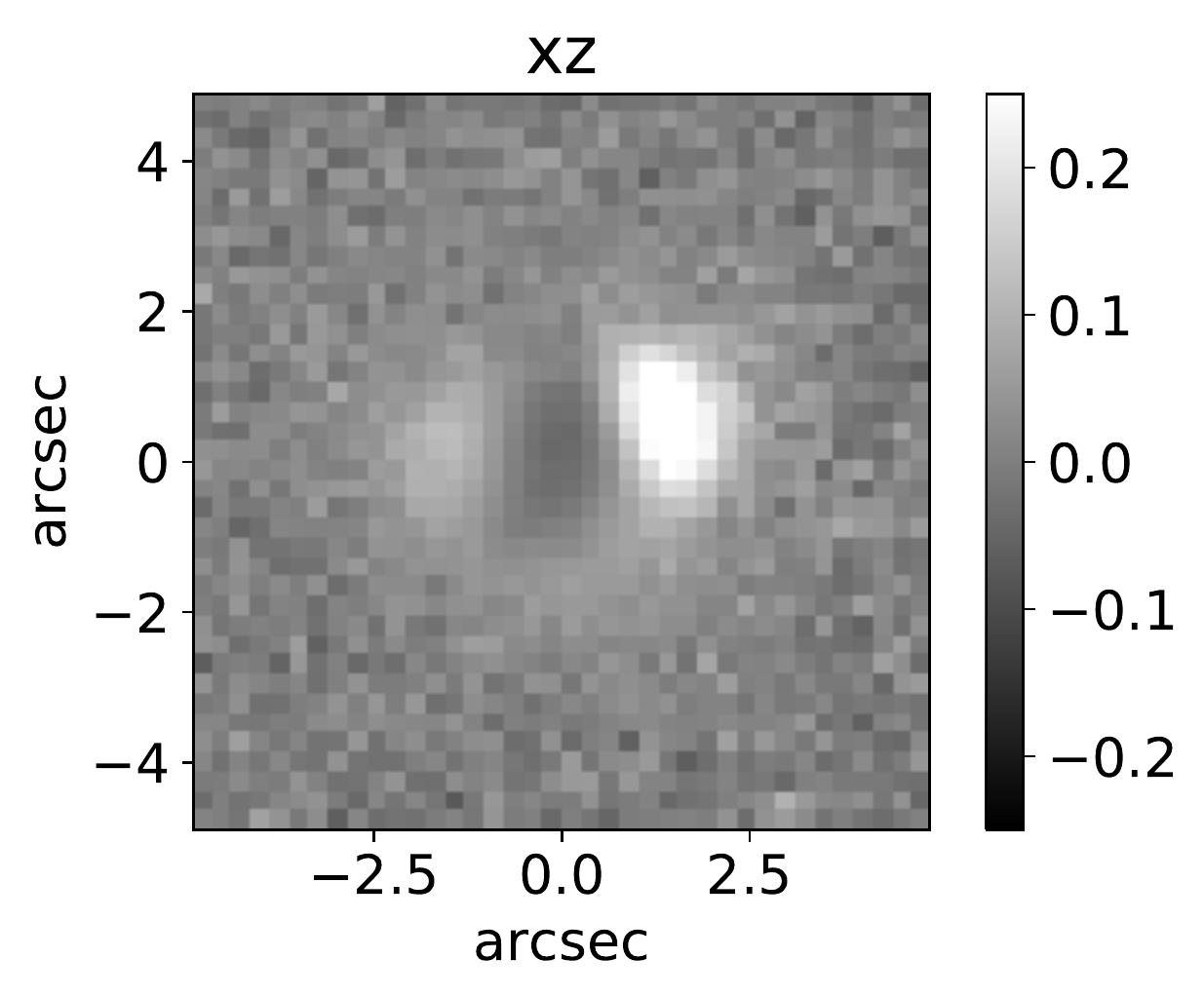} 
\end{array}$
\end{center}
\caption{Fractional residuals obtained after fitting and subtracting a PSF from the HCSC simulated for Pan-STARRS assuming V$_{k} = 500$ km s$^{-1}$ and d $ = 10$ kpc.}
\label{fig: hcss_fit_psf}
\end{figure*}

In the barely-resolved scenario, a first step towards a better understanding of the object nature might be the inspection of the residuals obtained after fitting and subtracting a PSF. We used \textsc{galfit} \citep{GALFIT2002,PengHIR10} to apply this approach on the rendered Pan-STARRS images for a cluster ejected with V$_{k}$ = 500 km s$^{-1}$ and located at d = $10$ kpc (bottom left panel in Fig.\ref{fig: hcss_mock_psnisp}, the adopted PSF was a star rendered with \textsc{skymaker}, using the same parameters adopted to render the cluster). The result is shown in Fig.\ref{fig: hcss_fit_psf}: the residuals show clear deviations from a pure point source, suggesting a more complex nature of the object. 
The resolved structure is due to the presence of an extended envelope, which might be the host of one or more RGB stars. Although a small fraction of the stars in the cluster are in such evolutionary phase (i.e. $<1$ per cent), because of their absolute magnitude, which can be as low as M$_{r} = -1$ (M$_{J} = -5$), they have a considerable weight in determining the appearance of the cluster, both in the well-resolved scenario (where the cluster appears as an unresolved nucleus embedded into an extended envelope), and in the barely-resolved case (where a single off-center star might be responsible for an elongated extension, resembling a slightly-resolved binary star or a compact galaxy and a foreground bright star).

The presence of the extended envelope depends from (at least) three factors: the age of the cluster, the kick velocity, and the metallicity. Ageing of a cluster is due to the combined action of dynamical ageing and ageing of the stellar population. Dynamical ageing is more important for low-mass clusters; e.g. \citealt{OLearyL12} found that clusters bound to BHs with mass M$_{\bullet} \geq 10^{7}$ M$_{\odot}$ lose a very small fraction of their mass over 10$^{10}$ yr, instead, the number of stars bound to a 10$^{4}$ M$_{\odot}$ BH decreases dramatically with time. A HCSC bound to a 10$^5$ M$_{\odot}$ BH is predicted to loose almost 90 per cent of its initial stars, however, also for V$_{k} = 250$ km s$^{-1}$ the HCSC should retain almost 3000 stars at the time of the kick, which translates into a cluster consisting of about 300 stars after 10$^{10}$ yr. The effect of the stellar population ageing is clear, instead, in Fig.\ref{fig: hcss_mock_ps1_tk} and \ref{fig: hcss_mock_nisp_tk}, where, in general, the cluster envelope fades away while stars grow old; however, depending on the initial age of the stellar population, the cluster might brighten up (e.g. because of stars going through the giant phase). The kick velocity, V$_{k}$, has perhaps the most important effect on the presenze of an envelope, as shown in Fig.\ref{fig: hcss_mock_psnisp}, with its size being dramatically reduced for V$_{k} > 250$ km s$^{-1}$. Finally, metallicity plays a mild role, with the number of bright stars decreasing at higher metallicity (high-metallicity stars loose more mass via stellar winds, e.g. \citealt[][]{TraniMB14}).

\begin{table}
\begin{center}
\caption{Logarithmic slope of the cumulative surface brightness associated with the profiles presented in Fig.\ref{fig: hcss_lightprof}. Assuming M$_{\bullet} = 10^{5}$ M$_{\odot}$, V$_{k}$ = 150 km s$^{-1}$, d = 10 kpc, unless otherwise specified. The range of values was derived from projections of the clusters on the \textit{xy, xz} and \textit{yz} planes.} 
\scalebox{0.65}{
\begin{tabular}{ll lllllll}
\hline
$\tau_{k}$					& $\tau_{\star}$ & $\Gamma_{1}$	& $\Gamma_{2}$	& $\Gamma_{3}$	& $\Gamma_{4}$ & $\Gamma_{5}$	& $\Gamma_{6}$ & $\Gamma_{7}$ \\
 (Gyr)					& (Gyr) \\
\hline
& &    \\
\multicolumn{8}{c}{Pan-STARRS}\\
\hline
						 &  	\multicolumn{8}{c}{Z = 0.002}\\
\hline
0.01						& 1		&	-	& 1.3				& 0.67 		& 0.43 - 0.45	& 0.26 - 0.28	& 0.11 		& 0.07	\\
0.1						& 1.1		&	-	& 1.3				& 0.67		& 0.36 - 0.4 	& 0.2 - 0.22 	& 0.08 - 0.09	& 0.05	\\
1						& 2		&	-	& 1.3				& 0.6 - 0.61	& 0.25 - 0.28 	& 0.13 - 0.14	& 0.05 		& 0.03 	\\
6						& 7		&	-	& 1.3				& 0.67		& 0.35 - 0.36	& 0.19 - 0.23	& 0.07 - 0.08	& 0.03 - 0.04\\

&&\\
						 &  	\multicolumn{8}{c}{Z = 0.02}\\
\hline
0.01						& 7		&	-	& 1.3				& 0.55 - 0.56 	& 0.21 - 0.22	& 0.12		& 0.05 - 0.06	& 0.03\\
0.1						& 7.1		&	-	& 1.3				& 0.46 - 0.47 	& 0.18		& 0.09 - 0.1	& 0.04		& 0.02\\
1						& 8		&	-	& 1.1-1.2 			& 0.38 - 0.39	& 0.14 - 0.15	& 0.08		& 0.03		& 0.02\\
6						& 13		&	-	& 0.97 - 1.13 		& 0.31 - 0.44	& 0.11 - 0.17	& 0.06 - 0.09	& 0.02 - 0.03	& 0.01 - 0.02\\
&&\\
						 &  	\multicolumn{8}{c}{Z = 0.02, V$_{k}$ = 500 km s$^{-1}$}\\
\hline
0.001					& 7			& 0.93				& 0.89 - 0.9 	& 0.25 - 0.26	& 0.09 		& 0.05 	& 0.02	& 0.01 \\
&&\\
\multicolumn{8}{c}{\textit{Euclid}/NISP}\\
\hline
						 &  	\multicolumn{8}{c}{Z = 0.002}\\
\hline
&&\\
0.01						& 1		&	-	& 1.04			& 0.57		& 0.38		& 0.24 - 0.25	& 0.1 - 0.11	& 0.06\\
0.1						& 1.1		&	-	& 0.98 - 0.99		& 0.38		& 0.18		& 0.1 - 0.11	& 0.04		& 0.02\\
1						& 2		&	-	& 0.87 - 0.89		& 0.25 - 0.29	& 0.09 - 0.11	& 0.06		& 0.02 - 0.03	& 0.01\\
6						& 7		&	-	& 1.04			& 0.55		& 0.25 - 0.27	& 0.15		& 0.06 - 0.07	& 0.04\\

&&\\
						 &  	\multicolumn{8}{c}{Z = 0.02}\\
\hline
0.01						& 7		&	-	& 1.01 - 1.03		& 0.36 - 0.4	& 0.15 - 0.16	& 0.08 - 0.09	& 0.03 - 0.04	& 0.02\\
0.1						& 7.1		&	-	& 0.97 - 1.03		& 0.34 - 0.38	& 0.16 - 0.17	& 0.09		& 0.03		& 0.02\\
1						& 8 		&	-	& 0.81 - 0.91		& 0.27 - 0.29	& 0.1 - 0.11	& 0.06		& 0.02		& 0.01\\
6						& 13		&	-	& 0.62 - 0.67		& 0.18 - 0.21	& 0.07 - 0.08	& 0.04		& 0.02		& 0.01\\
&&\\
						 &  	\multicolumn{8}{c}{Z = 0.02, d = 30 kpc, V$_{k}$ = 500 km s$^{-1}$}\\
\hline
0.001					& 7 		& 0.18	& 0.18			& 0.05		& 0.02		& 0.01	& - & - \\
&&\\
\hline
\hline
\end{tabular}}
\label{tab: logslope}
\end{center}
\end{table}

\subsection{Search strategies and challenges}

Searching for a subset of objects in a large data set requires the capability to remove false positives, and to identify the desired targets with the minimum amount of follow up observations.

The most distinctive characteristic of HCSCs is their high velocity-dispersion in conjunction with their moderate absolute magnitude; this combination of parameters separates them well from the population of globular clusters, ultra-compact dwarf galaxies, and elliptical galaxies \citep{MSK2009}. However, the spectroscopic information needed to populate such a diagram is not readily available for a large data set, and it is expensive to acquire. More realistically, a search for HCSCs would start with the mining of large public databases, with the aim of narrowing down the sample, and collect, in progressive steps and for a small subset of prime candidates, the necessary information to constrain their nature.

As shown in Sec. \ref{sec_colours} and \ref{discussion_colours}, in a colour-colour diagram HCSCs fall near or at the maximum of the distribution of stars and galaxies. Therefore, a selection based on optical colours alone will be heavily contaminated by false positives, and will not produce a sample of candidates small enough to allow the collection of additional data without investing substantial effort. It is clear, however, that a database search would greatly benefit from a multi-wavelength approach: probing a larger portion of the candidate's spectral energy distribution would set more stringent constraints. For example, our experiments show that combining optical and NIR constraints allows to reduce the sample size by a factor ten with respect to the selection based on optical colours alone. 
Alternatively, one could opt for a search of old high-metallicity clusters which, in colour-colour diagrams, lie outside the region containing the bulk of the population of stars and galaxies. 

When the candidates are not resolved, then kinematic and parallax information, in conjunction with absolute magnitude estimates will help constraining the nature of the candidates and remove a number of false positives: e.g. in a search targeting HCSCs in the Galaxy halo one could remove sources showing no evidence for proper motion, these are likely extragalactic sources; moreover, the proper motion of HCSCs should be peculiar with respect to the proper motion of other objects in the vicinity. When searching for unresolved clusters, many interlopers will be K-type giant stars, followed by G and M giants. Given a low-resolution broad-band spectrum, accurately flux-calibrated, then a blue excess in the spectrum of the candidates, at $\lambda < 5000\ \AA$, might help distinguishing between single stars and composite objects. 

For candidates which are barely resolved, then binary-stars (either visual or physical) will be an important population of interlopers. In that case, a shift in the photometric centroid from single-epoch observations at multiple wavelengths, and a periodic astrometric shift would allow to remove false positives. However, the orbital period of resolved binaries can be very long, precluding the detection of periodic astrometric shifts.

When the candidates are resolved, additional constraints can be placed on the light profile and, if bright off-center stars are present, then \textit{Gaia} might also detect the relative motion of the stars and allow to set dynamical constraints.

\section{Summary and Conclusions}  \label{sec_conclusions}
The possibility that hundreds of HCSCs populate galactic haloes is alluring. Their discovery would show that supermassive black holes do merge, and that gravitational waves can indeed remove BHs from their birthplaces (i.e. galactic nuclei). Their characterisation (i.e. measurement of velocity-dispersion, stellar mass, and effective radius) would allow to determine the distribution of GW recoils, would cast light on the assembly of the host galaxy, and on the distribution of stars in the nuclear environment at the time of merger. While several candidate HCSCs have been reported, none has been securely identified.
If predictions are correct, they have already been imaged in existing surveys. Identifying them is, however, non-trivial. 
In an effort to further our understanding of these objects and to ease the task of their identification, we expanded on the existing literature by producing broad-band spectra and photometric renditions of hypothetical HCSCs bound to recoiling BHs of mass $10^5$ M$_{\odot}$. We used the renditions to derive light profiles, and the broad-band spectra to compute the colours that HCSCs would display in a number of recent databases.

Photometric renditions were based on the dynamical simulations presented in \citet{MSK2009} and \citet{OLearyL12}, which we used to implement the spatial distribution of stars and the cluster evaporation. We produced images of the clusters as they would appear in the 3$\pi$ Pan-STARRS survey and in the Wide Field survey of the instrument NISP on board the forthcoming \textit{Euclid} space telescope. Images and light-profiles were derived for a range of kick velocities, distances, metallicities, dynamical ages and ages of the stellar population, with the inclusion of blue, yellow, and red-stragglers. While clusters ejected at moderate velocities (250 km s$^{-1}$) and located at a distance of a few tens of kpc are barely resolved by current surveys, they can be resolved by \textit{Euclid}, with both the instruments NISP and VIS (the Visible Imager, an instrument with a better resolution than NISP). When observed at a resolution of about 1$^{\prime\prime}$, typical of ground-based surveys, HCSCs located a few tens of kpc away and ejected with velocities above 250 km s$^{-1}$ would appear featureless and with light profiles resembling those of globular clusters. The inner cusp of HCSCs can be revealed, instead, with  sub-arcsecond resolutions, such as those achievable with \textit{Euclid}. The capability to resolve the light profile is important when implementing a search, as it allows to place additional constraints on the candidates (colours, we showed, have limited constraining power). It is, therefore, desirable to develop a library of models exploring the full parameter space of BH masses, stellar masses, and cluster age. 

    We used stellar evolutionary models to generate a set of broad-band synthetic spectra and the corresponding colours that HCSCs would display. While optical colours were previously derived by \citet[][]{MSK2009} and \citet[][]{OLearyL12}, we computed colours for additional bands, and for specific surveys, including recent ones, such as \textit{Gaia} and NGVS, among others. We covered, therefore, a larger portion of the spectrum, providing more stringent constraints for the selection of a sample. We show that most of the times, in a colour-colour diagram, HCSCs fall very close to the peak of the distribution of stars and galaxies. The exception are high-metallicity clusters (Z = 0.07) with an old stellar population ($\tau_{\star} = 13$ Gyr). While this combination is somewhat unusual, it might provide a good representation for old HCSCs: being dislodged galactic nuclei, leftovers of the galaxy-assembly process, it is not unreasonable to argue that they might resemble the nuclei of those high-metallicity quasars observed at high redshift.

The fact that most HCSCs fall very near the peak of the distribution of stars and galaxies, in colour-colour diagrams, implies that a search based on colours alone will be heavily contaminated by false positives. The set of simulated spectra allowed us to make a direct comparison with a library of observed stellar spectra to identify the most likely interlopers, and we found that HCSCs resemble, most of the times, K-type giant stars. However, often the spectra of HCSCs show a blue excess with respect to those of single stars. A possible distinctive signature which requires, however, the availability of spectra with carefully calibrated fluxes. 

The ever increasing availability of databases opens new opportunities to searching for HCSCs, and \textit{Euclid} will soon perform an unprecedented optical and NIR survey covering 40 per cent of the extragalactic sky. Results of the present paper can be used to select candidates across multiple databases, and to gain insights on their nature.
While we focussed on the properties of HCSCs bound to $10^5$ M$_{\odot}$ BHs (among the most massive and bright expected within the MW halo), it is desirable to produce a comprehensive library of models, exploring the full parameter space of BH masses (especially below $10^5$ M$_{\odot}$), stellar masses, and properties of the stellar population.

\section*{Acknowledgements}
We thank D. Merritt, O. R. Pols, V. Belokurov, V. H\'{e}nault-Brunet, K. M. L\'{o}pez, D. Rogantini, and A. Nanni for fruitful discussions. We thank D. Merritt for sharing the N-body data used in this work, L. Girardi for assistance with the use of \textsc{cmd}, and the referee for constructive comments which improved the clarity and content of the paper.
DL, PR, ZKR, and PGJ acknowledge funding from the European Research Council under ERC Consolidator Grant agreement no 647208 (PI: P. G. Jonker).

This publication made use of the SDSS, Pan-STARRS1, KIDS, VIKING, \textit{Gaia}, and 2MASS databases, and used observations obtained with MegaPrime/MegaCam.
This research also used the SVO Filter Profile Service (\url{http://svo2.cab.inta-csic.es/theory/fps/}) supported from the Spanish MINECO through grant AyA2014-55216.

\textit{Software:} This research made use of \textsc{astropy},\footnote{http://www.astropy.org} a community-developed core \textsc{python} package for Astronomy \citep{astropy13, astropy18}, \textsc{numpy} \citep{VanDerWalt11}, and \textsc{matplotlib} \citep{Hunter07}.

\section*{Notes}
A package for the reproduction of predicted colours and renditions is available on Zenodo (\url{https://doi.org/10.5281/zenodo.3763444}).



\bibliographystyle{mnras}
\bibliography{biblio} 


\appendix

\section{Best-fitting relations for the colour-colour loci}
\label{sec: fit_color}

\noindent SDSS:
\begin{align}
g - r &= (0.63 \pm 0.1) * (u - g) - (0.3 \pm 0.2)\\
r - i &= (0.44 \pm 0.04) * (g - r) - (0.02 \pm 0.04)\\
i - z &= (0.64 \pm 0.03) * (r - i) - (0.02 \pm 0.01)
\end{align}

\noindent KIDS + VIKING (AB mag):
\begin{align}
g - r &= (0.63 \pm 0.1) * (u - g) - (0.3  \pm 0.2)\\
r - i &= (0.39 \pm 0.04) * (g - r) - (0.02 \pm 0.04)\\
i - Z &= (0.6 \pm 0.02) * (r - i)  - (0.02 \pm 0.01)
\end{align}

\noindent MEGACAM:
\begin{align}
g - r &= (0.6 \pm 0.08) * (u - g) - (0.11 \pm 0.12)\\
r - i &= (0.47 \pm 0.04) * (g - r) - (0.02 \pm 0.04)\\
i - z &= (0.61 \pm 0.02) * (r - i) - (0.01 \pm 0.01)
\end{align}

\noindent Pan-STARRS1:
\begin{align}
r - i &= (0.51 \pm 0.04) * (g - r) - (0.02 \pm 0.04)\\ 
i - z &= (0.52 \pm 0.02) * (r - i) - (0.02 \pm 0.01)\\
z - y & = (0.59 \pm 0.02) * (i - z) + 0.01
\end{align}

\noindent VIKING:
\begin{align}
&Y - J = (1.2 \pm 0.03) * (Z - Y) + (0.17 \pm 0.01)\\
&J - H = (1.25 \pm 0.12) * (Y - J) + (0.05 \pm 0.06)\\
&H - K_{s} = (0.39 \pm 0.07) * (J - H) - (0.1 \pm 0.04)
\end{align}

\noindent NGVS (AB mag):
\begin{align}
(i - Ks) &= (0.38 \pm 0.07) * (u - i) - (0.61 \pm 0.2)
\end{align}

\noindent \textit{Gaia}:
\begin{align}
G - RP &= (-0.36 \pm 0.05) * (BP - G)^2 + \\\nonumber & (1.18 \pm 0.07) * (BP - G) + (0.12 \pm 0.02)
\end{align}

\noindent 2MASS:
\begin{align}
H - K_{s} &= (0.36 \pm 0.06) * (J - H) - (0.11 \pm 0.04)
\end{align}

\section{Source selection for colour-colour plots}
\label{sec: app_color_selection}

SDSS: we used the CasJobs\footnote{\url{https://skyserver.sdss.org/CasJobs/login.aspx}} interface to query SDSS-DR15. Magnitudes and extinction corrections were obtained from the tables (``views'') Stars and Galaxy. Within the colour ranges of interest we obtained 10000 galaxies and 10000 stars.

KIDS: we used the TAPVizieR\footnote{\url{http://tapvizier.u-strasbg.fr/adql/}} online service to query the third data release of the KIDS catalog. Candidate stars and galaxies were distinguished on the basis of the parameter \textit{mClass} (null for galaxies and equal to 5 for stars). Magnitude uncertainties were selected to be in the range (0,1). Plotted colours are homogenised and extinction-corrected GAaP (Gaussian Aperture and Photometry) colours from the database. Within the colour range examined here, we obtained 10000 galaxies and approximately 8000 stars.

MEGACAM: we queried the Canada-France-Hawaii Telescope
Legacy Survey (CFHTLS) catalog\footnote{\url{http://www.cadc-ccda.hia-iha.nrc-cnrc.gc.ca/en/megapipe/cfhtls/cq.html}} selecting star and galaxy candidates from the CFHTLS Wide fields. We selected magnitudes (\textit{mag auto}) in the range [0,21] as the star/galaxy classification becomes less reliable for fainter objects, and magnitudes uncertainties were selected to be in the range [0,0.5]. The search was restricted to areas which are not masked (the keyword \textit{dubious} was set to zero), and the value of the keyword \textit{flags} was required to be no larger than 10. Candidate stars and galaxies were distinguished on the basis of the keyword \textit{class star}, which we required to be at least 0.9 for stars, and no larger than 0.1 for galaxies. Within the desired colour range we obtained 10000 galaxies and 10000 stars.

Pan-STARRS: we queried the Pan-STARRS1 Catalog Archive Server Jobs System (CasJobs\footnote{\url{http://mastweb.stsci.edu/ps1casjobs/home.aspx}}) to select candidate galaxies and stars. To separate galaxies from stars we used the PSF likelihood (in the range [-0.1,0.1] for galaxies, and in the range [0.9,1.1], in absolute value, for stars), and the empirical separation based on the discrepancy between the PSF and Kron magnitudes (iPSFMag-iKronMag > 0.05 to select galaxies, and iPSFMag-iKronMag < 0.05 to select stars, \citealt{FarrowCM14}). The search was restricted to objects with magnitude and magnitude uncertainty larger than zero. To derive colours we used Kron magnitudes for galaxies, PSF magnitudes for stars. Within the colour ranges taken into account in this paper, we obtained 10000 galaxies and about 9000 stars.
 
VIKING: we used the TAPVizieR online service to query the second data release of the Viking catalog. To distinguish stars from galaxies we used the keyword \textit{Mclass}, which takes the value of 1 for candidate galaxies and $-1$ for candidate stars. Objects were selected to have magnitude uncertainties in the range (0,0.5) for the bands Z and Y, and (0,1) for the Ks band. Colours were computed using Petrosian magnitudes for candidate galaxies and aperture magnitudes (\textit{ap3}) for candidate stars.
Within the wanted colour ranges, we selected 10000 galaxy candidates and 10000 star candidates.

\textit{Gaia}: We selected spectroscopically-confirmed stars, QSOs, and galaxies in SDSS with a counterpart in the \textit{Gaia}-DR2 catalog within a radius of 0\farcs1. The number of galaxies in the \textit{Gaia} database falling in the colour-colour region of Fig.\ref{fig: colors_sim1} is negligible. The number of stars and galaxies plotted is approximately $10^{5}$.

2MASS: Point sources and extended sources have been selected via the CasJobs interface from the table PhotoObjAll and PhotoXSC, respectively. 
Point sources have been selected to satisfy the following criteria: i) photometric quality flag \textit{ph\_qual} = AAA (that is measurements with SNR$\geq 10$ and photometric uncertainty $cmsig \leq 0.1$); ii) reduced chi-square goodness of fit for the profile-fit photometry in the range $0.7 <$ \textit{m\_psfchi} $< 1.3$, where \textit{m} represents the magnitudes \textit{j, h, k}. We derived colours from ``default magnitudes'' (\textit{rd\_flag} $= 2$) derived from profile-fitting measurements. 
Extended sources were selected from the table PhotoXSC to satisfy the following constraints: i) galaxy score \textit{g\_score}$< 1.4$; ii) confusion flag \textit{m\_flg\_i20e} = 0 (where \textit{m} represents the magnitudes \textit{j, h, k}). Within the colour ranges considered here we obtained 10000 galaxies and 10000 stars.

\section{\textsc{skymaker} configurations files}
\label{sec: con_files}
Parameters adopted in the \textsc{skymaker} configuration file used to render the Pan-STARRS images are given below. Exposure time and wavelength from \citet{SchlaflyFJ12}, mirror diameters from \citet{HodappSK04}. Following \citet{Roddier81}, the simulated exposure times are deemed long enough to require the parameter ``seeing type'' to be set equal to ``long exposure''. 
\begin{lstlisting}
--------------------------------------------

IMAGE_TYPE      SKY          
MAG_LIMITS 	15, 21.8	
PIXEL_SIZE	0.25	  # arcsec/pixel
GAIN  		1.0	  # e-/ADU
READOUT_NOISE 	10.5	  # e-
SATUR_LEVEL 	888949.1  # ADU
EXPOSURE_TIME 	40	  # sec 
MAG_ZEROPOINT 	29	  # ADU/sec
PSF_OVERSAMP 	10
SEEING_TYPE 	LONG_EXPOSURE 
SEEING_FWHM 	1.2	  # arcsec
M1_DIAMETER 	1.8	  # meters
M2_DIAMETER 	0.9	  # meters           
WAVELENGTH      0.617	  # microns
BACK_MAG 	22	  # mag/arcsec2

============================================
\end{lstlisting}

Parameters used to render the NISP \textit{J}-band images are given below. Exposure time from \citet{Carry18}, wavelength and readout noise from \citet{LaureijsAA11}, mirror diameters from the ESA \textit{Euclid} webpages. Parameters followed by an asterisk ($*$) are uncertain.
\begin{lstlisting}
--------------------------------------------

IMAGE_TYPE      SKY          
MAG_LIMITS 	17*, 24 
PIXEL_SIZE	0.3	# arcsec/pixel
GAIN  		1.0	# e-/ADU
READOUT_NOISE 	4.5	# e-, predicted 
			# upper limit
SATUR_LEVEL 	6553500 # ADU
EXPOSURE_TIME 	116	# sec 
MAG_ZEROPOINT 	29.1	# ADU/sec
PSF_OVERSAMP 	2
SEEING_TYPE 	NONE    # diffraction 
			# limited
M1_DIAMETER 	1.2	# meters
M2_DIAMETER 	0.35	# meters
ARM_COUNT       3       # number of 
			# spider-arms
ARM_THICKNESS   10      # millimiters
WAVELENGTH      1.26	# microns
BACK_MAG 	23*	# mag/arcsec2

============================================
\end{lstlisting}

\section{Additional figures}
\begin{figure*}
\begin{center}$
\begin{array}{cccc}

\includegraphics[trim= 0cm 0cm 0cm 0cm, clip=true, scale=0.25]{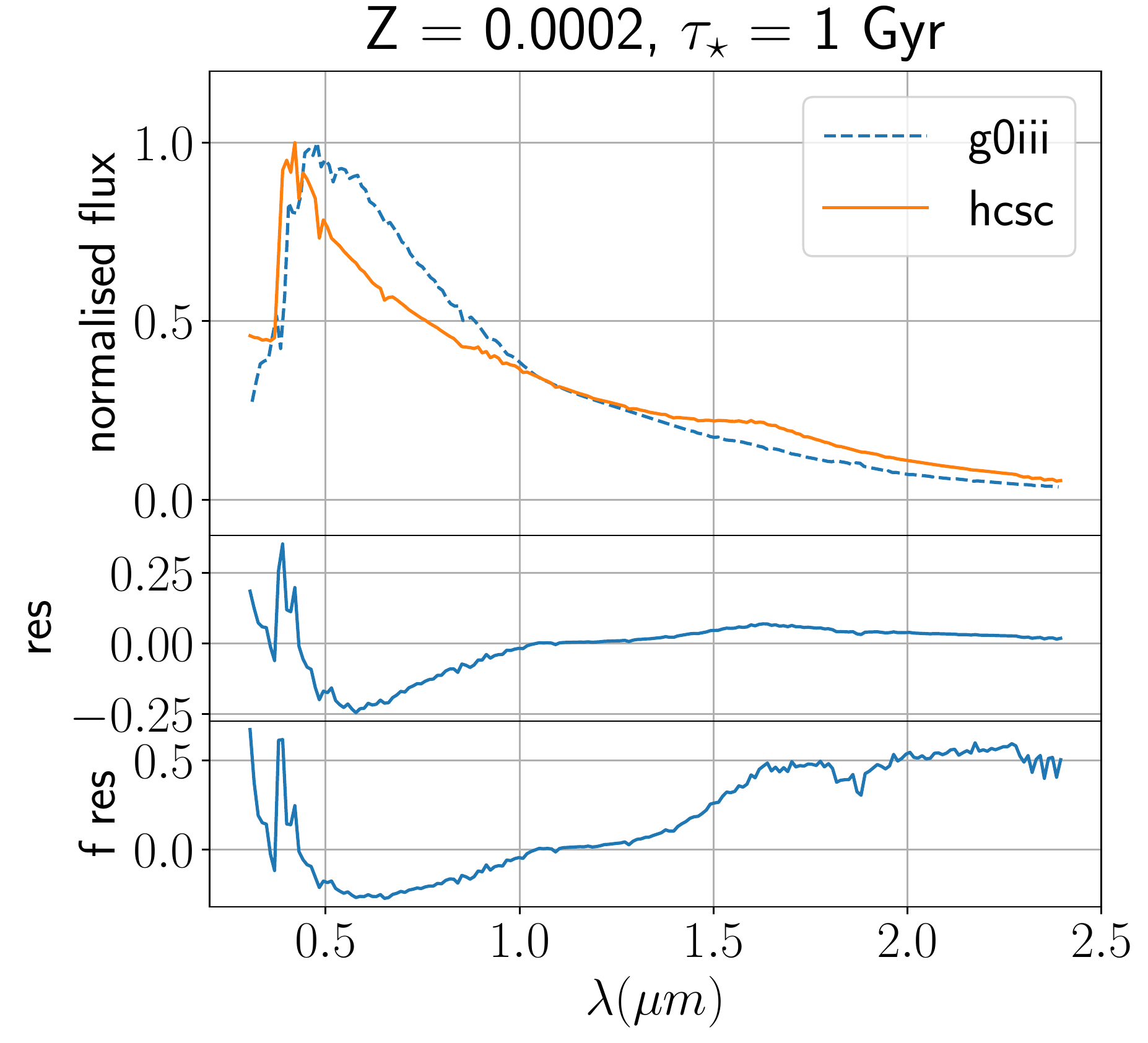}&
\includegraphics[trim= 0cm 0cm 0cm 0cm, clip=true, scale=0.25]{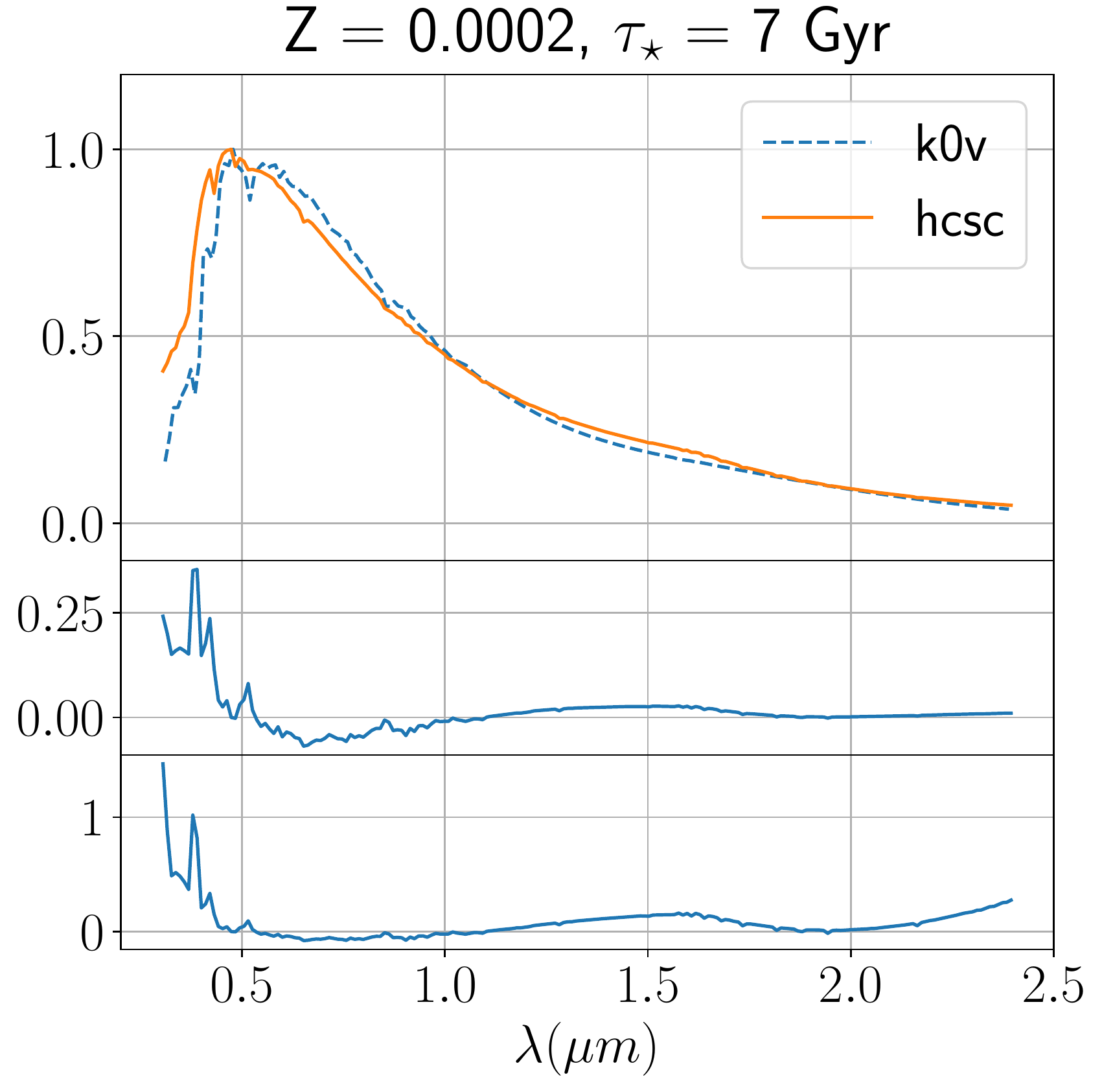}&
\includegraphics[trim= 0cm 0cm 0cm 0cm, clip=true, scale=0.25]{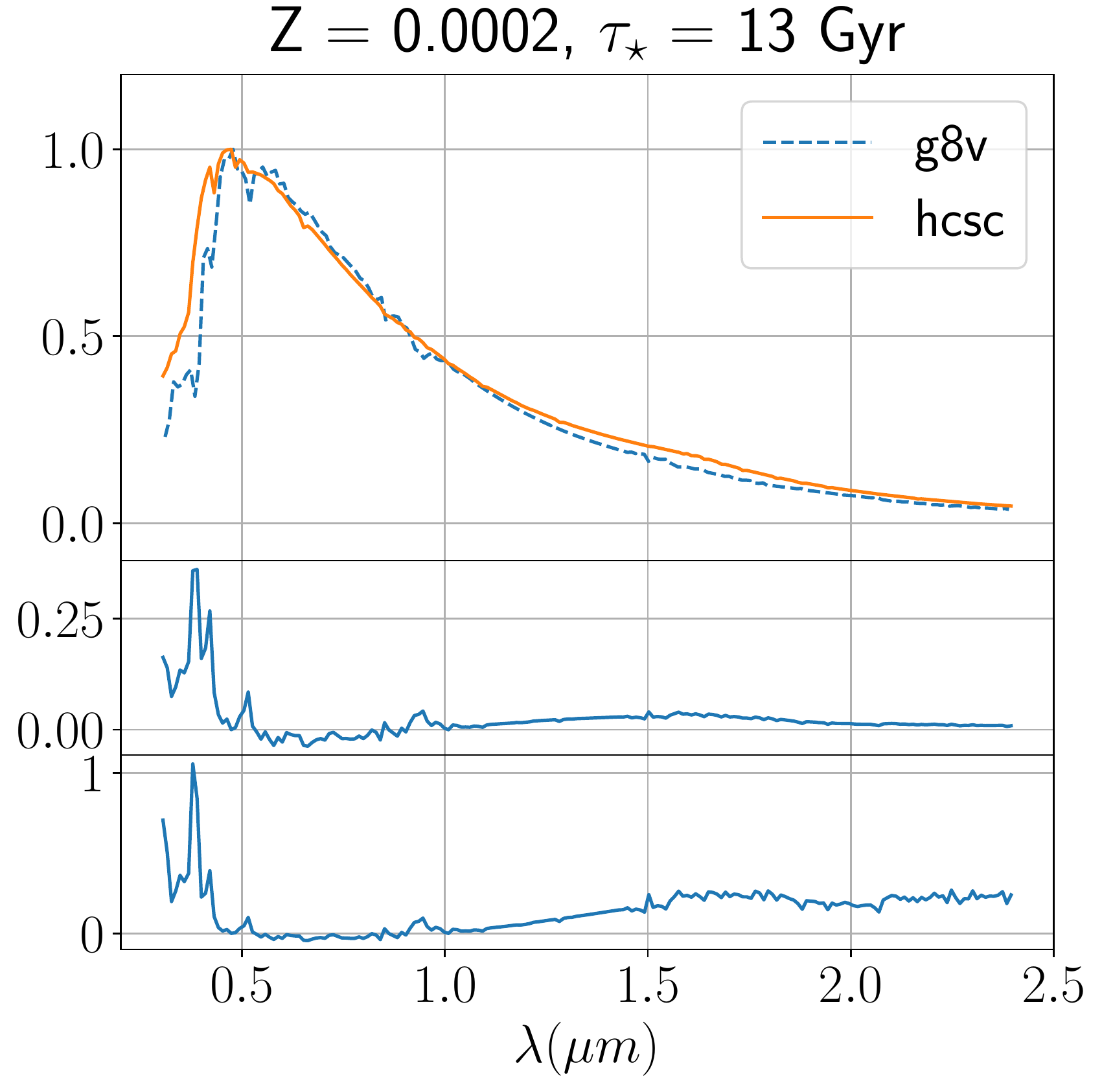}\\

\includegraphics[trim= 0cm 0cm 0cm 0cm, clip=true, scale=0.25]{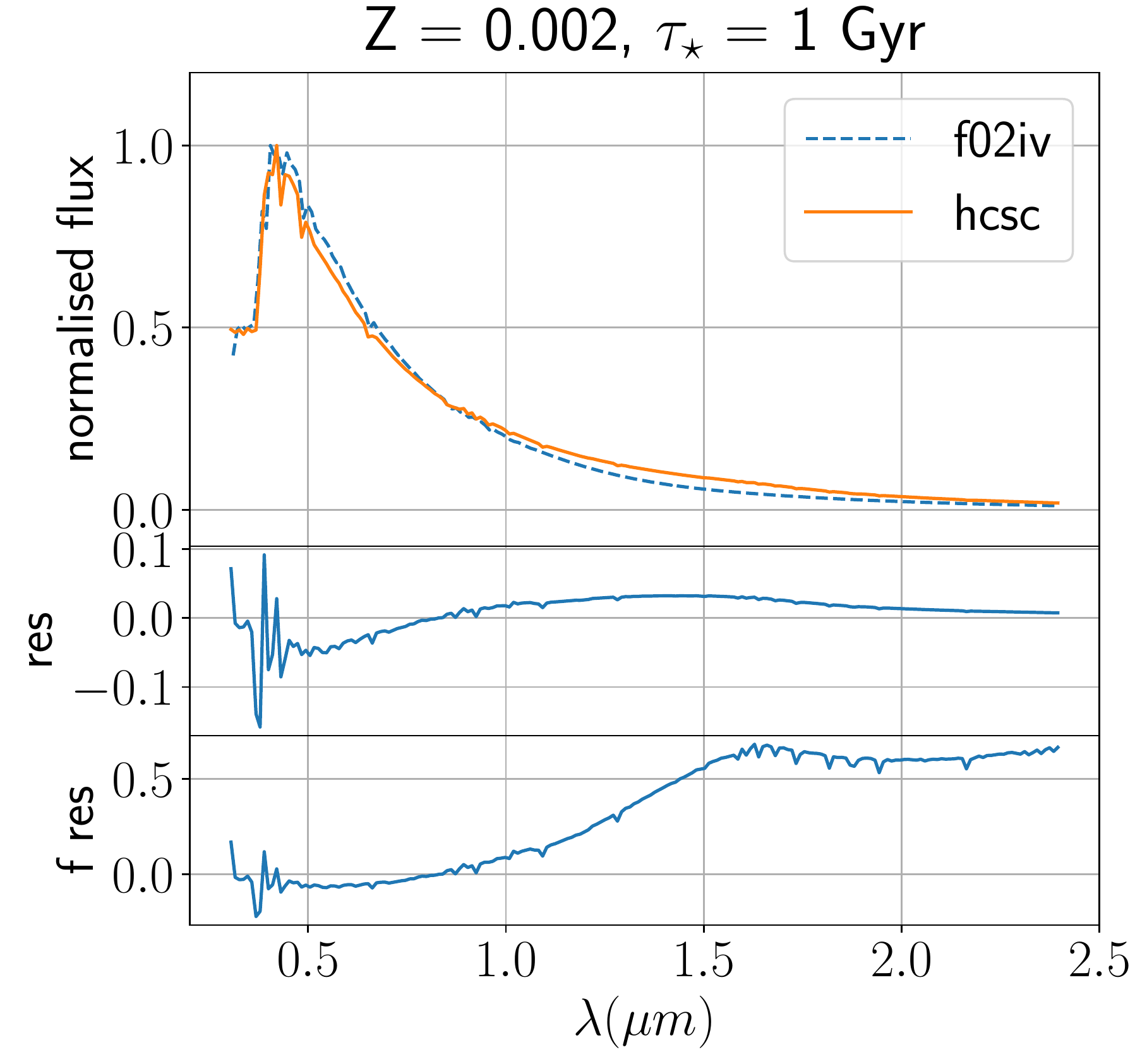}&
\includegraphics[trim= 0cm 0cm 0cm 0cm, clip=true, scale=0.25]{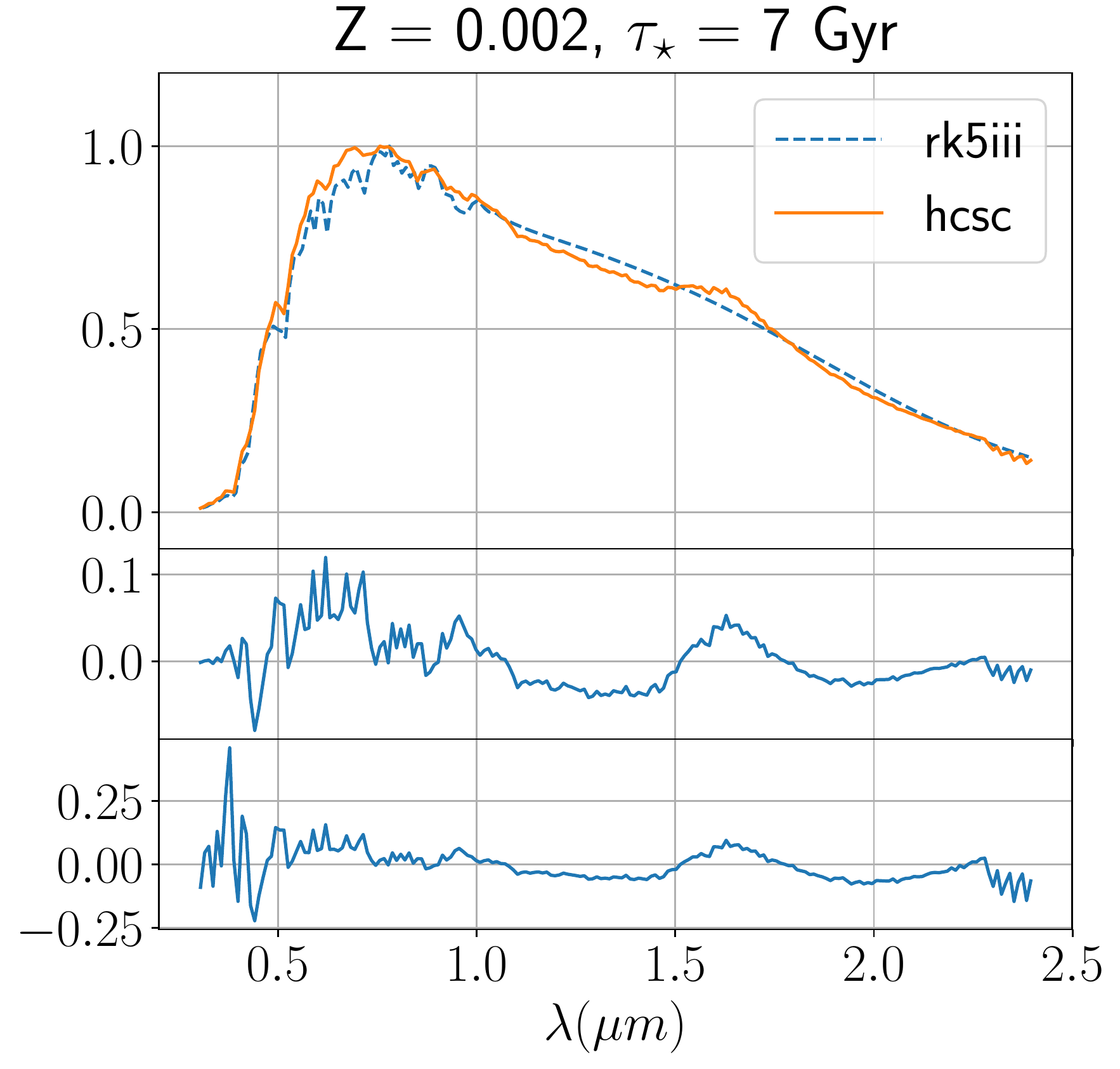}&
\includegraphics[trim= 0cm 0cm 0cm 0cm, clip=true, scale=0.25]{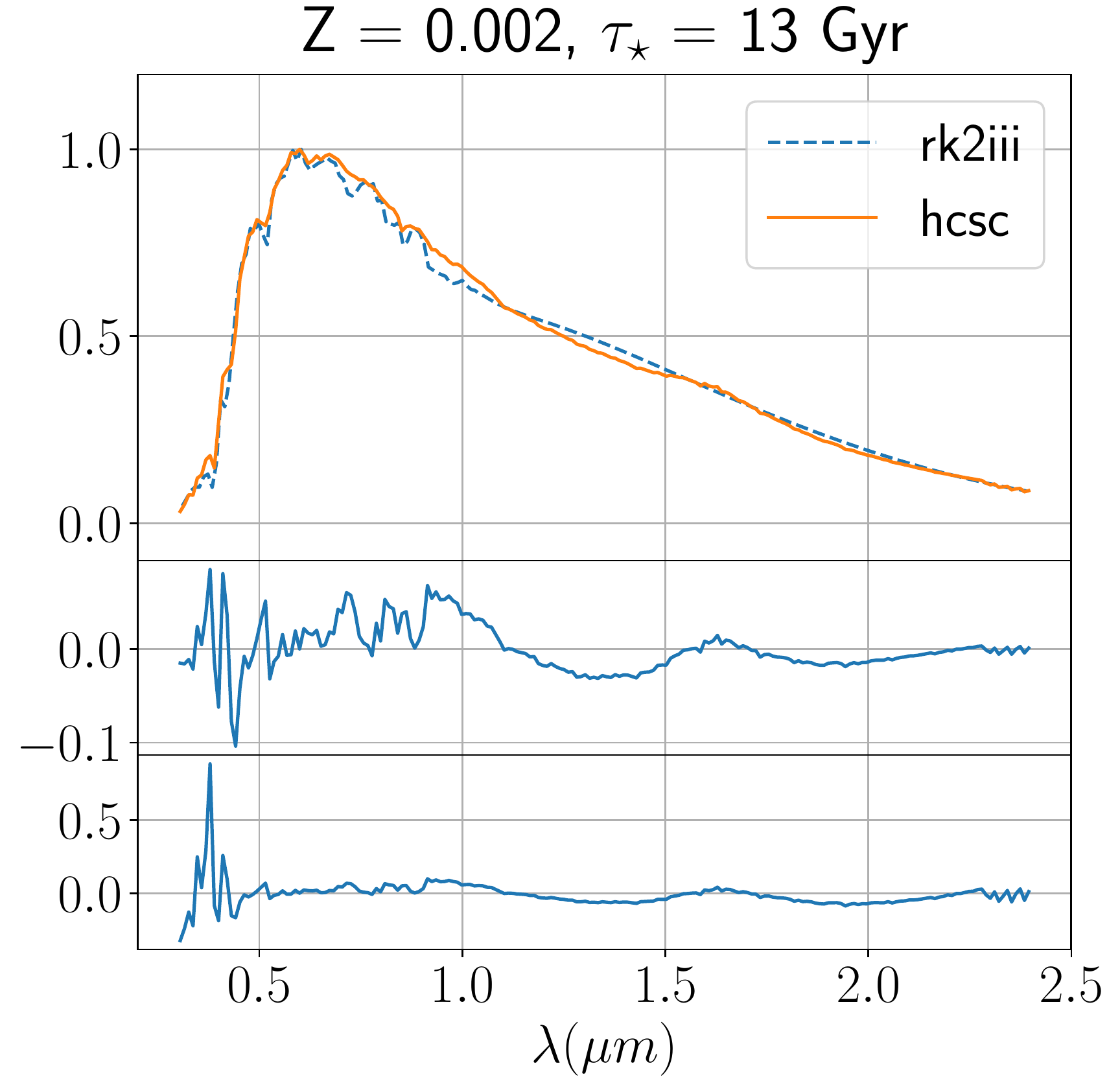}\\

\includegraphics[trim= 0cm 0cm 0cm 0cm, clip=true, scale=0.25]{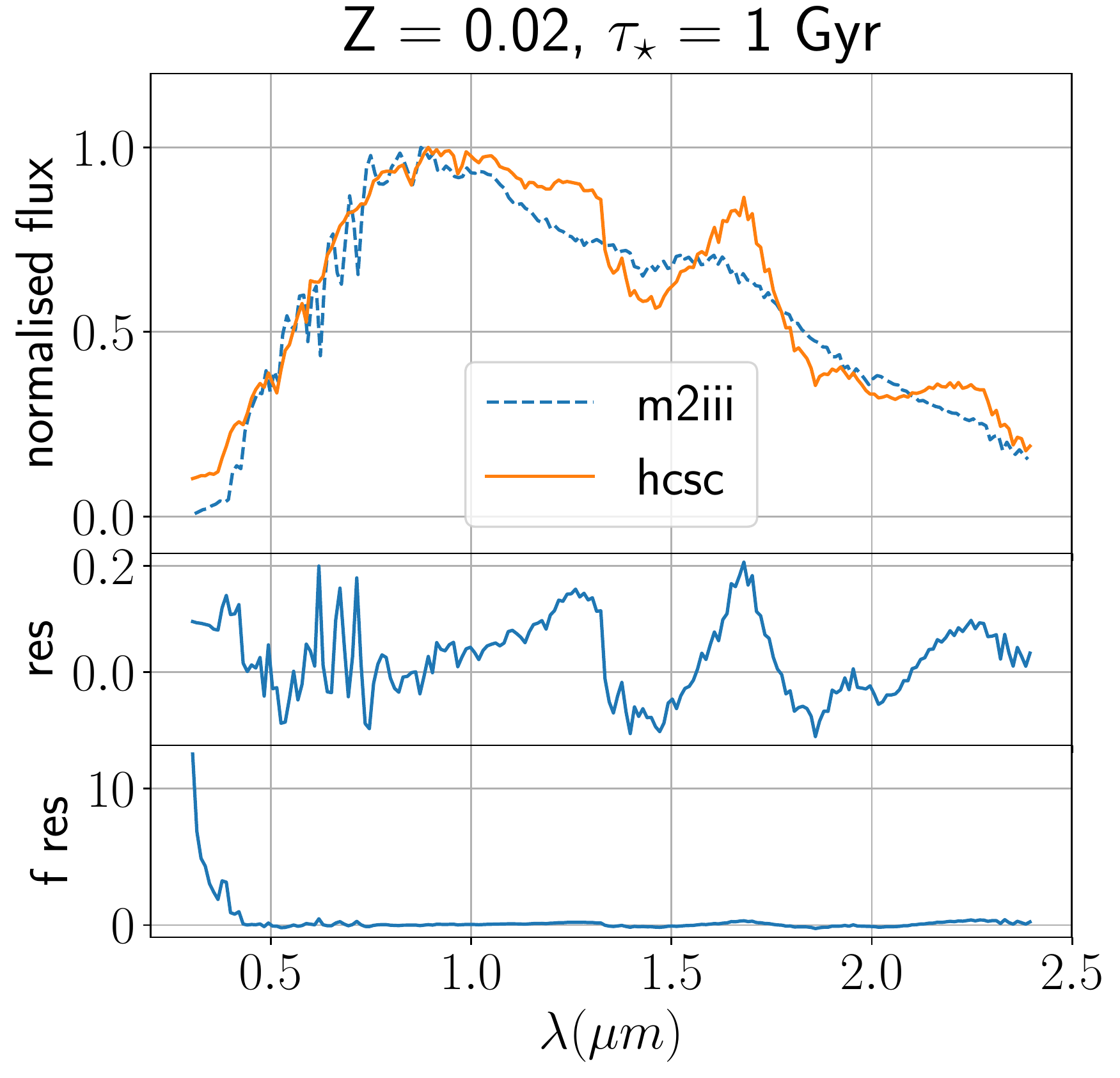}&
\includegraphics[trim= 0cm 0cm 0cm 0cm, clip=true, scale=0.25]{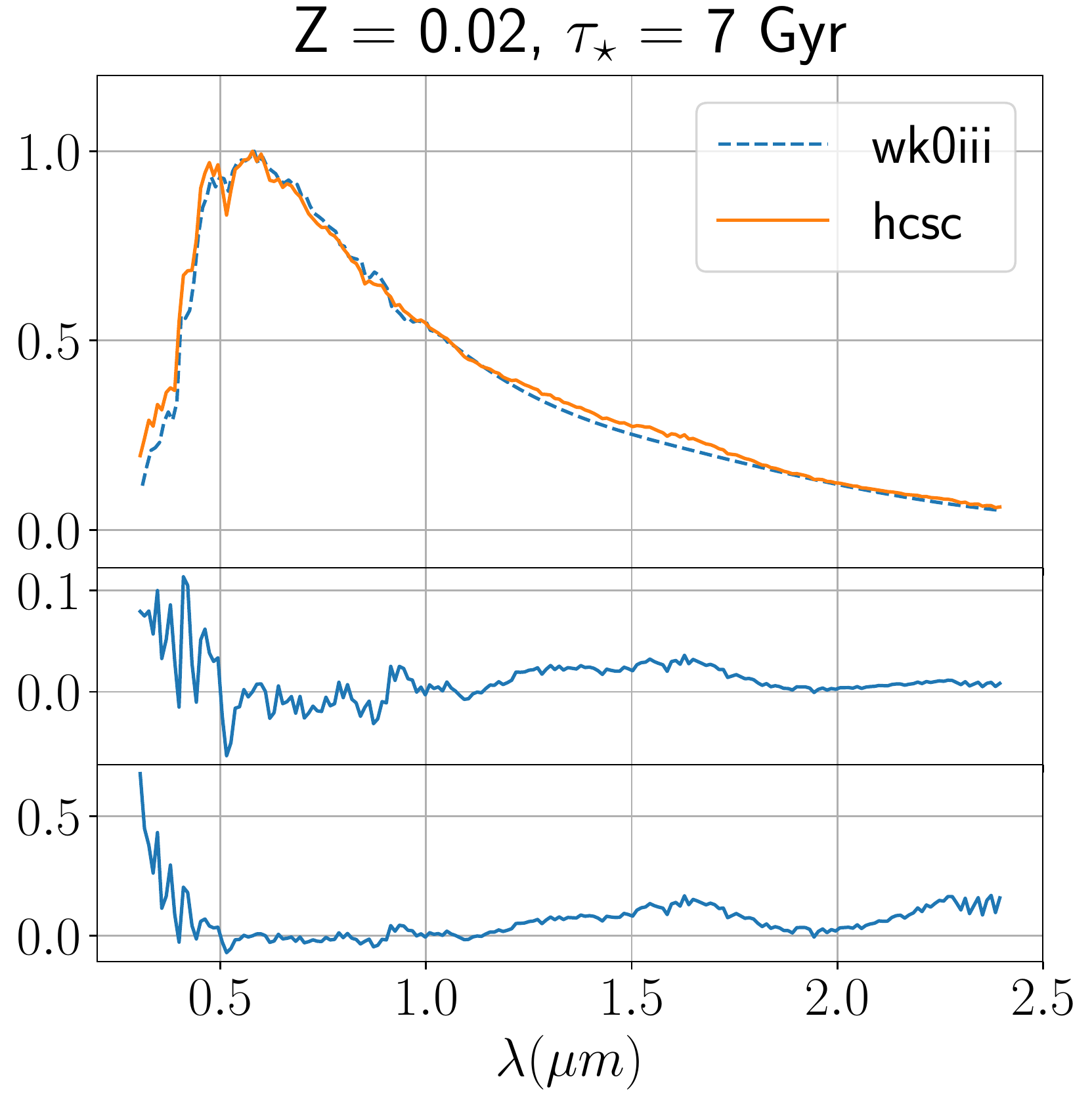}&
\includegraphics[trim= 0cm 0cm 0cm 0cm, clip=true, scale=0.25]{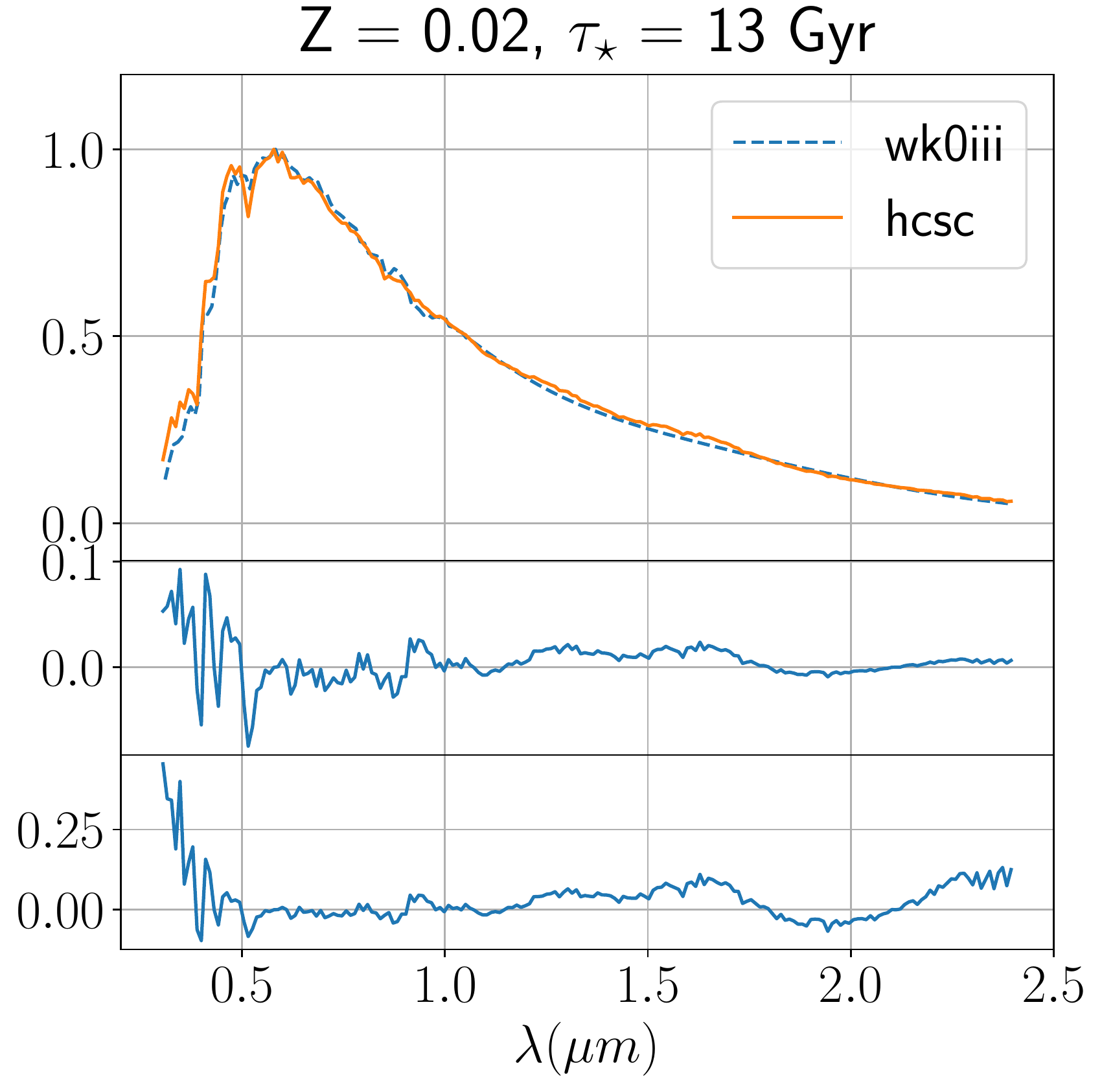}\\

\includegraphics[trim= 0cm 0cm 0cm 0cm, clip=true, scale=0.25]{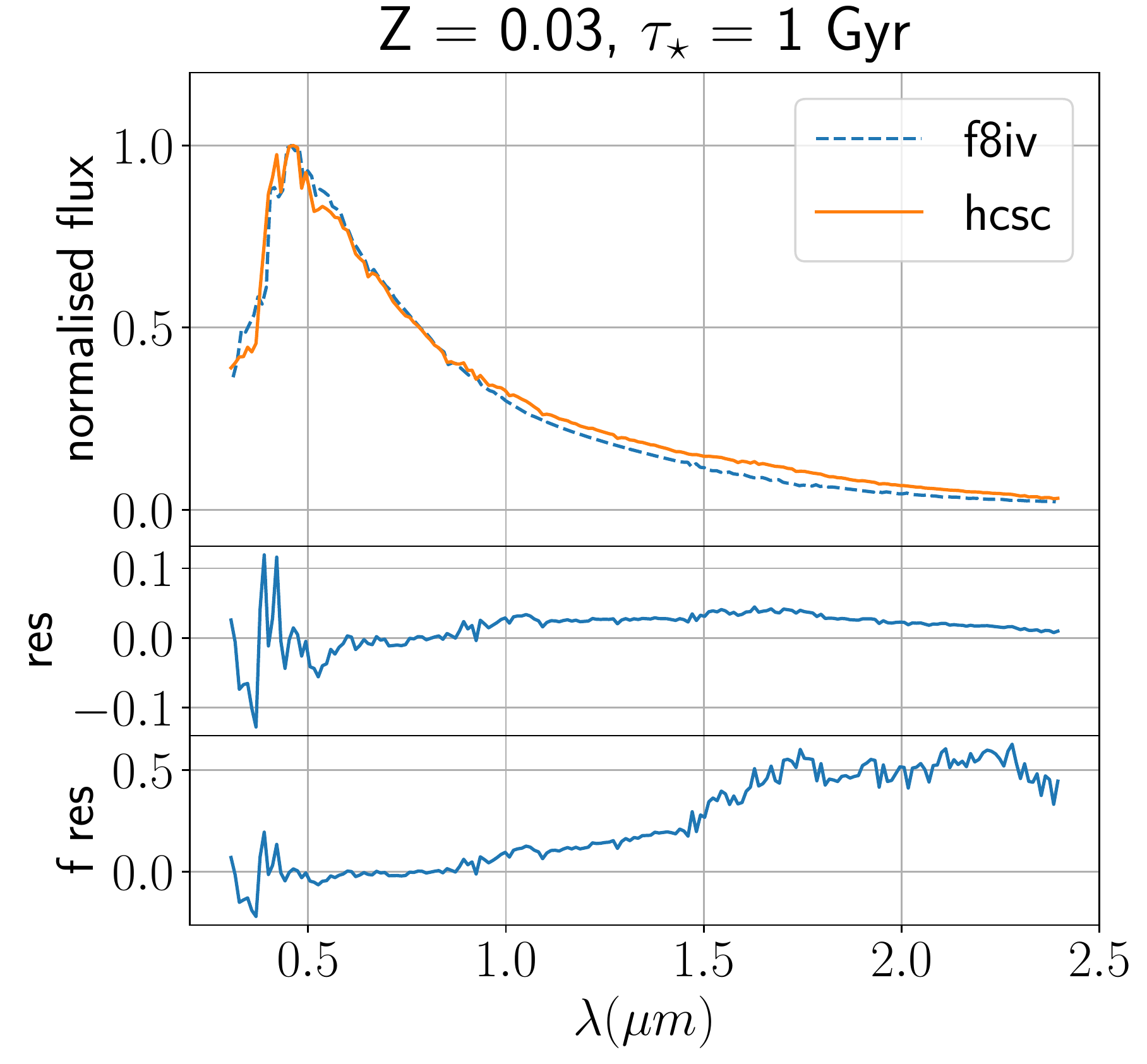}&
\includegraphics[trim= 0cm 0cm 0cm 0cm, clip=true, scale=0.25]{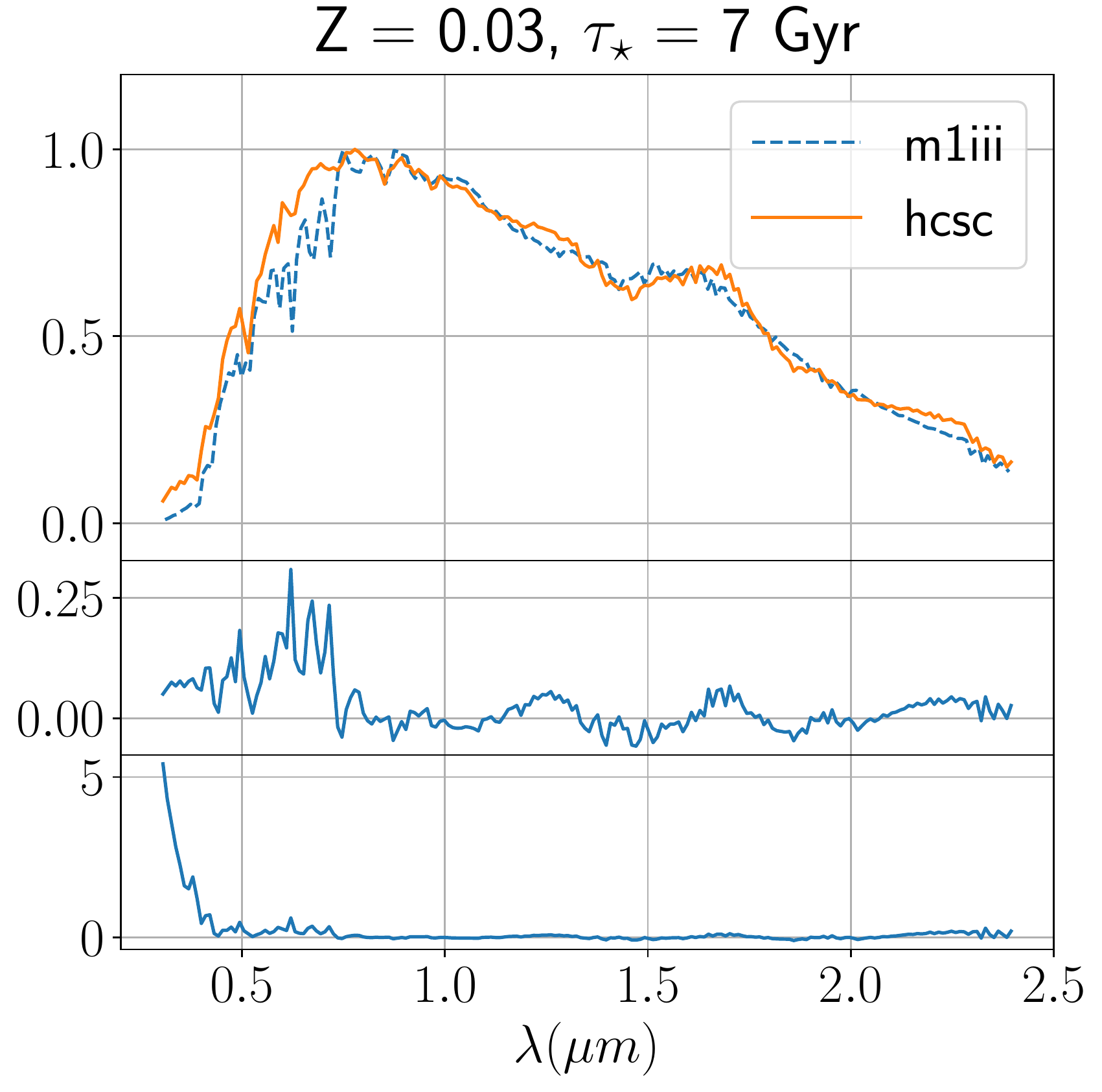}&
\includegraphics[trim= 0cm 0cm 0cm 0cm, clip=true, scale=0.25]{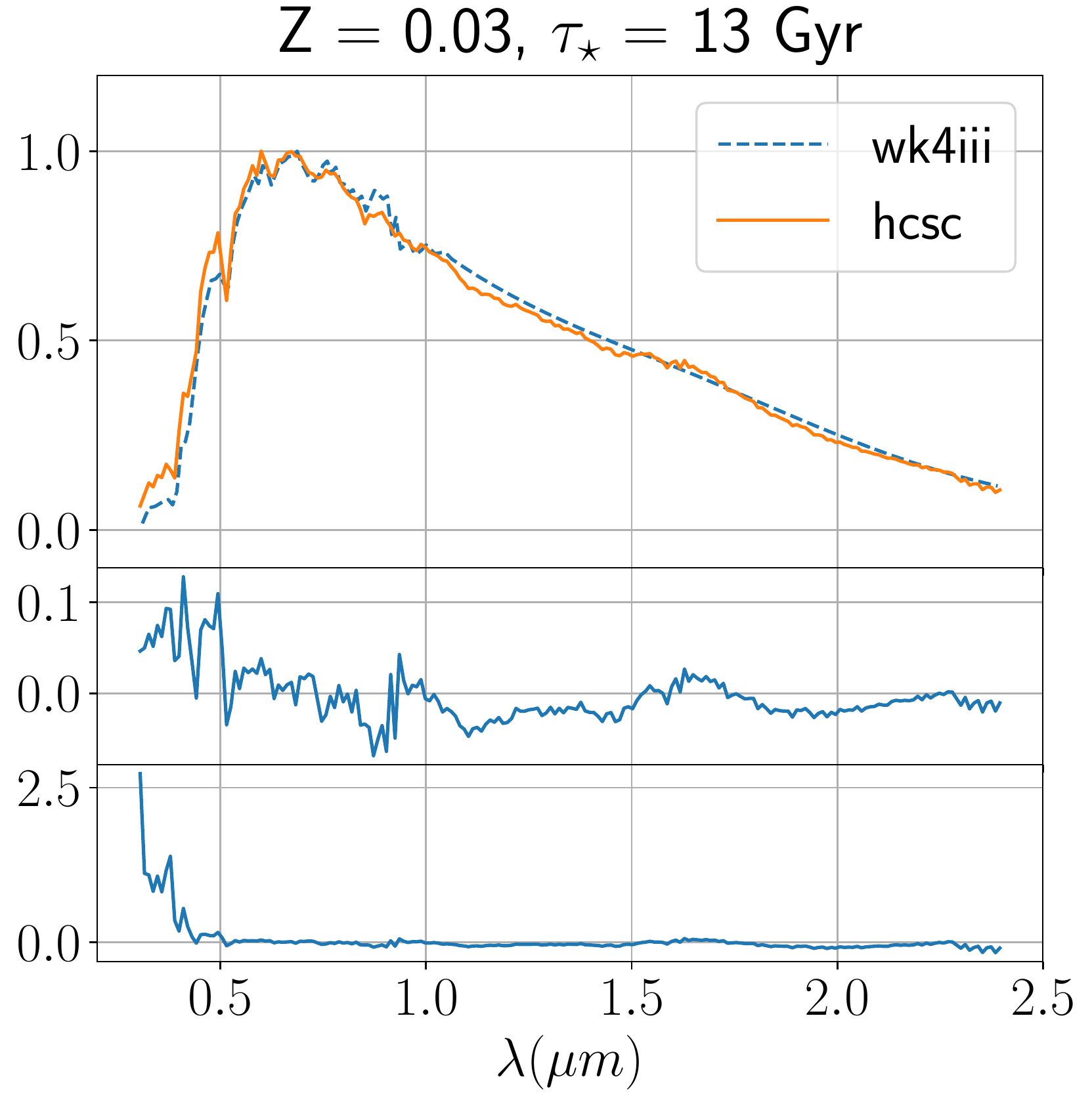}\\

\includegraphics[trim= 0cm 0cm 0cm 0cm, clip=true, scale=0.25]{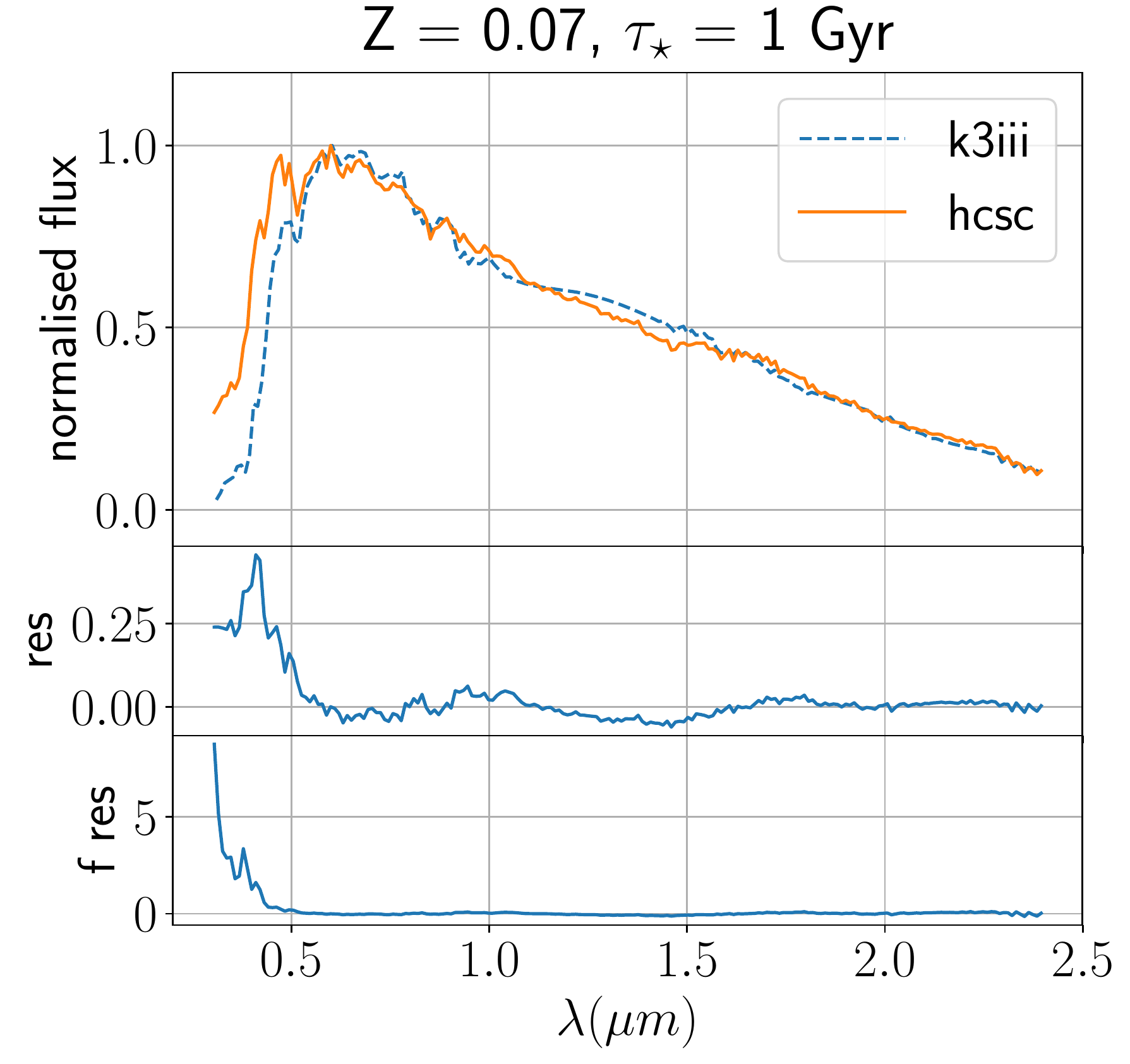}&
\includegraphics[trim= 0cm 0cm 0cm 0cm, clip=true, scale=0.25]{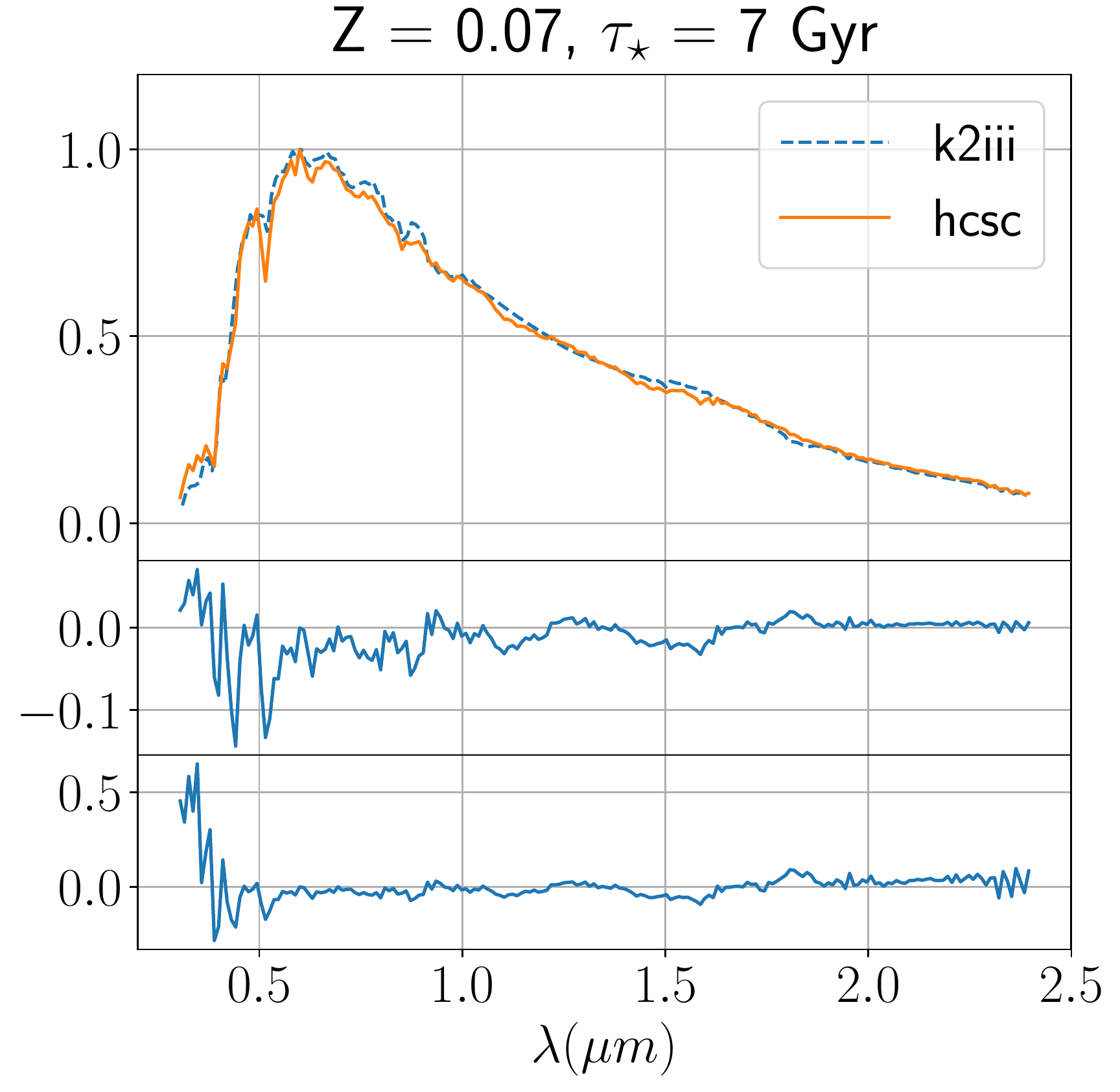}&
\includegraphics[trim= 0cm 0cm 0cm 0cm, clip=true, scale=0.25]{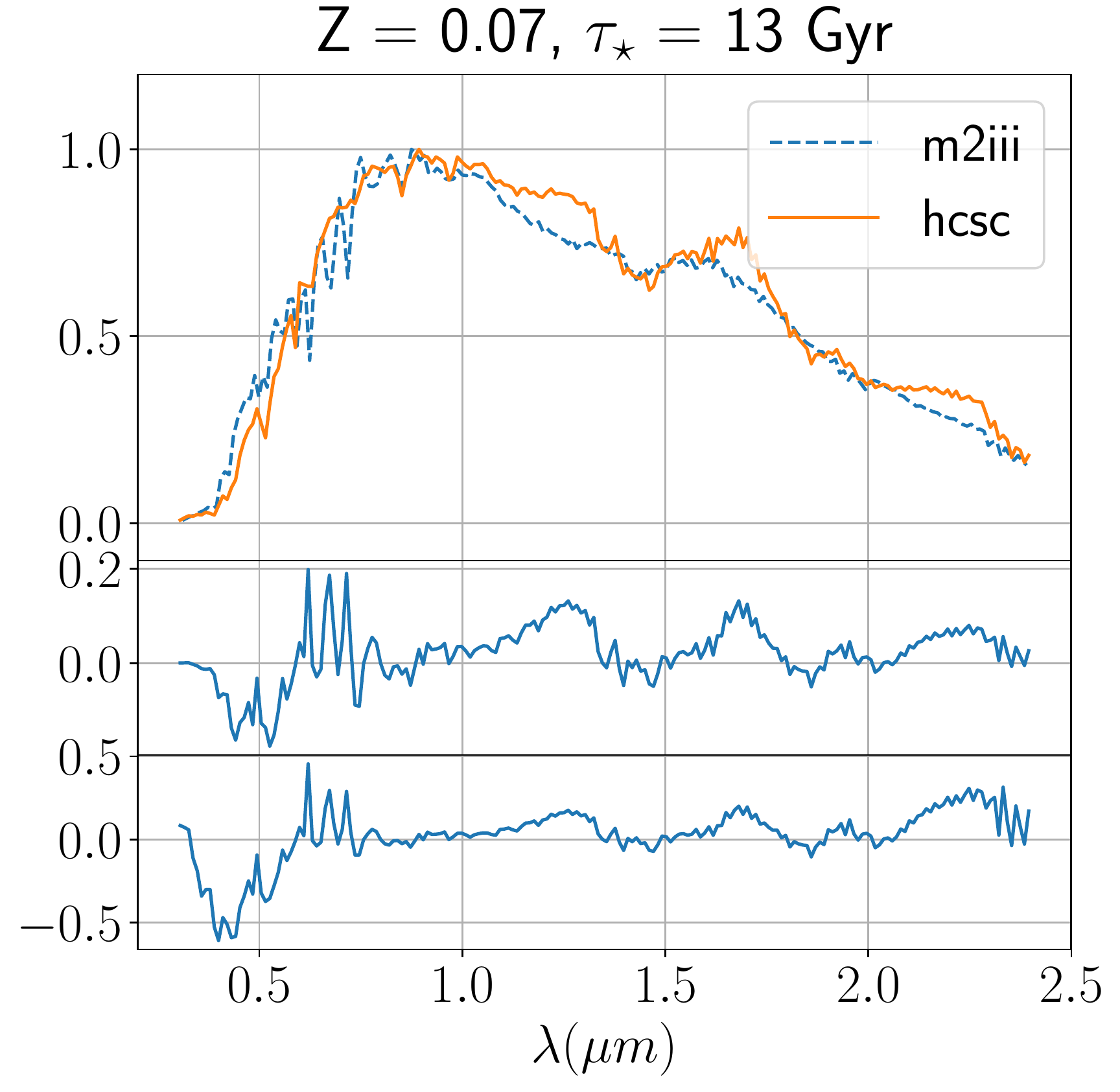}\\

\end{array}$
\end{center}
\caption{HCSC model spectra with metallicity and stellar population age as indicated in the plot title, and best-matching stellar spectra from the Pickles Atlas. Residual (model-star) and fractional residual (model-star)/star are plotted in the bottom panels. Stellar names are indicated following the scheme \textit{xxy}, with \textit{xx} the spectral type, and \textit{y} the luminosity class. The nomenclature \textit{rxxy} and \textit{wxxy} refers to metal-rich and metal-weak stars, respectively.}
\label{fig: hcss_stellar_library}
\end{figure*}

\begin{figure*}
\begin{center}$
\begin{array}{ccc}

\includegraphics[trim= 0cm 0cm 0cm 0cm, clip=true, scale=0.68]{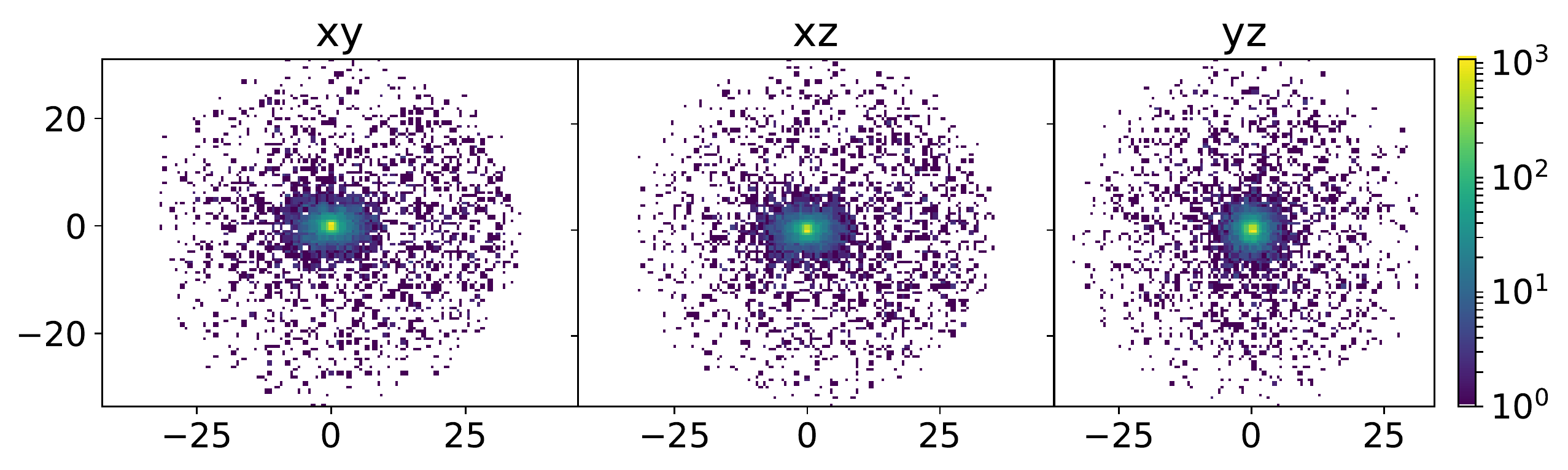} 

\end{array}$
\end{center}
\caption{Positions of stars within the HCSC as produced by \citet{MSK2009} at t = 100 $\times$ GM$_{\bullet}$/V$^{3}_{k}$; the kick was in the $-$X direction (horizontal negative values in the left and central panel) at t = 0. Units of length are GM$_{\bullet}$/V$^{2}_{k}$. The colour indicates the number of stars at a given location.}
\label{fig: hcss_nbody}
\end{figure*}

Synthetic spectra along with best-matching spectra from the Pickles library are shown in Fig.\ref{fig: hcss_stellar_library}. Positions of stars within the HCSC as derived by \citet{MSK2009} at t = 100 $\times$ GM$_{\bullet}$/V$^{3}_{k}$ are reproduced in Fig.\ref{fig: hcss_nbody}.

\bsp	
\label{lastpage}
\end{document}